\documentclass[11pt,a4paper]{article}
\usepackage{bm}
\usepackage{amsmath}
\usepackage{amsfonts}
\usepackage{amssymb}
\usepackage{bm}
\usepackage{authblk}
\usepackage{sidecap}
\usepackage[pdftex]{graphicx} 
\usepackage{fullpage}
\usepackage{subfigure}
\usepackage{wrapfig}
\usepackage{hyperref} 
\usepackage[numbers,sort&compress]{natbib}
\usepackage{hypernat}
\DeclareMathOperator{\sech}{sech}
\DeclareMathOperator{\cosech}{cosech}
\DeclareMathOperator{\Erfc}{Erfc}
\makeatletter

\newcommand{\vast}{\bBigg@{3.0}}

\newcommand{\Vast}{\bBigg@{4}}

\begin{document}
\title{Analytical Study of Giant Fluctuations and Temporal Intermittency in an Aggregation Model\\}
\author{Himani Sachdeva} 
\author{Mustansir Barma}  
\affil{Tata Institute of Fundamental Research, Homi Bhabha Road, Mumbai, India}
\date{}

\maketitle
\begin{abstract}
We study analytically giant fluctuations and temporal intermittency in a stochastic one-dimensional model with diffusion and aggregation of masses in the bulk, along with 
influx of single particles and outflux of aggregates at the boundaries. 
We calculate various static and dynamical properties of the total mass in the system for both biased and unbiased movement of particles and different boundary 
conditions. These calculations show that (i) in the unbiased case, the total mass has a non-Gaussian distribution and shows giant fluctuations which scale as system size (ii) 
in all the cases, the system shows strong intermittency in time, which is manifested in the anomalous scaling of the dynamical structure functions of the total
 mass. The results are derived by taking a continuum limit in space and agree well with numerical simulations performed on the discrete lattice.  
The analytic results obtained here are typical of the full phase of a more general model with fragmentation, which was studied earlier using numerical simulations.   
\end{abstract}

\section{Introduction}
\label{sec:sec1}
\paragraph*{}
Macroscopic observables of equilibrium systems typically have narrow distributions characterised by Gaussian fluctuations, except at a critical point. 
In non-equilibrium systems, by contrast, broad distributions and large fluctuations are fairly common \cite{dey}, and can arise in
 a variety of systems such as self-propelled particles \cite{chate,ramaswamy,narayan}, granular gases \cite{goldhirsch}, passive sliders on fluctuating interfaces \cite{das} and the 
Burgers fluid \cite{bec}.
Some of these systems also exhibit intermittency, which arises from extreme variations over short length or time scales, leading to the breakdown of self-similarity in 
space or time \cite{frisch}.    
\paragraph*{}
In this paper, we present an analytical study of a simple stochastic model which shows both giant number fluctuations and strong temporal intermittency in steady state. 
The system that we study consists of masses on a one-dimensional lattice. These masses diffuse and aggregate on contact to form larger masses which also diffuse, aggregate, 
and so on, leading to a steady transfer of mass from smaller to larger aggregates. Steady state is maintained by allowing for influx of single particles and outflux of aggregates 
at the boundaries. We choose boundary conditions that result in net transport of particles through the system, so there is a current in the steady state.
\paragraph*{}
We calculate analytically various static and dynamical properties of the system for both diffusive (unbiased) and driven (biased) movement of particles, and further
consider two different kinds of boundary conditions for the diffusive case.
These calculations establish the following results: (i) for diffusive movement of particles, the total mass i.e. the total number of particles in the system 
has a non-Gaussian probability distribution and shows \emph{giant fluctuations} proportional to system size. 
(ii) For both diffusive and driven movement, the system shows anomalous (non self-similar) dynamics, manifested most clearly in \emph{temporal intermittency} of the total mass. 
Intermittency, which is a typical feature of turbulent 
systems, is quantified in our study by calculating dynamical structure functions of the total mass, in analogy with structure functions of the velocity field in fluid turbulence \cite{frisch}.
\paragraph*{}
This model provides a simple case where anomalous scaling exponents of structure functions can be calculated analytically, a task which is generally rather
 difficult for most turbulent systems. 
Moreover, unlike other turbulent systems such as the Burgers fluid, the structure functions considered here are \emph{temporal} structure functions and are related to 
various dynamical correlation functions of the total mass in the system. Calculation of dynamical quantities can be somewhat difficult and has been done earlier only for 
the case of aggregation with spatially uniform injection \cite{majumdar2,rajesh}. 
\paragraph*{} 
Another point of interest is the relevance of the present results to other aggregation models. The simplest of these models 
 also show non-trivial features such as non-mean-field or fluctuation-dominated behaviour in lower dimensions \cite{kang}, self-organised criticality in the presence of injection 
\cite{takayasu} and turbulence-like (multi-scaling) properties of mass distributions \cite{connaughton}. The present work demonstrates a different sort of turbulence-like
 behaviour in an aggregation model, which is characterised in terms 
of temporal intermittency of the total number of particles. It also suggests the possibility of probing intermittency in other aggregation models by measuring or calculating temporal structure
 functions of particle number.
\paragraph*{}
Finally, this model represents a special limit of a more general model which also allows for fragmentation of single particles from aggregates \cite{sachdeva}. The general model was studied 
using numerical simulations and  found to undergo a phase transition from a normal phase to an aggregation-dominated phase characterised by giant fluctuations and temporal intermittency
as the fragmentation rate is decreased \cite{sachdeva}. The zero fragmentation limit of the general model analysed in this paper is an instructive limit, as it exhibits 
all the features which characterise the full aggregation-dominated phase, but is also simple enough to be analytically tractable. The analytical approach used in this paper 
and some of the results for the zero-fragmentation limit were briefly outlined in \cite{sachdeva}.
\paragraph*{}
The rest of the paper is organised as follows: in sec. \ref{sec:sec2.1} we define the general model and the various limits  of the model that arise on considering different
 types of bias and boundary conditions. Section \ref{sec:sec2.2} summarises the main results
 of the paper and sec. \ref{sec:sec2.3} contains a brief discussion of related models. In sections \ref{sec:sec3} and \ref{sec:sec4}, we present detailed calculations of static
 and dynamical properties for the case with diffusive movement of particles for two different boundary conditions. Section \ref{sec:sec5} contains the analysis of the case with driven
 (biased) movement of particles. In sec. \ref{sec:sec6}, we conclude with a discussion of some issues and open questions.
\section{Model, results and related models}
\label{sec:sec2}
\subsection{Model}
\label{sec:sec2.1}
We study a one-dimensional open system with diffusion and aggregation of particles in the bulk and influx and outflux at the boundaries. The model is defined as follows:
Let $m_{i}$ denote the number of particles on site $i$. Starting with an empty lattice of $L$ sites at $t=0$, a site $i$ is chosen at random, and one of the following moves occurs 
(see Fig. 1):
\begin{figure} [h]
\centering
{\includegraphics[width=0.7\textwidth]{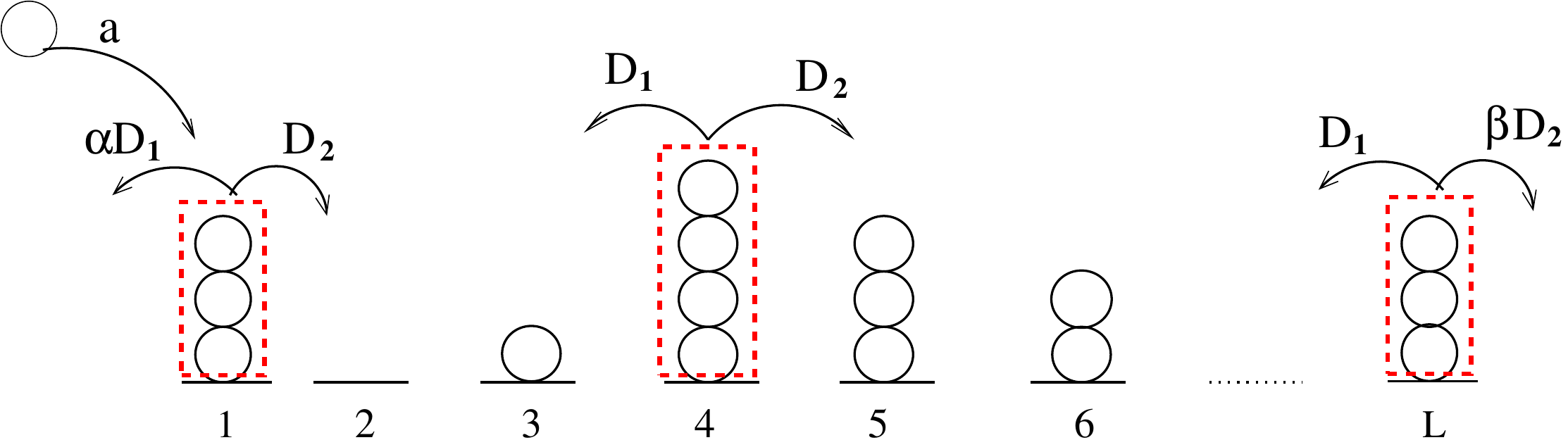}
\label{model}}
\caption {Illustration of allowed moves in the model: A unit mass can be injected into the system at site $1$ with rate $a$. The entire mass at a site can move as a whole 
to the right or left with rates $D_{2}$ and $D_{1}$ respectively. The entire mass at sites $L$ or $1$ can hop out of the system with rates $\beta D_{2}$ and $\alpha D_{1}$ respectively.}
\end{figure}
\begin{enumerate}
\renewcommand{\theenumi}{\roman{enumi}}
\item Influx: A single particle of unit mass is injected at the first site ($i=1$)  at rate $a$: $m_{1}\rightarrow m_{1}+1$.
\item Diffusion and aggregation: With rate $D_{1}$ (or $D_{2}$ ), the full stack on site i (i.e., all particles on the site collectively) hops to site $i-1$ (or $i+1$) and adds to the mass
 already there:  $m_{i\pm 1}\rightarrow m_{i\pm 1}+m_{i}$,  $m_{i}\rightarrow 0$. 
\item Outflux of mass from boundaries: With rate $\beta D_{2}$ (or $\alpha D_{1}$), the entire mass at site $L$ (or site $1$) exits the system: $m_{L}\rightarrow 0$ (or $m_{1}\rightarrow 0$).
\end{enumerate}
\paragraph*{}
The model has been defined with general asymmetric diffusion rates $D_{1}$ and $D_{2}$ to the left and right. 
However, in this paper, we consider only the two extreme cases $D_{1}=D_{2}=D$ and $D_{1}=0, D_{2}=D$. These correspond respectively to  purely diffusive (unbiased) movement
 and purely driven (fully biased) movement of aggregates. 
The behaviour of the system with both drive and diffusion is expected to be qualitatively similar to that of the fully driven case.
The other source of variation in this model comes from the parameters $\alpha$ and $\beta$ which control the outflux of aggregates from the two boundaries. 
In the interest of simplicity, we consider only two special but representative scenarios: one in which outflux occurs only from site $L$ (corresponding to $\alpha=0, \beta=1$) and the 
other where outflux occurs from both site $1$ and $L$ at the same rate (corresponding to $\alpha=1, \beta=1$). The purpose of considering these two cases is to explore
the effect of having influx and outflux at the same boundary as opposed to spatially separated influx and outflux. These two variations give rise to the 
following three cases:
\begin{enumerate}
\renewcommand{\theenumi}{\Alph{enumi}}
\item Diffusive movement, outflux only from right boundary: $D_{1}=D_{2}=D$ and $\alpha=0, \beta=1$.
\item Diffusive movement, outflux from both boundaries: $D_{1}=D_{2}=D$ and $\alpha=1, \beta=1$.
\item Driven movement, outflux from right boundary: $D_{1}=0, D_{2}=D$ and $\beta=1$.
\end{enumerate}
In all three cases, the influx is only at the left boundary. Each of these cases is analysed separately in sections \ref{sec:sec3}-\ref{sec:sec5}.
\paragraph*{}
The outflux of aggregates of all sizes from the system boundaries allows the total number of particles in the system
(or any part of the system) to attain a stationary i.e. time-independent distribution at long enough times. This is quite different from systems with only influx where the total
 number of particles keeps growing indefinitely \cite{takayasu}. In fact, for our system, the total number of particles $M=\sum\limits_{i=1}^{L}m_{i}$ (henceforth referred to as the total
 mass) turns out to have interesting static and dynamical properties in the stationary state. In particular, we calculate:
\begin{enumerate}
 \item The probability distribution $P(M)$ of the total mass $M$ in the system in steady state. Equivalently, one can specify the distribution by the moments
$\langle M^{n}\rangle$. 
 \item Dynamical structure functions:
\begin{equation}
 S_{n}(t)=\langle [M(t)-M(0)]^{n}\rangle
\label{eq2.1.1}
\end{equation}
which monitor time correlations of the total mass $M(t)$ at time $t$. Here $t=0$ is any arbitrary time instant after the system has attained steady state. 
\end{enumerate}
The motivation for analysing structure functions comes from studies of turbulent systems where they are used to quantify the breakdown of self-similarity due to intermittency.
 Intermittency, here, refers to a behaviour of a system characterised by periods of quiescence (little or no activity) interspersed with extreme changes over very short time scales \cite{frisch}.
For time scales shorter than a typical time-scale $\tau$ (which characterises the lifetime of the largest structures in the system), the statistical properties of intermittent signals are dominated 
by these extreme events.
This is also reflected in the small $t$ behaviour of structure functions defined in eq. \eqref{eq2.1.1}: for self-similar signals, $S_{n}(t)$ scales as $S_{n}(t) \sim t^{\gamma n}$ as
 $t/\tau \rightarrow 0$, where $\gamma$ is a constant; structure functions of intermittent signals deviate from this scaling form. 
Other useful measures of intermittency are the flatness $\kappa(t)$ and hyperflatness $h(t)$ , defined respectively as \cite{frisch}:
\begin{subequations}
\begin{equation}
\kappa(t)=S_{4}(t)/S^{2}_{2}(t)  
\label{eq2.1.2a}
\end{equation}
\begin{equation}
h(t)=S_{6}(t)/S^{3}_{2}(t)  
\label{eq2.1.2b}
\end{equation}
\end{subequations}
Note that $\kappa(t)$ is just the kurtosis (upto an additive constant) of the time-dependent probability distribution of $\Delta M(t)$. For intermittent signals, $\kappa(t)$ and $h(t)$
diverge as $t/\tau\rightarrow 0$.
\subsection{A brief statement of the main results}
\label{sec:sec2.2}
We now summarise the main results derived in this paper for the static and dynamical properties of total mass $M$ in the three cases (A)-(C):   
\begin{enumerate}
\renewcommand{\theenumi}{(\Alph{enumi})}
 \item \emph{Diffusive movement, outflux only from right boundary}:\\
\emph{Statics:} For injection rates that scale as $1/L$ with system size $L$, we explicitly calculate the distribution $P(M)$ and show that it has
 a non-Gaussian tail of the form $P(M)\sim (1/M_{0})\exp(-M/M_{0})$ where $M_{0} \propto L$. In keeping with this, the total mass has `giant' root mean square (rms) 
fluctuations $\Delta M$ that scale as $\Delta M\propto L$, implying that $\Delta M/\langle M\rangle$ is finite even as $L\rightarrow \infty$. Further, the mass in any 
part of the system also shows a similar non-Gaussian distribution tail and giant rms fluctuations.\\ 
\emph{Dynamics:} We calculate the structure functions $S_{n}(t)$ [defined in eq. \eqref{eq2.1.1}]
 for $n=2,3,4$ and find that they obey the scaling form: $S_{n}(t)=L^{n}\ensuremath{\mathcal{F}}_{n}(t/L^{2})$. The functions $\ensuremath{\mathcal{F}}_{n}(x)$ are
 proportional to $x$ for $x\ll 1$ independently of $n$, thus
exhibiting the extreme anomalous scaling associated with strong intermittency. The flatness [as defined in eq. \eqref{eq2.1.2a}] also diverges as $t\rightarrow 0$, 
with the divergence becoming stronger for larger $L$.
\item \emph{Diffusive movement, outflux from both boundaries:}\\
\emph{Statics:} In this case, we are not able to calculate the full distribution, but instead obtain various moments of the total mass and also of mass 
in any fraction of the lattice. This model, with $\ensuremath{\mathcal{O}\!(1)}$ injection rates, shows the same scaling of mass moments with $L$ as in case (A): in particular, the rms fluctuations are giant, scaling as  $L$ rather 
than $\sqrt{L}$ with system size.\\ 
\emph{Dynamics:}
We explicitly calculate $S_{2}(t)$ and show that it behaves as
$S_{2}(t)\propto-L^{2}[\frac{t}{L^{2}}\log\left(\frac{A_{1}Dt}{L^{2}}\right)]$ (where $A_{1}$ is a numerical constant) as $t\rightarrow 0$.
In this case also, structure functions show the anomalous time-dependence associated with intermittency. However, their functional form for small $t$ is different from 
that in case (A) due to the presence of additional multiplicative log terms.
\item \emph{Driven movement, outflux from right boundary:}\\
\emph{Statics:} In this case, the total system mass has a Gaussian distribution and rms fluctuations that scale as $\Delta M \propto \sqrt{L}$. \\
\emph{Dynamics:} We calculate dynamical structure functions $S_{n}(t)$ for $n=2,3,4$ and find that at short times, 
$t\ll \sqrt{L}$, they scale as $S_{n}(t) \sim L^{n/2}\ensuremath{\mathcal{G}}_{n}\left(\frac{Dt}{\sqrt{L}}\right)$ where 
$\ensuremath{\mathcal{G}}_{n}(x)\sim x$ as $x\rightarrow 0$. Thus, this system also shows strong intermittency, but at much shorter [$\ensuremath{\mathcal{O}\!(\sqrt{L})}$] time scales than the 
diffusive systems.
\end{enumerate}
The calculations for cases (A) and (B) are performed in the continuum approximation, wherein we replace recursion relations involving discrete lattice sites
with differential equations in a continuous spatial coordinate and retain only the lowest order derivatives in the differential equations. 
We check that this approximation is self-consistent, and also find that the resulting expressions for various quantities are in excellent agreement with numerics for the discrete lattice.
In case (C), where the continuum approximation is not good, we work with exact recursion relations.
\begin{figure}[h]
\subfigure[] {
\centering
\includegraphics[width=0.3\textwidth]{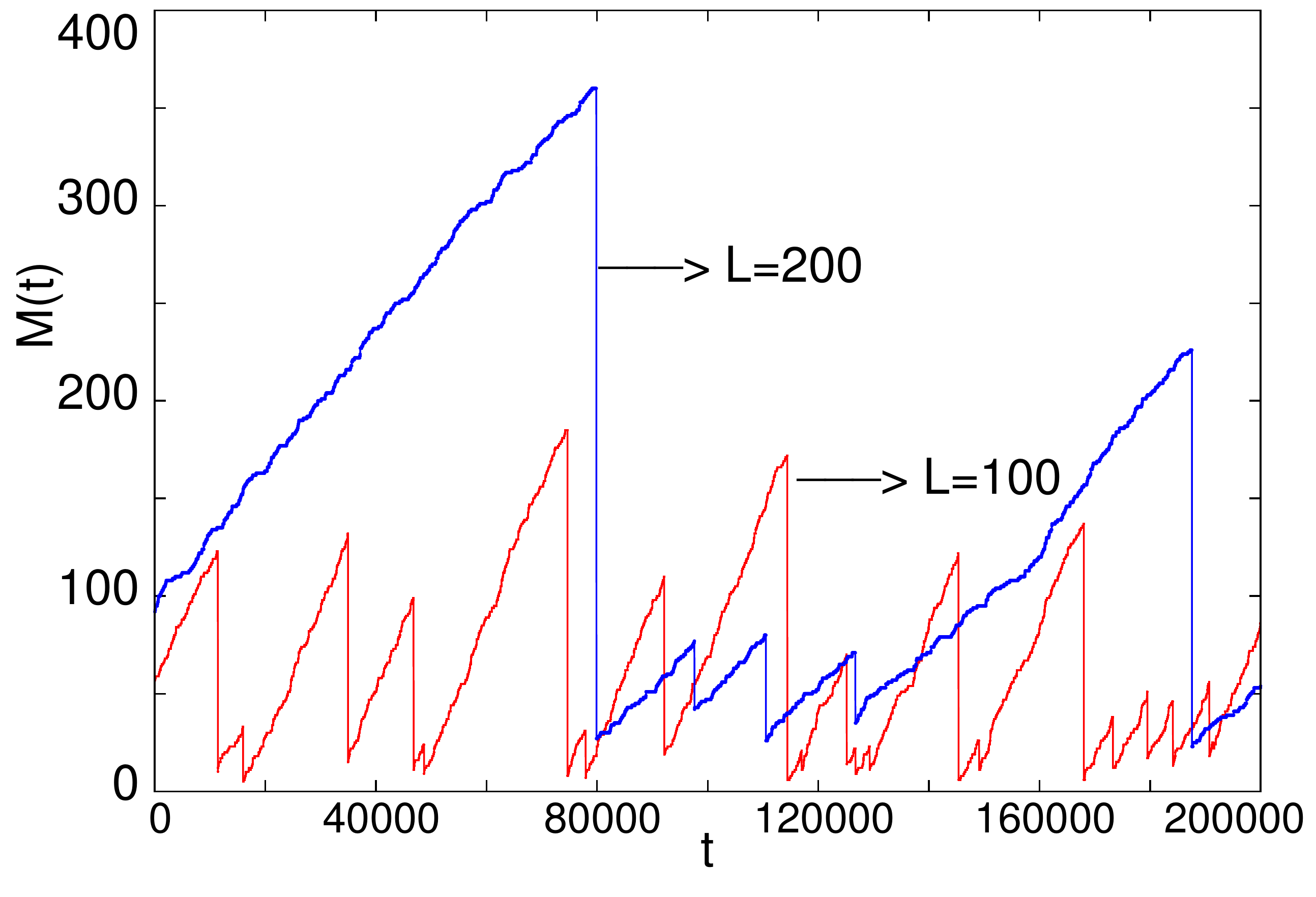}
\label{fig2a}
}
\subfigure[]{
\includegraphics[width=0.3\textwidth]{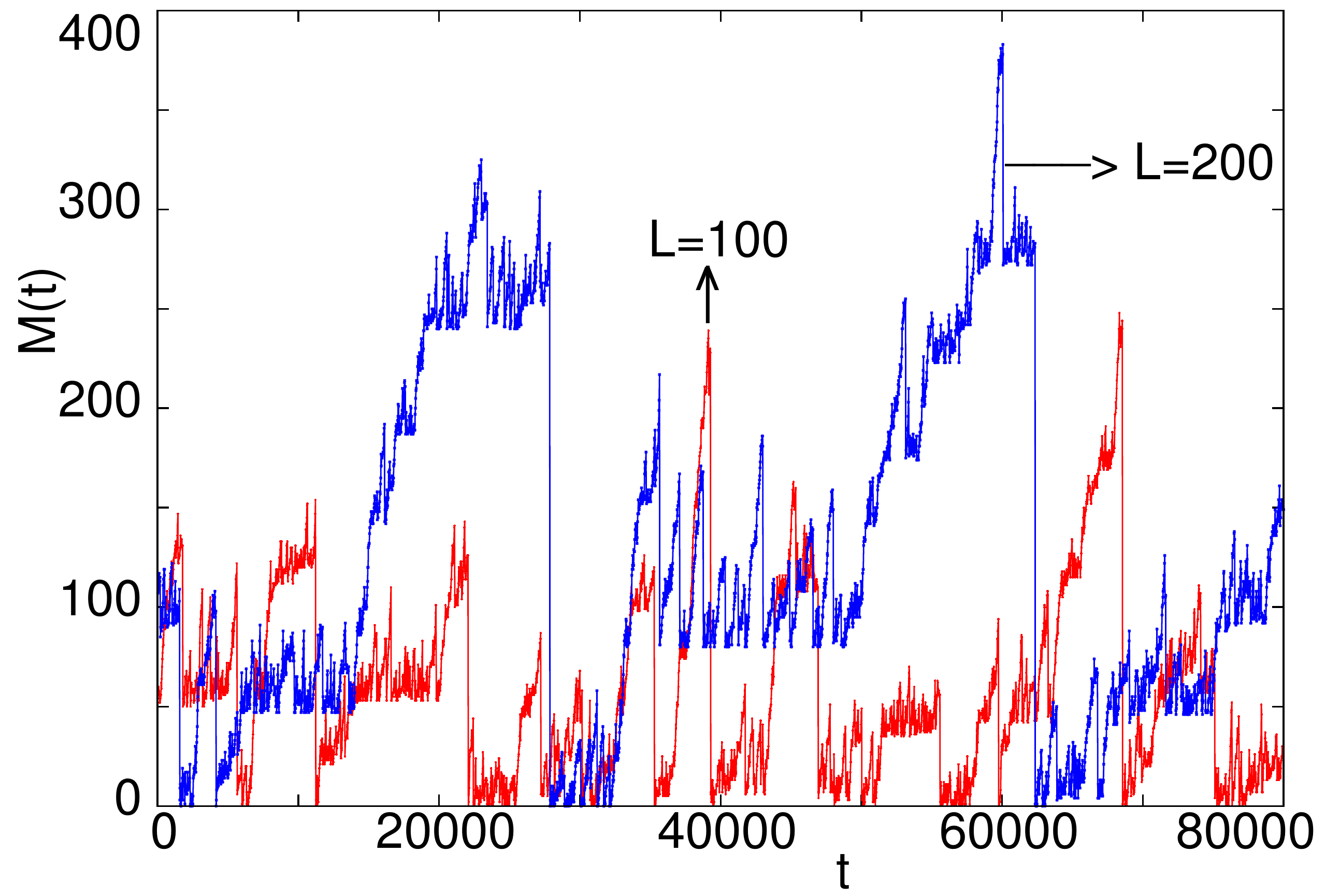}
\label{fig2b}}
\subfigure[]{
\includegraphics[width=0.3\textwidth]{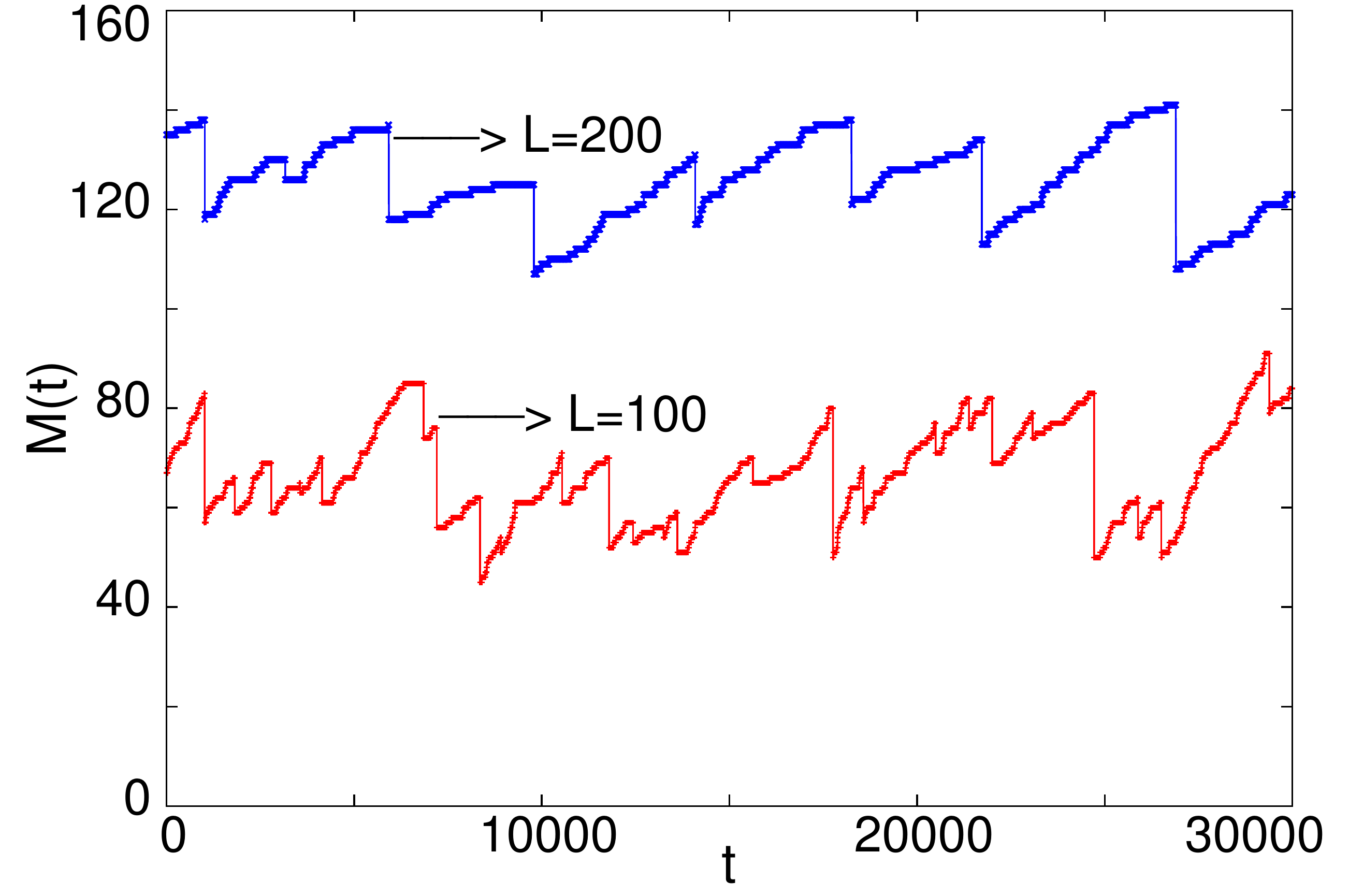}
\label{fig2c}}
\caption{ A typical realisation of $M(t)$ vs. $t$ for two different system sizes $L$ for (a) Case A: diffusive movement; outflux only from right boundary 
(b) Case B: diffusive movement; outflux from both boundaries (c) Case C: driven movement; outflux from right boundary. Sharp drops in $M(t)$ dominate the small $t$ 
behaviour of structure functions in all three cases, giving rise to temporal intermittency. 
}
\label{fig2}
\end{figure}
\paragraph*{}  
While the calculations presented in sections \ref{sec:sec3}-\ref{sec:sec5} are somewhat involved, the physical origin of some of these anomalous features can be understood 
quite simply. Both  large fluctuations and intermittency of the total mass have their origin in the exit of large aggregates from the system.
 The typical size of the exiting aggregates sets the scale of the rms fluctuations of the total mass: in the diffusive models (A) and (B), these aggregates can be macroscopic, with average
mass scaling as system size $L$, thus giving rise to $\ensuremath{\mathcal{O}\!(L)}$ fluctuations of the total mass. In the driven case (C), exiting aggregates typically
have $\ensuremath{\mathcal{O}\!(\sqrt{L})}$ mass leading to $\sqrt{L}$ fluctuations of the total mass. Temporal intermittency is also a consequence of
the sudden changes in $M(t)$ due to the exit of a large aggregate. This can be seen clearly in figure \ref{fig2} which shows a typical time series $M(t)$ for each of the three cases (A)-(C).
The intervals between large exit events sets the characteristic time scale $\tau$ which appears in the scaling of the structure functions. For the diffusive systems, this is
$\ensuremath{\mathcal{O}\!(L^{2})}$ while for driven systems, it is $\ensuremath{\mathcal{O}\!(\sqrt{L})}$. For $t$ much smaller than this characteristic time $\tau$, 
higher moments of $M(t)-M(0)$ are dominated by the large crashes in fig. \ref{fig2}, giving rise to the anomalous scaling of structure functions at these time scales.  
 
\subsection{Related Models}
\label{sec:sec2.3}

\paragraph*{}
Theoretical studies of aggregation span a variety of models which may be used to address different sorts of questions.
The simplest description of aggregation dynamics is provided by the mean field Smoluchowski equation which describes the time evolution of the mass distribution
 of aggregates, ignoring spatial distributions altogether (see \cite{leyvraz}). It provides a useful framework for studying aggregation phenomena in high dimensions.
However, in low dimensions, it is important to explicitly account for the movement of  aggregates in space as fluctuations are strong, leading to non mean field behaviour.
In many studies, aggregates are assumed to be point objects and the main focus is on studying the kinetics of the aggregation
 process and distribution of aggregate masses, both with \cite{takayasu,doering,cheng,derrida,hinrichsen,howard} and without influx \cite{kang,spouge,doering2}. In some of these studies, aggregation is
treated as an $A+A\rightarrow A$ reaction \cite{doering2,doering,derrida,hinrichsen} while in others, the mass of the reacting aggregates is also taken into account
 \cite{kang,takayasu,spouge,cheng}. Below we present a more detailed discussion of some of these models and their relation to the present work.
\paragraph*{}
(i) Takayasu et al studied diffusion and aggregation with spatially uniform injection of particles \cite{takayasu} and found that at large times, the interplay of injection and 
aggregation gives rise to a quasi-stationary state in which the mean mass at any site grows indefinitely, while the mass distribution of aggregates approaches a 
stationary form with a power law decay. In 1D, they calculated the power law exponent exactly and found that it is different from the mean field value. This quasi-stationary 
phase survives even if the model is generalised to allow for evaporation of single particles from aggregates, as long as the evaporation rate is less than a critical
value \cite{majumdar1,majumdar3}. 
The model we study in the present paper is quite different in that it has \emph{spatially localised} influx and outflux and constant average mass, but 
it too shows a broad-tailed distribution of aggregate sizes. However, this distribution is not scale free; instead it has a 
characteristic width which diverges with system size $L$. 
\paragraph*{}
(ii) Models of aggregation with localised injection have been studied for finite and infinite systems, and different sorts of boundaries \cite{cheng,derrida,hinrichsen}. These
 models lack translational invariance and are characterised by position-dependent distributions of mass. For example, Cheng et al considered diffusing and aggregating
 masses in an infinite system with a source of particles at the origin and explicitly calculated the steady state distribution of aggregate size as a function of distance
 from the source \cite{cheng}. Further, they argued that in the presence of a sink, the distribution of masses exiting the system has an 
exponential tail which depends on system size. The present work also considers aggregation on a finite system with a source and a sink and finds that this exponential tail
 is universal, in that it appears in the mass distribution at all points arbitrarily far away from the sink. 
\paragraph*{}
(iii) In models with localised injection of particles, bias in movement of aggregates can lead to behaviour which is
 qualitatively different from the unbiased case. This behaviour was mentioned in \cite{cheng}, and studied in detail by Jain et al in \cite{jain} and
Reuveni et al in \cite{reuveni}, where it was referred to as the Asymmetric Inclusion Process. These studies analyse various mass distributions of the system and show that their spatial dependence is quite different from 
that in the unbiased case. In this paper, we have considered this model again (see sec. \ref{sec:sec5}), but have gone beyond earlier treatments by calculating 
time-dependent properties and demonstrating temporal intermittency. 
\paragraph*{}
(iv)  Some studies have investigated the effect of outflux of large masses on aggregation by introducing models where aggregates with mass higher than some cutoff value
 are removed from the system \cite{racz,vicsek,ball}.
These systems typically show non-monotonic and, in some cases, even oscillatory time evolution. By contrast, the system we study in this paper does not have a fixed cutoff scale. 
Instead, the mass of the exiting clusters follows a broad distribution with a characteristic mass scale that emerges from the interplay of the time taken to form an aggregate
 and the typical residence time of an aggregate in the system.
\paragraph*{}
(v) Real space condensation \cite{majumdar_rev} has been studied in several mass exchange models with conserved total mass and periodic boundary conditions,
 e.g., the zero range process and its variants \cite{evans_rev} and the aggregation-chipping model \cite{majumdar1}. In these systems, when the total mass exceeds a 
critical value, the excess mass collects into a single macroscopic aggregate or `condensate' with a well-defined mean mass and relatively small fluctuations.
The large fluctuations in the open system we study also have their origin in the presence of such a macroscopic aggregate with mean size proportional to $L$. However, in our case the
 condensate mass itself has a broad distribution \cite{sachdeva}, resulting in a broad distribution of various quantities including the total mass.   
\paragraph*{}
(vi) The model studied in this paper is a special limit of a more general model which also allows for fragmentation of single particles from aggregates at a constant rate
\cite{sachdeva}. This system was studied numerically in \cite{sachdeva} and found to undergo a phase transition on varying the fragmentation rate. At fragmentation rates
lower than a critical value (determined by the injection rate), the system shows all the features of the zero-fragmentation limit, notably, 
giant fluctuations and temporal intermittency of the total mass. 
However, the analytic approach used in this paper cannot be extended to this general model with fragmentation. 
\paragraph*{}
(vii) Connaughton et al demonstrated that the Takayasu model with aggregation and injection shows turbulence-like behaviour in the sense of multi-scaling of $n$-point mass-mass 
correlation functions \cite{connaughton}. They calculated these $n$-point functions using field theoretic techniques and found that for $d\leq 2$, these have an anomalous 
dependence on mass that deviates from the prediction of a Kolmogorov-like, self-similarity hypothesis. The turbulence-like behaviour that we observe in the present work 
is somewhat different, being related to the temporal intermittency of the total mass. It may, however, be interesting to explore whether there is a connection between 
these two characterisations of turbulent behaviour in the context of aggregation.
\paragraph*{}
Dimensionality plays an important role in aggregation phenomena. Most of the models discussed above have an upper critical dimension $d_{c}=2$, below which the system
 shows diffusion-limited behaviour with spatial (anti-) correlations between particles being strong enough
to cause deviation from mean field predictions. This raises the question whether there is an upper critical dimension in the present model too,
 beyond which there is no intermittency.
We  are not able to address this question using the analysis of this paper. More generally, while several techniques 
for finding exact solutions of such models exist in 1D \cite{takayasu,spouge,doering}, they cannot be generalised to higher dimensions. 
The only studies in higher dimensions that we are aware of, analyse reaction-diffusion models by studying the corresponding field theory using renormalisation group techniques
\cite{connaughton,howard,RG}. 

\section{Case A: Diffusive movement; outflux only from right boundary}
\label{sec:sec3}   
\paragraph*{}
In this section, we analyse the case where single particles are injected onto site $1$ at rate $a$, aggregates diffuse to the left or right with symmetric rates $D$, and aggregates
exit from the opposite end i.e. site $L$, also with rate $D$. 
\paragraph*{}
This section is organised as follows: In sec. \ref{sec:sec3.1}, we calculate the steady state probability distribution of mass in any region of the system and recover the total
mass distribution and single site mass distributions as special cases of this. This calculation establishes that all these distributions have a universal exponential tail with a characteristic 
scale proportional to system size $L$. Section \ref{sec:sec3.2} deals with the calculation of dynamical quantities, specifically temporal structure functions $S_{n}(t)$ of the total 
mass. Structure functions are obtained by finding correlation functions which relate the total mass in the system at a particular time instant to the mass in any sub-part of the system
at another instant. This calculation demonstrates that structure functions show strongly intermittent behaviour, $S_{n}(t)\propto t$, at short times.
\paragraph*{}
The analytical approach used in these calculations is based on a `closure' property: the probability distributions of mass in any single continuous region of the system 
satisfy recursion relations that involve \emph{only single-region} probability distributions \cite{doering,takayasu,majumdar2}. This property also forms the basis of 
the dynamical calculations in sec. \ref{sec:sec3.2}. The only approximation  involved in these calculations is that we take a continuum limit in space,
so that recursion relations involving discrete lattice sites become partial differential equations in a continuous spatial variable. The validity of this approximation is examined
in the course of the analysis. 
\subsection{Statics}
\label{sec:sec3.1}
The average mass $\langle m_{i}\rangle$ at a site $i$, obeys the continuity equation:
\begin{subequations}
\begin{equation}
\frac{d\langle m_{i}\rangle}{dt}=D[\langle m_{i+1}\rangle+\langle m_{i-1}\rangle-2\langle m_{i}\rangle] 
\label{eq3.1.1a}
\end{equation}
\begin{equation}
\frac{d\langle m_{1}\rangle}{dt}=a+D[\langle m_{2}\rangle-\langle m_{1}\rangle]  
\label{eq3.1.1b}
\end{equation}
\begin{equation}
\frac{d\langle m_{L}\rangle}{dt}=D[\langle m_{L-1}\rangle-2\langle m_{L}\rangle]  
\label{eq3.1.1c}
\end{equation}
\label{eq3.1.1}
\end{subequations}
In steady state, all time derivatives are zero and $\langle m_{i}\rangle$ is given by:
\begin{equation}
 \langle m_{i}\rangle = \frac{a(L+1)}{D}\left(1-\frac{i}{L+1}\right); \qquad \langle M\rangle =\sum\limits_{i=1}^{L} \langle m_{i}\rangle=\frac{aL(L+1)}{2D}
\label{eq3.1.2}
\end{equation}
Note that the mean value of the total mass in this model is super-extensive [$\langle M\rangle \propto L^{2}$]. To ensure extensivity, we consider rates of the sort 
$a=\tilde{a}/L$, where $\tilde{a}$ is $\ensuremath{\mathcal{O}\!(1)}$, so that the injection rate scales as the inverse of the system size. With such injection rates, 
the total mass becomes an extensive quantity and the average particle current $D[\langle m_{i}\rangle-\langle m_{i+1}\rangle]$ in the system becomes proportional to $1/L$,
 as is typical of diffusive systems obeying Fourier's law. This also facilitates comparison with the case studied in section \ref{sec:sec4}, where the total mass is 
naturally extensive. 
\paragraph*{}
Unlike $\langle m_{i}\rangle$ or $\langle M\rangle$ , higher moments of mass cannot be derived from the continuity equation. To derive $\langle M^{n}\rangle$, we follow the approach
 used by Takayasu et al \cite{takayasu} and write the evolution equations for:
\begin{equation}
 M_{i,j}=\sum\limits_{l=i+1}^{j} m_{l} \qquad \text{and} \quad P_{i,j}(M,t)=\text{Prob}[M_{i,j}=M \text{ at time } t]
\label{eq3.1.3}
\end{equation}
The aim is to solve for  $P_{i,j}(M)$, which is the steady state probability distribution of mass in the region [$i+1$, $j$].
 The distribution $P(M)$ of total mass $M$ can be recovered from the solution as the special case $i=0, j=L$,  while 
the mass distribution $p_{k} (m_{k})$ at the $k^{th}$ site corresponds to the case $i=k-1,j=k$. 
\paragraph*{}
The evolution equations for the time-dependent probability distribution $P_{i,j}(M,t)$ are as follows:
\begin{subequations}
\begin{equation}
\begin{split}
\frac{\partial{P_{i,j}(M,t)}}{\partial{t}}=D\{P_{i,j+1}(M,t)+P_{i,j-1}(M,t)+P_{i-1,j}(M,t)&+P_{i+1,j}(M,t)-4P_{i,j}(M,t)\} \\
& i>0,j>i+1,j<L\\
\end{split}
\label{eq3.1.4a}
\end{equation}
\begin{equation}
\begin{split}
\frac{\partial{P_{0,j}(M,t)}}{\partial{t}}=a\{[1-\delta_{M,0}]P_{0.j}(M-1,t)-P_{0,j}(M,t)\}+D\{P_{0,j+1}&(M,t) +P_{0,j-1}(M,t)-2P_{0,j}(M,t)\} \\
& i=0,j>i+1,j<L\\
\end{split}
\label{eq3.1.4b}
\end{equation}
\begin{equation}
\frac{\partial{P_{i,L}(M,t)}}{\partial{t}}=D\{P_{i,L-1}(M,t)+P_{i-1,L}(M,t)+P_{i+1,L}(M,t)-3P_{i,L}(M,t)\}\qquad i>0,j>i+1,j=L
\label{eq3.1.4c}
\end{equation}
\begin{equation}
\frac{\partial{P_{i,i+1}(M,t)}}{\partial{t}}=D\{2\delta_{M,0}+P_{i,i+2}(M,t)+P_{i-1,i+1}(M,t)-4P_{i,i+1}(M,t)\} \qquad i>0,j=i+1,j<L
\label{eq3.1.4d}
\end{equation} 
\label{eq3.1.4}
\end{subequations}
Equation \eqref{eq3.1.4a} reflects how the mass in the region [$i+1$, $j$] evolves by exchange of aggregates between sites $i$ and $i+1$ at one end and 
$j$ and $j+1$ at the other. If this region includes the first site [eq. \eqref{eq3.1.4b}], then the mass can also change due to injection of single particles. 
Equation \eqref{eq3.1.4c} takes into account the effect of the sink next to site $L$ and eq. \eqref{eq3.1.4d} describes the special case of a single site. 
\paragraph*{}
In steady state, all the time derivatives are equal to zero. The steady state probability distributions $P_{i,j}(M)$ can be obtained by solving for the corresponding generating
 functions $F_{i,j}(z)=\sum\limits_{M=0}^{\infty} P_{i,j}(M)z^{M}$. From eq. \eqref{eq3.1.4b} it can be seen that $P_{0,j}(M)$ does not depend on $P_{i,j}(M)$ for $i>0$.
Thus, $F_{0,j}(z)$ can be solved for independently of other values of $i$. The equations satisfied by $F_{0,j}(z)$ are:
\begin{subequations}
\begin{equation}
F_{0,j+1}(z)+F_{0,j-1}(z)-2F_{0,j}(z)=\frac{a}{D}[1-z]F_{0,j}(z) \qquad 1<j<L
\label{eq3.1.5a}
\end{equation}
\begin{equation}
 F_{0,0}(z)=1, \qquad \qquad F_{0,L+1}(z)=F_{0,L}(z)
\label{eq3.1.5b}
\end{equation}
\label{eq3.1.5}
\end{subequations}
To solve eq. \eqref{eq3.1.5}, we take the continuum limit in space: $j/L\rightarrow y$, $F_{0,j}(z)\rightarrow Q(y,z)$, and Taylor expand $F_{0,j\pm 1}(z)$ such that 
$F_{0,j\pm 1}(z) \rightarrow Q(y,z) \pm \frac{1}{L}\frac{\partial Q}{\partial y}+\frac{1}{2L^{2}}\frac{\partial^{2} Q}{\partial y^{2}}+ ..$ . Retaining only the leading order term in
$1/L$, we find:
\begin{subequations}
 \begin{equation}
 \frac{\partial^{2}{Q(y,z)}}{\partial{y^{2}}} =\beta (1-z)Q(y,z) \qquad \text{where} \quad \beta=\frac{aL^{2}}{D}=\frac{\tilde{a}L}{D}
\label{eq3.1.6a}
 \end{equation}
\begin{equation}
Q(y=0,z)=1,  \qquad \qquad \left.\frac{\partial{Q}}{\partial{y}}\right|_{y=1}=0
\label{eq3.1.6b}
 \end{equation}
\label{eq3.1.6}
\end{subequations}

\begin{figure} [!][h]
\subfigure[] {
\centering
\includegraphics[width=0.6\textwidth]{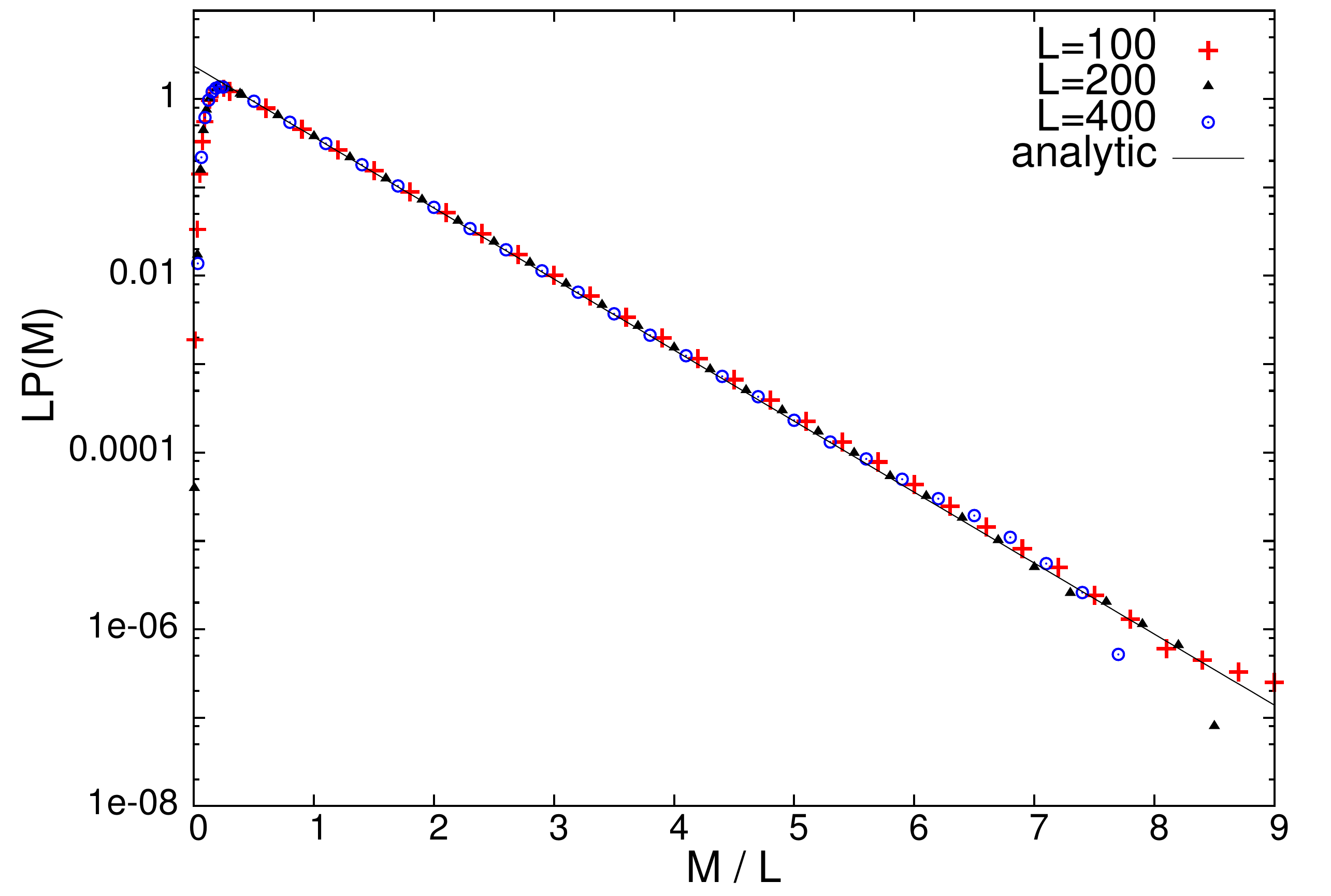}
\label{fig3a}
}\\
\subfigure[]{
\includegraphics[width=0.48\textwidth]{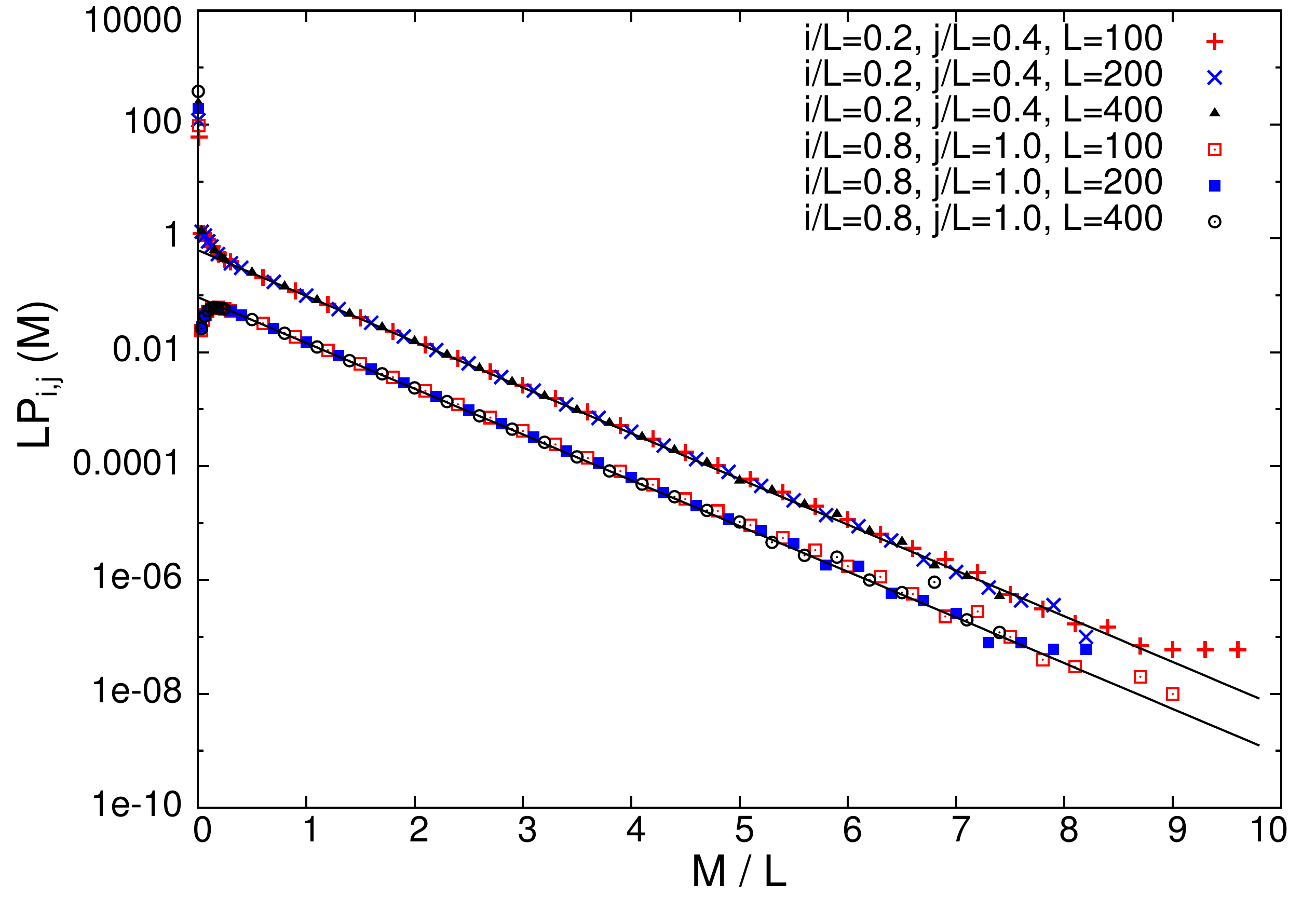}
\label{fig3b}}
\subfigure[]{
\includegraphics[width=0.48\textwidth]{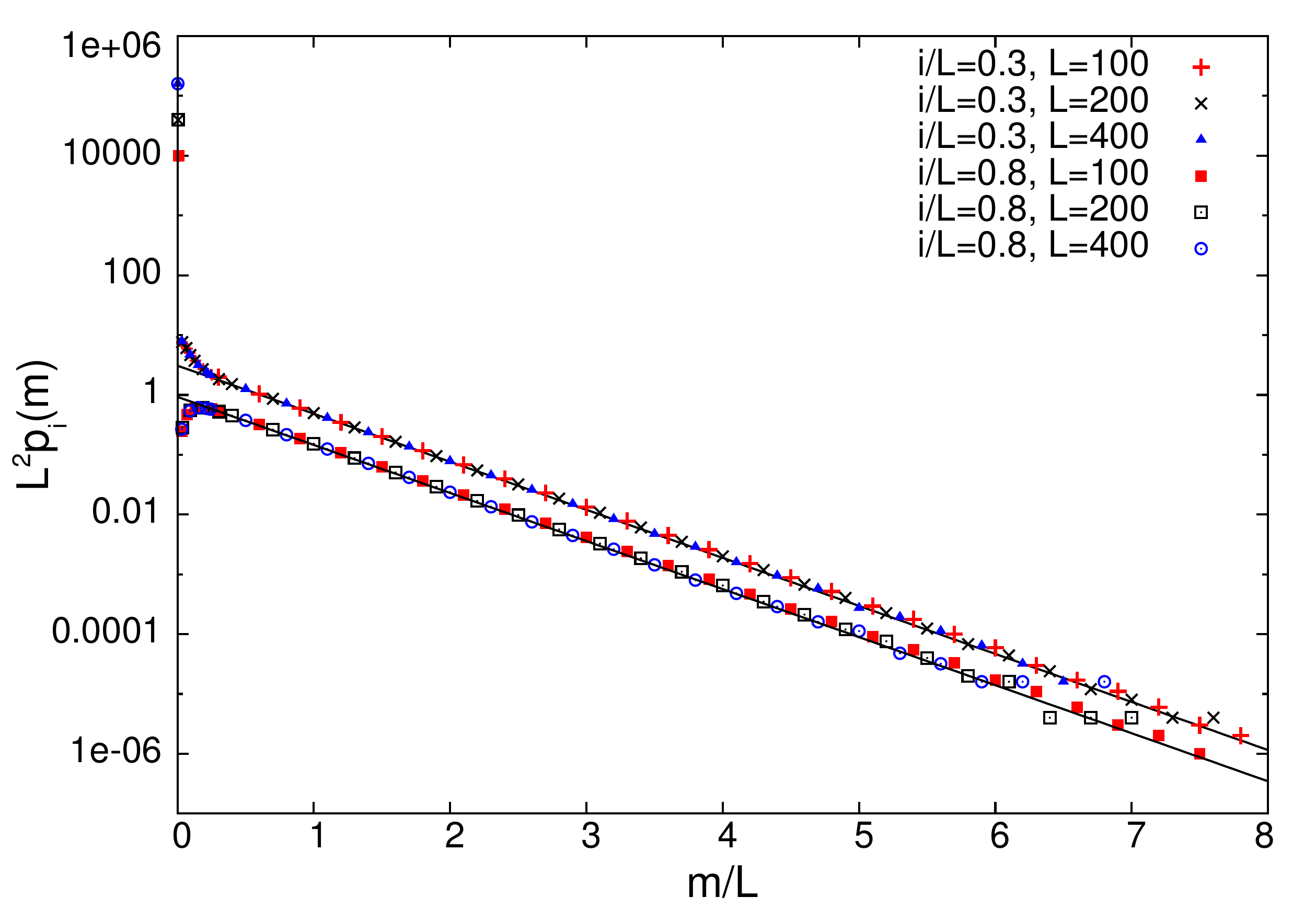}
\label{fig3c}}
\caption{Comparison of numerics (data points) and analytical expressions (solid lines) for various steady state mass distributions with $a/D=4/3$ fixed and different $L$:\newline
(a) Total mass: Data collapse of tails of the probability distribution $P(M)$ for different $L$ on plotting $LP(M)$ vs. $M/L$. \newline
(b) Mass in a region [$i+1$, $j$]: Data collapse of tails on plotting $LP_{i,j}(M)$ vs. $M/L$ for a fixed $i/L$ and $j/L$. \newline 
(c) Mass at a single site $i$: Data collapse of tails on plotting $L^{2}p_{i}(m)$ vs. $m/L$ for a fixed $i/L$.
}
\label{fig3}
\end{figure}

Equation \eqref{eq3.1.6} can be solved to obtain:
\begin{equation}
 Q(y,z)=\frac{\cosh\left[\sqrt{\beta(1-z)}(1-y)\right]}{\cosh\left[\sqrt{\beta(1-z)}\right]}
\label{eq3.1.7}
\end{equation}
\paragraph*{}
To check the self-consistency of the continuum approximation, note that the next correction to the leading order term in eq. \eqref{eq3.1.6a} would be
 $\sim (1/L^{2})\partial^{4}Q/\partial y^{4}$, which from eq.\eqref{eq3.1.7} is $\beta/L^{2}$ times the leading order term. Thus, for injection rates of the sort $a=\tilde{a}/L$, the correction term is $1/L$ times the leading order term, whereas 
for $\ensuremath{\mathcal{O}\!(1)}$ injection rates, all higher order terms are comparable to the leading order term. The continuum limit is thus, self-consistent only for injection 
rates of the sort $a=\tilde{a}/L$, and all the results we derive in this section are valid only for such rates \footnote{Although the exact results derived in this section are not valid 
for $\ensuremath{\mathcal{O}\!(1)}$ injection rates, numerics show that the scaling of the moments of mass and the structure functions with $\beta$, as derived here, 
holds in that case too.}
\paragraph*{}
The generating function for the distribution of total system mass can be obtained by setting $y=1$ in eq. \eqref{eq3.1.7}. 
This can be inverted for large $M$ and large $L$ (details in appendix \ref{appendix1.1}), 
to give the following expression for the tail of the probability distribution $P(M)$:
 \begin{equation}
  P(M)\sim \frac{\pi}{\beta}\exp\left(-\frac{\pi^{2}M}{4\beta}\right)
\label{eq3.1.8}
 \end{equation}
where $\beta \propto L$ for $\ensuremath{\mathcal{O}\!(1/L)}$ injection rates. 
\paragraph*{}
We now turn to the calculation of $F_{i,j}(z)$ for $i>0$. As for the $i=0$ case, this becomes easier in the continuum limit: $i/L\rightarrow x$, $j/L\rightarrow y$,
$F_{i,j}(z)\rightarrow F(x,y,z)$, $\langle M_{i,j}^{n}\rangle\rightarrow \langle M_{xy}^{n}\rangle$. Then, it follows from eq. \eqref{eq3.1.4}, that in steady state,
 $F(x,y,z)$ satisfies:
\begin{subequations}
 \begin{equation}
 \frac{\partial^{2}{F(x,y,z)}}{\partial{x^{2}}}+\frac{\partial^{2}{F(x,y,z)}}{\partial{y^{2}}}=0, \qquad x<y
\label{eq3.1.9a}
\end{equation}
\begin{equation}
 F(x,x,z)=1,  \qquad  \qquad \left.\frac{\partial{F}}{\partial{y}}\right|_{y=1}=0, \qquad  \qquad F(0,y,z)=Q(y,z)
\label{eq3.1.9b}
\end{equation}
\label{eq3.1.9}
\end{subequations}
This is just the Laplace equation on a triangle with vertices $(0,0)$, $(0,1)$, $(1,0)$, and mixed, inhomogeneous boundary conditions. It can be solved by mapping onto the
 Laplace equation on a square with suitably chosen boundary 
conditions \cite{damle,prager} (details in appendix \ref{appendix2.1}), to give:
\begin{equation}
 F(x,y,z)=1+2\sum\limits_{n=0}^{\infty} \frac{\beta(1-z)}{\alpha_{n}(\alpha_{n}^{2}+\beta(1-z))}\left[\frac{\sin[\alpha_{n}x]\cosh[\alpha_{n}(1-y)]-\sin[\alpha_{n}y]\cosh[\alpha_{n}(1-x)]}{\cosh\alpha_{n}}\right]
\label{eq3.1.10}
\end{equation}
where $\alpha_{n}=(n+\frac{1}{2})\pi$.\\ \\
$F(x,y,z)$ can be inverted to give $P_{i,j}(M)$:
\begin{equation}
 P_{i,j}(M)=2\beta^{M}\sum\limits_{n=0}^{\infty} \frac{\alpha_{n}}{(\alpha_{n}^{2}+\beta)^{M+1}}\left[\frac{\sin[\alpha_{n}\frac{j}{L}]\cosh[\alpha_{n}(1-\frac{i}{L})]-
\sin[\alpha_{n}\frac{i}{L}]\cosh[\alpha_{n}(1-\frac{j}{L})]}{\cosh\alpha_{n}}\right], \quad M>0
\label{eq3.1.11}
\end{equation}
For large $M$ and large $L$, just the first term ($n=0$) in the series in eq. \eqref{eq3.1.11} is enough to give a good approximation for $P_{i,j}(M)$, so that:
\begin{equation}
 P_{i,j}(M)\sim\frac{\pi}{\beta}\left[\frac{\sin[\frac{\pi j}{2L}]\cosh[\frac{\pi}{2}(1-\frac{i}{L})]-
\sin[\frac{\pi i}{2L}]\cosh[\frac{\pi}{2}(1-\frac{j}{L})]}{\cosh\frac{\pi}{2}}\right] \exp\left(-\frac{\pi^{2}M}{4\beta}\right)
\label{eq3.1.12}
\end{equation}
The asymptotic expression for the probability distribution $p_{i}(m)$ of mass at a single site $i$ can be obtained similarly:
\begin{equation}
 p_{i}(m)\sim\frac{\pi^{2}}{2\beta L}\left[\frac{\sin[\frac{\pi i}{2L}]\sinh[\frac{\pi}{2}(1-\frac{i}{L})]+
\cos[\frac{\pi i}{2L}]\cosh[\frac{\pi}{2}(1-\frac{i}{L})]}{\cosh\frac{\pi}{2}}\right] \exp\left(-\frac{\pi^{2}m}{4\beta}\right)
\label{eq3.1.13}
\end{equation}  
\paragraph*{}
Figure \ref{fig3} shows the mass distributions $P(M)$, $P_{i,j}(M)$ and $p_{i}(m)$, as obtained from numerics, along with the asymptotic analytic expressions of eqs. \eqref{eq3.1.8}, \eqref{eq3.1.12}
and \eqref{eq3.1.13}. Note that all distributions, whether of the total system mass or of the mass at a single site, show the same dependence on mass $m$: 
for large $m$, the tail goes as $\sim(1/M_{0})\exp[-m/M_{0}]$ 
with $M_{0}\sim (4/\pi^{2})\beta$. This reflects the fact that a large fluctuation in the mass in any region of the system arises due to the visit of a macroscopic 
aggregate with a characteristic mass scale $M_{0} \propto L$. The universal tail distribution $\sim(1/M_{0})\exp[-m/M_{0}]$ of mass in different regions just reflects the distribution of mass in this
macroscopic aggregate. How often a particular region or site is visited by the aggregate is encoded in the non-universal pre-factor of the distributions. For the single-site 
distribution $p_{i}(m)$, for example, this prefactor is of the form $\sim (1/L)c(i/L)$. The $1/L$ factor reflects the fact that the aggregate is shared among $L$ sites,
 whereas the function $c(i/L)$ encodes the spatial inhomogeneity of the system. The occurrence of this kind of $L$-dependent tail in all mass distributions implies that 
even a single site far from the boundaries `knows' about the size of the system, through the visits of the macroscopic aggregate. 

\paragraph*{}
Moments and cumulants of the mass in any region of the system can also be obtained from the generating functions $F(x,y,z)$ and $Q(y,z)$. For example, for the total system mass $M$, we have:
\begin{subequations}
\begin{equation}
\langle M\rangle=\frac{\beta}{2}, \qquad  
\langle M^{2}\rangle=\frac{5}{12}\beta^{2}+\frac{\beta}{2}, \qquad \langle M^{3}\rangle=\frac{61}{120}\beta^{3}+\frac{5}{4}\beta^{2}+\frac{\beta}{2}
\label{eq3.1.14a}
\end{equation}
\begin{equation}
\langle[M-\langle M\rangle]^{2}\rangle=\frac{\beta^{2}}{6}+\frac{\beta}{2},  \qquad
\langle[M-\langle M\rangle]^{3}\rangle=\frac{2}{15}\beta^{3}+\frac{\beta^{2}}{2}+\frac{\beta}{2},   \qquad \beta=\tilde{a}L/D
\label{eq3.1.14b}
\end{equation}
\label{eq3.1.14}
\end{subequations}
The rms fluctuations $\Delta M$ of the total mass are anomalously large, scaling as system size $L$, rather than $\sqrt{L}$, as expected in a system with normal fluctuations.
Thus, $\Delta M/\langle M\rangle$ remains finite even
in the thermodynamic limit. Similarly, by calculating $\langle M_{i,j}^{n}\rangle$ from $F(x,y,z)$ it is possible to establish that the mass in any region [$i+1$, $j$] of the system 
also shows giant fluctuations which scale as $L$ for fixed $i/L$ and $j/L$. 
\paragraph*{}
The moment-generating function $F(x,y,z)$ can be used to calculate many other quantities as well. For example, the 2-point correlation function $\langle m_{i}m_{j}\rangle$ can be obtained from
the second moments using:
\begin{equation}
\langle m_{i}m_{j}\rangle=\frac{1}{2}[\langle M_{i,j-1}^{2}\rangle + \langle M_{i-1,j}^{2}\rangle-\langle M_{i,j}^{2}\rangle-\langle M_{i-1,j-1}^{2}\rangle] 
\label{eq3.1.15}
\end{equation}
The fluctuations of the particle current $J_{i}$ in the $i^{th}$ bond of the lattice can also be calculated using:
\begin{equation}
 \langle J_{i}^{2}\rangle = \langle [D(m_{i}-m_{i+1})]^{2}\rangle=D^{2}[2\langle M_{i-1,i}^{2}\rangle+2\langle M_{i,i+1}^{2}\rangle-\langle M_{i-1,i+1}^{2}\rangle]
\end{equation}
Finding higher moments of the current is more difficult, as it involves quantities like $\langle m_{i}^{2}m_{i+1}\rangle$ for which we need to calculate the 
multi-region moments like $\langle M_{i,j}^{2}M_{j,k}\rangle$. In principle, this can be done by generalising the previous analysis, thus allowing for the calculation of
various statistics of the particle current.  
\paragraph*{}
In summary, the main result of this section is that the distribution of mass in any region of the system, from the full system, down to a single site, has a universal non-Gaussian tail which 
depends explicitly on system size. Thus, the system size $L$  enters in a central way into the properties of the system, in contrast to the usual finite size 
corrections which become increasingly irrelevant at large system sizes. 
An alternative way to see this is that $\Delta M/\langle M\rangle$ etc. remains finite even as $L\rightarrow \infty$, implying that the system is not self averaging even
 in the thermodynamic limit.

\subsection{Dynamics}
\label{sec:sec3.2}
The exit of a macroscopic aggregate from the system can cause the total mass $M$ to drop sharply [fig. \ref{fig2a}]. The occurrence of very sharp changes over relatively 
small time scales ($t\ll \tau$) is characteristic of turbulent signals and is known as intermittency. As discussed in sec. \ref{sec:sec2.1}, intermittency can be systematically probed 
by analysing structure functions:
\begin{equation}
 S_{n}(t)=\langle[M(t)-M(0)]^{n}\rangle
\label{eq3.2.1}
\end{equation}
where $t=0$ is any time instant after the system attains steady state.
In general, structure functions behave as $S_{n}(t)\sim t^{u(n)}$ as $t\rightarrow 0$, where $u(n)$ is typically sublinear in $n$ for intermittent signals. 
The main goal of this section is to calculate structure functions for various $n$ for this model and hence, find the $n$-dependence of the exponents $u(n)$. 
\paragraph*{}
Structure functions are related to autocorrelation functions of the total mass. For example,
\begin{equation}
 S_{2}(t)=2\langle M^{2}\rangle-2\langle M\rangle^{2}- 2[\langle M(0)M(t)\rangle-\langle M(0)\rangle\langle M(t)\rangle]
\label{eq3.2.2}
\end{equation}
The correlation function $\langle M(0)M(t)\rangle$ depends on the function $\langle M_{0,L}(0)M_{0,L-1}(t-1)\rangle$ which in turn also depends on 
$\langle M_{0,L}(0)M_{0,L-2}(t-2)\rangle$ and so on. Thus, 
it becomes necessary to define and solve for spatio-temporal correlation functions of the sort: 
\begin{subequations}
\begin{equation}
 C_{2}(j,t)=\langle M_{0,L}(0)M_{0,j}(t)\rangle-\langle M_{0,L}(0)\rangle\langle M_{0,j}(t)\rangle
\label{eq3.2.3a}
\end{equation}
\begin{equation}
 C_{31}(j,t)= \langle M_{0,L}^{2}(0)M_{0,j}(t)\rangle-\langle M_{0,L}^{2}(0)\rangle\langle M_{0,j}(t)\rangle
\label{eq3.2.3b}
\end{equation}
\begin{equation}
 C_{32}(j,t)= \langle M_{0,L}(0)M_{0,j}^{2}(t)\rangle-\langle M_{0,L}(0)\rangle\langle M_{0,j}^{2}(t)\rangle
\label{eq3.2.3c}
\end{equation}
\begin{equation}
 C_{41}(j,t)= \langle M_{0,L}^{3}(0)M_{0,j}(t)\rangle-\langle M_{0,L}^{3}(0)\rangle\langle M_{0,j}(t)\rangle
\label{eq3.2.3d}
\end{equation}
\begin{equation}
 C_{42}(j,t)= \langle M_{0,L}^{2}(0)M_{0,j}^{2}(t)\rangle-\langle M_{0,L}^{2}(0)\rangle\langle M_{0,j}^{2}(t)\rangle
\label{eq3.2.3e}
\end{equation}
\begin{equation}
 C_{43}(j,t)= \langle M_{0,L}(0)M_{0,j}^{3}(t)\rangle-\langle M_{0,L}(0)\rangle\langle M_{0,j}^{3}(t)\rangle
\label{eq3.2.3f}
\end{equation}
\label{eq3.2.3}
\end{subequations}

By solving for the above correlation functions and setting $j=L$, structure functions of various orders can be obtained from: 
\begin{subequations}
\begin{equation}
 S_{2}(t)=2\langle M_{0,L}^{2}\rangle-2\langle M_{0,L}\rangle^{2}- 2C_{2}(L,t)
\label{eq3.2.4a}
\end{equation}
\begin{equation}
 S_{3}(t)= 3C_{31}(L,t)-3C_{32}(L,t)
\label{eq3.2.4b}
\end{equation}
\begin{equation}
 S_{4}(t)= 2\langle M_{0,L}^{4}\rangle-8\langle M_{0,L}^{3}\rangle\langle M_{0,L}\rangle+6\langle M_{0,L}^{2}\rangle^{2}-4C_{41}(L,t)+6C_{42}(L,t)-4C_{43}(L,t)
\label{eq3.2.4c}
\end{equation}
\label{eq3.2.4}
\end{subequations}

Now we turn to the actual computation of the correlation functions defined in eq. \eqref{eq3.2.3}. This becomes easier in the spatial continuum limit 
($j/L\rightarrow y$, $M_{0,j}\rightarrow M_{y}$) as the continuum versions
of the correlation functions in eq. \eqref{eq3.2.3} satisfy relatively simple partial differential equations in $y$ and $t$. 
Motivated by eq. \eqref{eq3.2.4}, we write down equations for the following functions: 
\begin{subequations}
\begin{equation}
 G_{2}(y,t)=2C_{2}(j=yL,t)
\label{eq3.2.5a}
\end{equation}
\begin{equation}
 G_{3}(y,t)= 3C_{31}(yL,t)-3C_{32}(yL,t)
\label{eq3.2.5b}
\end{equation}
\begin{equation}
 G_{4}(y,t)= 4C_{41}(yL,t)-6C_{42}(yL,t)+4C_{43}(yL,t)
\label{eq3.2.5c}
\end{equation}
\label{eq3.2.5}
\end{subequations}

To derive the time-evolution of the above functions, note that the time evolution of $M_{0,j}$, as captured by eq. \eqref{eq3.1.4b},  can be re-expressed as follows:
\begin{equation}
M_{0,j}(t+\Delta t)=\begin{cases}
M_{0,j+1}(t), & \text{with probability $D\Delta t$}\\
M_{0,j-1}(t), & \text{with probability $D\Delta t$}\\
M_{0,j}(t)+1, & \text{with probability $a\Delta t$}\\
M_{0,j}(t), & \text{with probability $1-(a+2D)\Delta t$}.
\end{cases}
\label{eq3.2.6}
\end{equation}
The time-dependent equation satisfied by $\langle M_{0,L}(0)M_{0,j}(t)\rangle$ can be obtained by multiplying eq. \eqref{eq3.2.6} by $M_{0,L}(0)$ and then taking an average over 
different realisations of the stochastic time evolution. Similarly, the equation for $\langle M_{0,L}(0)M^{2}_{0,j}(t)\rangle$ is obtained by taking the square of 
eq. \eqref{eq3.2.6}, multiplying it by $M_{0,L}(0)$ and taking averages. In this way,
we can write time-evolution equations satisfied by each of the correlation functions in eq. \eqref{eq3.2.3}. 
Taking the continuum limit ($j/L\rightarrow y$) in space, finally yields the following partial differential equations for $G_{n}(y,t)$:
\begin{subequations}
\begin{equation}
\begin{split}
&\frac{\partial{G_{2}(y,t)}}{\partial{t}}=\frac{D}{L^{2}}\frac{\partial^{2}{G_{2}(y,t)}}{\partial{y^{2}}}\\
&G_{2}(0,t)=0, \qquad \left.\frac{\partial{G_{2}}}{\partial{y}}\right|_{y=1}=0.
\end{split}
\label{eq3.2.7a}
\end{equation}
\begin{equation}
\begin{split}
&\frac{\partial{G_{3}(y,t)}}{\partial{t}}=\frac{D}{L^{2}}\left[\frac{\partial^{2}{G_{3}(y,t)}}{\partial{y^{2}}}-3\beta G_{2}(y,t)\right]\\
&G_{3}(0,t)=0, \qquad \left.\frac{\partial{G_{3}}}{\partial{y}}\right|_{y=1}=0.
\end{split}
\label{eq3.2.7b}
\end{equation}
\begin{equation}
\begin{split}
&\frac{\partial{G_{4}(y,t)}}{\partial{t}}=\frac{D}{L^{2}}\left[\frac{\partial^{2}{G_{4}(y,t)}}{\partial{y^{2}}}+4\beta G_{3}(y,t)-6\beta G_{2}(y,t)\right]\\
&G_{4}(0,t)=0, \qquad \left.\frac{\partial{G_{4}}}{\partial{y}}\right|_{y=1}=0.
\end{split}
\label{eq3.2.7c}
\end{equation}
\label{eq3.2.7}
\end{subequations}
To solve these equations, an additional boundary condition is needed which is the value of the correlation functions at $t=0$.
These are given by:
\begin{subequations}
 \begin{equation}
\begin{split}
G_{2}(y,0)&=2\langle M_{y}M_{1}\rangle-2\langle M_{y}\rangle\langle M_{1}\rangle=-[\langle(M_{1}-M_{y})^{2}\rangle-\langle M_{1}^{2}\rangle-\langle M_{y}^{2}\rangle]
-2\langle M_{y}\rangle\langle M_{1}\rangle\\
&=\langle M_{1}^{2}\rangle+\langle M_{y}^{2}\rangle-\langle M_{y1}^{2}\rangle-2\langle M_{y}\rangle\langle M_{1}\rangle
\end{split}
\label{eq3.2.8a}
\end{equation}
\begin{equation}
\begin{split}
G_{3}(y,0)&= 3\langle M_{1}^{2}M_{y}\rangle-3\langle M_{1}M_{y}^{2}\rangle-3\langle M_{1}^{2}\rangle\langle M_{y}\rangle+3\langle M_{1}\rangle\langle M_{y}^{2}\rangle\\
&=\langle M_{1}^{3}\rangle-\langle M_{y}^{3}\rangle-\langle M_{y1}^{3}\rangle-3\langle M_{1}^{2}\rangle\langle M_{y}\rangle+3\langle M_{1}\rangle\langle M_{y}^{2}\rangle
\end{split}
\label{eq3.2.8b}
\end{equation}
\begin{equation}
\begin{split}
G_{4}(y,0)&=4\langle M_{1}^{3}M_{y}\rangle-6\langle M_{1}^{2}M_{y}^{2}\rangle+4\langle M_{1}M_{y}^{3}\rangle-4\langle M_{1}^{3}\rangle\langle M_{y}\rangle+
6\langle M_{1}^{2}\rangle\langle M_{y}^{2}\rangle-4\langle M_{1}\rangle\langle M_{y}^{3}\rangle\\
&=\langle M_{1}^{4}\rangle+\langle M_{y}^{4}\rangle-\langle M_{y1}^{4}\rangle-4\langle M_{1}^{3}\rangle\langle M_{y}\rangle+6\langle M_{1}^{2}\rangle\langle M_{y}^{2}\rangle
-4\langle M_{1}\rangle\langle M_{y}^{3}\rangle.
\end{split}
\label{eq3.2.8c}
\end{equation}
\label{eq3.2.8}
\end{subequations}
 Thus, all correlation functions at $t=0$ can be expressed in terms of $\langle M_{y}^{n}\rangle$ and $\langle M_{y1}^{n}\rangle$  which can be obtained respectively from the
moment generating functions $Q(y,z)$ [eq. \eqref{eq3.1.7}] and $F(x,y,z)$ [eq. \eqref{eq3.1.10}].
\paragraph*{}
Each of the equations in eq. \eqref{eq3.2.7} is a heat equation in one spatial dimension with an inhomogeneous source term. It can be solved for the initial conditions given 
in eq. \eqref{eq3.2.8} using standard techniques (details in appendix \ref{appendix2.2}) \cite{haberman}. The structure functions $S_{n}(t)$
can be now obtained from $G_{n}(y,t)$ using eq. \eqref{eq3.2.4}:
\begin{subequations}
\begin{equation}
S_{2}(t)=\sum\limits_{n=0}^{\infty}4(-1)^{n}\left(\frac{2\beta^{2}}{\alpha_{n}^{5}}\sech{\alpha_{n}}+\frac{\beta}{\alpha_{n}^{3}}\right)\left(1-\exp\left[-\alpha_{n}^{2}\frac{Dt}{L^{2}}\right]\right)
\qquad \text{where} \quad  \alpha_{n}=\left(n+\frac{1}{2}\right)\pi
\label{eq3.2.9a}
\end{equation}
\begin{equation}
S_{3}(t)=\sum\limits_{n=0}^{\infty}6(-1)^{n}\beta^{2}\left\{\frac{4}{\alpha_{n}^{5}}\left(\left[1-\beta \frac{Dt}{L^{2}}\right]\sech{\alpha_{n}}-1\right)+\frac{1}
{\alpha_{n}^{3}}\left(1-2\frac{Dt}{L^{2}}\right)\right\}\exp\left[-\alpha_{n}^{2}\frac{Dt}{L^{2}}\right]
\label{eq3.2.9b}
\end{equation}
\begin{equation}
\begin{split}
S_{4}(t)=&\vast\{\sum\limits_{n=0}^{\infty}(-1)^{n}\left(\frac{-96\beta^{4}\sech{\alpha_{n}}}{\alpha_{n}^{9}}+\beta^{3}\left[\frac{-10}{\alpha_{n}^{3}}+\frac{48}{\alpha_{n}^{5}}+\frac{-144}{\alpha_{n}^{7}}\right]
-\beta^{2}\left[\frac{56\sech{\alpha_{n}}}{\alpha_{n}^{5}}+\frac{12}{\alpha_{n}^{3}}\right]-\frac{4\beta}{\alpha_{n}^{3}}\right) \\
&\times\left(1-\exp\left[-\alpha_{n}^{2}\frac{Dt}{L^{2}}\right]\right)\vast\}\\
&+\sum\limits_{n=0}^{\infty}24\beta^{3}\left(\frac{2\sech{\alpha_{n}}}{\alpha_{n}^{5}}-\frac{4}{\alpha_{n}^{5}}-\frac{Dt}{L^{2}}\left[\frac{1}{\alpha_{n}^{3}}+
\frac{2\beta\sech{\alpha_{n}}}{\alpha_{n}^{5}}\right]\right)\frac{Dt}{L^{2}}\exp\left[-\alpha_{n}^{2}\frac{Dt}{L^{2}}\right]
\end{split}
\label{eq3.2.9c}
\end{equation}
\label{eq3.2.9}
\end{subequations}
\paragraph*{}
To study intermittency, we need to analyse the small $t$ behaviour of the above expressions. Note that $t$ always appears in the exponentially decaying terms in the combination $t/\tau$ where
 $\tau=L^{2}/D$. The time scale $\tau$ is proportional to $L^{2}$ which, up to a constant, is the mean time interval between successive exit events involving a macroscopic aggregate. This, thus, 
provides a natural time scale for the problem,  so that the small $t$ behaviour can be obtained by Taylor expanding the exponentials in eq. \eqref{eq3.2.9} 
in powers of $t/\tau$ and retaining only the lowest order term. This yields the following small $t$ ($t\ll L^{2}$) expressions for the structure functions:
\begin{subequations}
\begin{equation}
S_{2}(t)\sim 4\beta\left(0.2053\beta+\frac{1}{2}\right)\frac{Dt}{L^{2}}
\label{eq3.2.10a}
\end{equation}
\begin{equation}
S_{3}(t)\sim-\beta^{2}(\beta+2.46378)\frac{Dt}{L^{2}}
\label{eq3.2.10b}
\end{equation}
\begin{equation}
S_{4}(t)\sim \beta(1.6214\beta^{3}-8.9\beta^{2}-0.25108\beta-2)\frac{Dt}{L^{2}}
\label{eq3.2.10c}
\end{equation}
\label{eq3.2.10}
\end{subequations}
In the limit $L\rightarrow \infty$, only leading order terms in $L$ are important, so that:
\begin{equation}
 S_{n}(t) \sim c_{n}(-1)^{n} \left(\frac{aL^{2}}{D}\right)^{n} \left(\frac{Dt}{L^{2}}\right) = c_{n}(-1)^{n}(a\tau)^{n-1}at
\label{eq3.2.11}
\end{equation}
where $c_{n}$ are the $\ensuremath{\mathcal{O}\!(1)}$ coefficients of $\beta^{n}$ in eq. \eqref{eq3.2.10}. As is clear from the above expression, the anomalous time scale in the problem, which results in 
breakdown of self-similarity is simply $\tau \propto L^{2}$. Since we have assumed $\ensuremath{\mathcal{O}\!(1/L)}$ injection rates, $a\tau$ is however, proportional to $L$, resulting
in the following scaling of $S_{n}(t)$ with $L$:
\begin{equation}
 S_{n}(t) = L^{n}f_{n}\left(\frac{Dt}{L^{2}}\right)
\label{eq3.2.12}
\end{equation}
\begin{figure}[h]
\subfigure[] {
\centering
\includegraphics[width=0.5\textwidth]{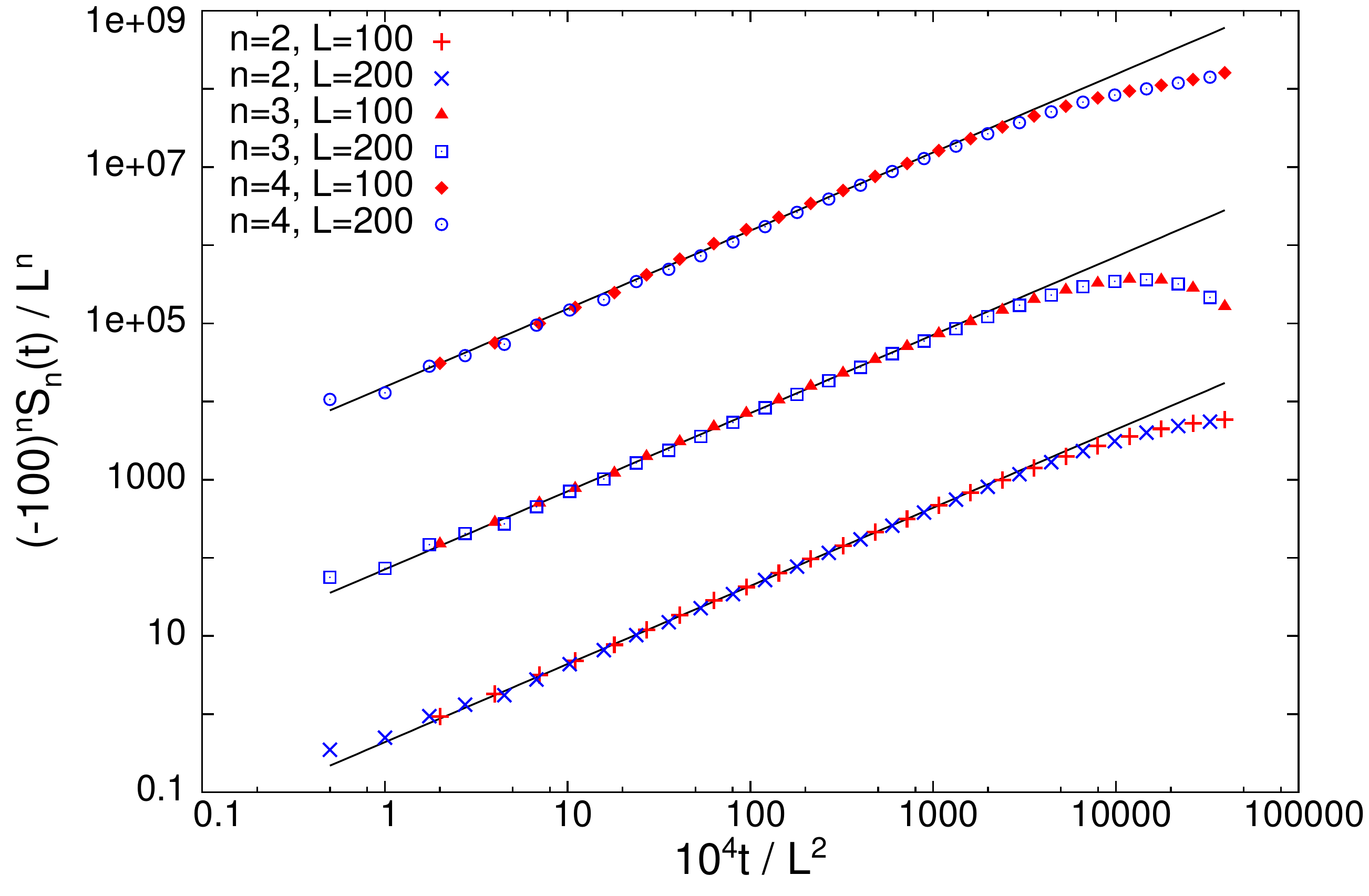}
\label{fig4a}
}
\subfigure[]{
\includegraphics[width=0.48\textwidth]{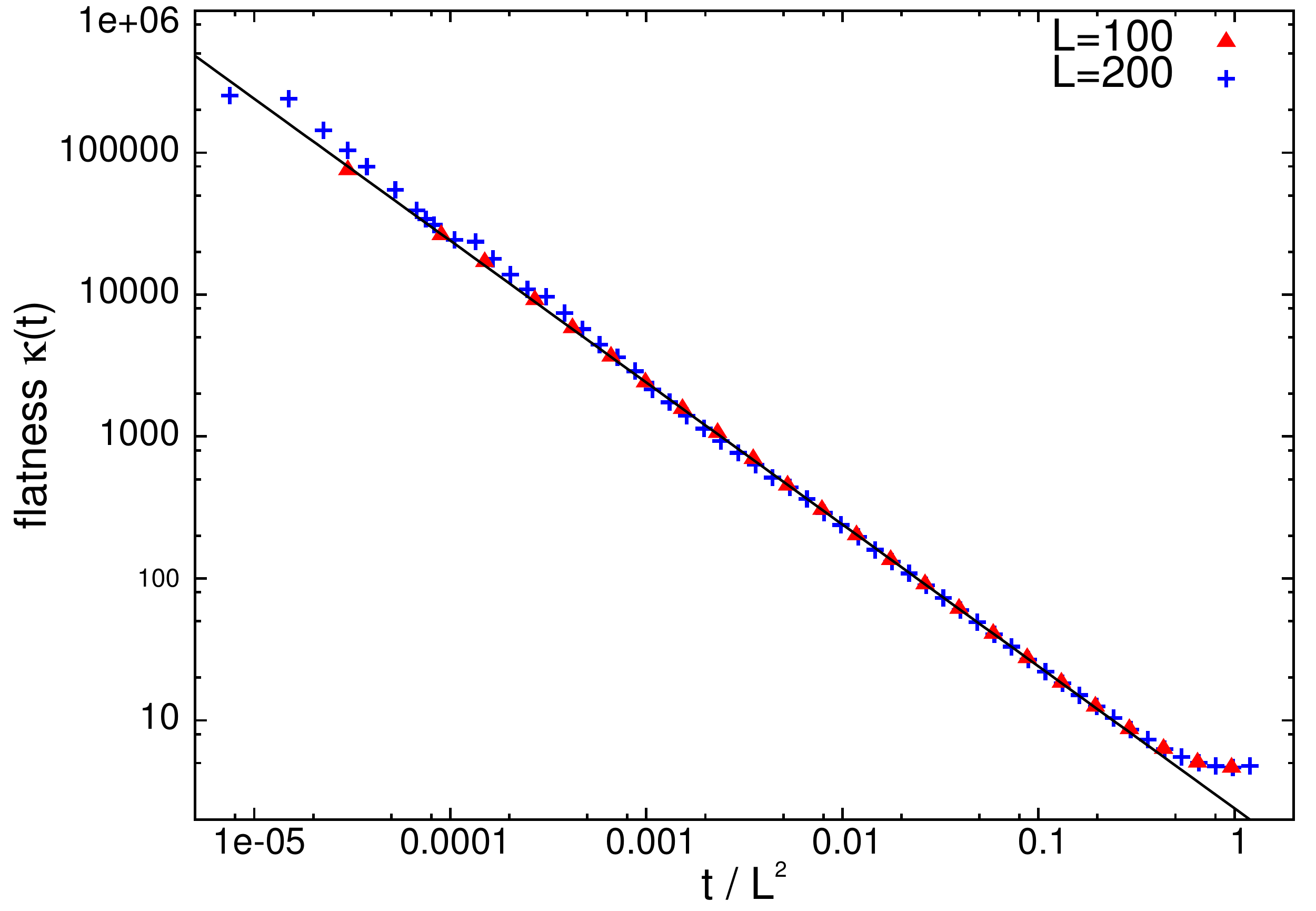}
\label{fig4b}}
\caption{(a) Structure functions:  $(-1)^{n}S_{n}(t)/\tilde{L}^{n}$
 vs. $t/\tilde{L}^{2}$ where $\tilde{L}= L/100$ (rescaling of $L$ done to display all three structure functions in the same plot clearly.) 
Solid lines representing the small $t$ analytical expressions of eq. \eqref{eq3.2.10} are in very good agreement with data points obtained from numerical simulations.\newline
(b) Flatness $\kappa(t)$ vs. $Dt/L^{2}$: Data collapse for different $L$. Analytical prediction for small $t$ as given by eq. \eqref{eq3.2.13} (solid line) agrees well with
numerics.}
\label{fig4}
\end{figure}

\paragraph*{}
The above calculation, thus demonstrates the intermittent behaviour of $M(t)$ by establishing that $S_{n}(t)\propto t^{u_{n}}$ for small $t$, where $u_{n}$ is \emph{not linear} in $n$, as would be 
expected for a self-similar signal.
In fact, the structure functions show an extreme form of anomalous scaling where the exponent $u_{n}$ is actually independent of $n$. This extreme scaling 
is referred to as \emph{strong} intermittency and is also seen in Burgers turbulence, where spatial structure functions of the velocity field exhibit similar behaviour. 
The Burgers equation which describes the time evolution of a fully compressible fluid admits solutions with shocks (discontinuities) in the velocity field in the inviscid limit.
 These shocks dominate the behaviour of higher order structure functions, giving rise to anomalous scaling. This is qualitatively similar to the system we study,
 where the behaviour of higher order structure functions in time is dominated by the occasional large crashes of total mass [see fig. \ref{fig2a}] that occur when a macroscopic aggregate exits the 
system.
\paragraph*{}
As discussed in sec. \ref{sec:sec2.1} , another useful measure of intermittency is the flatness $\kappa(t)$ which was defined in eq. \eqref{eq2.1.2a}.  From eq. \eqref{eq3.2.10}, it 
follows that for small $t$, the flatness is given by:
\begin{equation}
\kappa(t)=\frac{S_{4}(t)}{(S_{2}(t))^{2}}\sim 2.4\left(\frac{Dt}{L^{2}}\right)^{-1}
\label{eq3.2.13}
\end{equation}
 to leading order in $L$.
Thus, the flatness diverges as a power law as $t/L^{2}\rightarrow 0$. This sort of $L$-dependent divergence of flatness is reminiscent of intermittency in fluid turbulence, where the divergence of flatness becomes 
stronger as Reynolds number increases. Figure \ref{fig4} shows a comparison of numerical data with the analytical expressions for structure functions and flatness derived within the continuum 
approximation. The two are in very good agreement.   
\paragraph*{}
\begin{wrapfigure}{r}{0.5\textwidth}
\begin{center}
\vspace{1.5cm}
\includegraphics[width=0.48\textwidth]{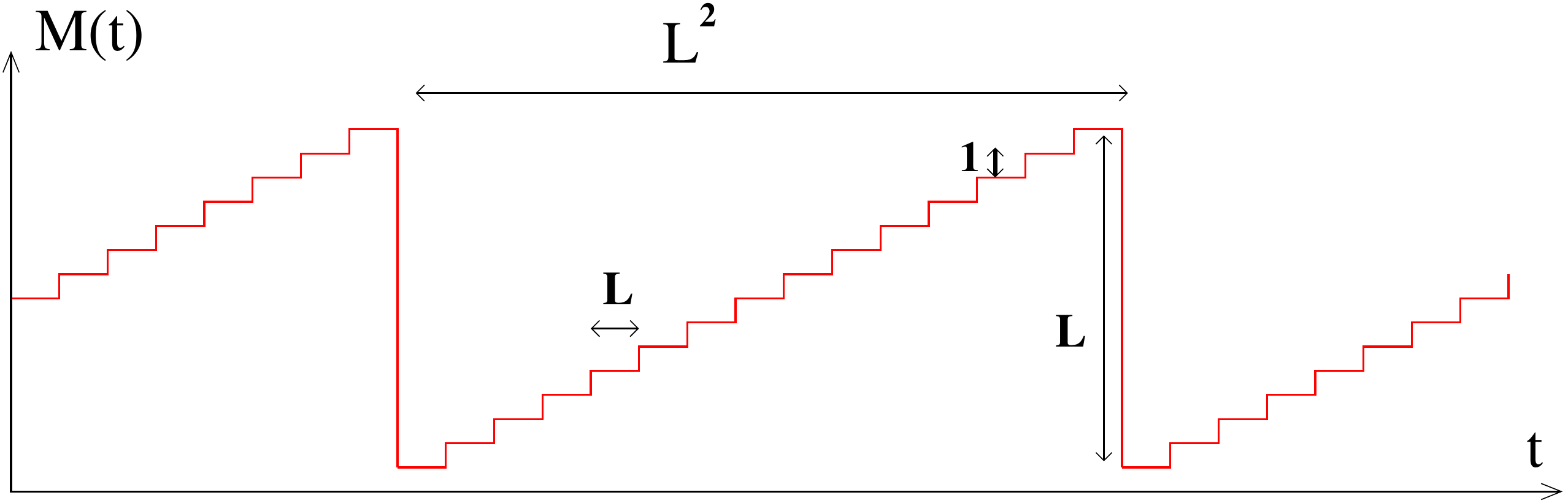}
\end{center}
\caption{A toy model approximating the time series $M(t)$ vs. $t$ for case (A)}
\label{fig5}
\end{wrapfigure}
\paragraph*{}
We conclude this section with two important points. For Burgers turbulence, it is known that structure functions show self-similar scaling
 for $n\leq 1$ and anomalous scaling for $n\geq 1$. In the context of the present work, this raises the question whether generalised structure functions
$\tilde{S_{n}}(t)=\langle|M(t)-M(0)|^{n}\rangle$, defined for general (possibly non-integer) $n$, also show this kind of bifractal behaviour. 
To answer this question, it is necessary to calculate structure functions of non-integer order. This, however, cannot be done using the preceding analysis.
 To get an insight into the behaviour of these generalised structure functions, let us consider a toy model in which the time evolution of $M(t)$ takes place
in the regular, deterministic manner illustrated in fig. \ref{fig5}.  The deterministic increase of $M$ by 1 after every $L$ steps in the figure is akin
to the injection of unit mass with rate $1/L$ in the real time series $M(t)$, while the decrease by $L$ after every $L^{2}$ steps, represents the exit of the macroscopic 
aggregate. For the regular pattern in fig. \ref{fig5}, we can analyse the probability distribution of $M(t)-M(0)$ for a randomly chosen $t=0$, and from this obtain 
$\tilde{S_{n}}(t)=\langle|M(t)-M(0)|^{n}\rangle$. 
This simple analysis shows that $\tilde{S_{n}}(t)\propto t$ for $n\geq 1$ when $t\ll L^{2}$, whereas $\tilde{S_{n}}(t)\propto t^{n}$ for $n\leq 1$ and $L\ll t\ll L^{2}$. 
For $t\ll L$ however, $\tilde{S_{n}}(t)\propto t/L$ when $n\leq 1$.
Thus, except for the $t\ll L$ behaviour when $n\leq 1$, this is analogous to the behaviour of structure functions in Burgers turbulence. The anomalous  $t\ll L$ behaviour
 is simply a consequence of $1/L$ injection rates.
The predictions of this simple sawtooth-like picture are borne out by numerical results for $\tilde{S_{n}}(t)$ for non-integer $n$. 
 
\paragraph*{}
The second comment concerns the intermittency properties of mass in a part of the system. The passage of
 the macroscopic aggregate through a region of the system can cause the total mass in that region to show a sharp change. Thus, we expect that the mass in any region of
the system to also be intermittent.
 Numerics show that this is indeed the case: the structure functions of mass in a region of the system show
 strong intermittency, but they now scale as $S_{n}(t)\propto \sqrt{t}$ rather than as $t$. The analysis of this section can be extended in a straightforward way to compute these structure 
functions also. However, the algebra becomes more cumbersome and we have not carried out these calculations. 
 
\section{Case B: Diffusive movement; outflux from both boundaries}
\label{sec:sec4} 
In this section, we consider diffusive movement of masses with equal rates $D$ to the left and right, injection of single particles at site $1$ at rate $a$ and exit of aggregates from 
both boundaries, i.e., from site $1$ as well as site $L$ at rate $D$. As in section \ref{sec:sec3.1}, it is possible to calculate $\langle m_{i}\rangle$ and $\langle M\rangle$ from the
 continuity equation: 
\begin{equation}
 \langle m_{i}\rangle = \frac{a}{D}\left(1-\frac{i}{L+1}\right), \qquad \langle M\rangle =\sum\limits_{i=1}^{L} \langle m_{i}\rangle=\frac{aL}{2D}
\label{eq4.0.1}
\end{equation}
\paragraph*{}
In contrast to model A, in this case, the total mass is extensive and the average particle current through the system is proportional to $1/L$, even with 
$\ensuremath{\mathcal{O}\!(1)}$ injection rates. The $1/L$ current arises because, on an average, all but a fraction $1/L$ of the influx $a$ onto site $1$ is canceled out by the outflux $D\langle m_{1}\rangle$ out of site $1$.
In case A (sec. \ref{sec:sec3}), where no exit is allowed from site $1$, injection with $\ensuremath{\mathcal{O}\!(1/L)}$ rates was required to ensure
$1/L$ current through the system and extensivity of total mass. Is model A with $1/L$ injection rates, identical to model B?
While the \emph{average} value of the particle current and total mass is identical for both cases, it turns out that other properties show significant differences. 
The origin of these differences lies in the fact that in model A, the $1/L$ current is a consequence of Poissonian injection at rate $1/L$, whereas in model B, it is $1/L$
 due to the almost complete cancellation of in and out currents at the left boundary over a sufficiently long time spanning many injection and exit events, which include the
 exit of very large mass aggregates. 
\paragraph*{}
To analyse this case, we follow the general approach used in the previous section. First, in sec. \ref{sec:sec4.1}, we write recursion relations satisfied by the steady state 
distributions $P_{i,j}(M)$ and solve for various moments $\langle M_{i,j}^{n}\rangle$ of the distributions. This also allows us to obtain the moments $\langle M^{n}\rangle$ of the
total mass and demonstrate the occurrence of giant fluctuations $\Delta M \propto L$. The behaviour of dynamical structure functions of the total mass is discussed in sec. \ref{sec:sec4.2};
we explicitly calculate $S_{2}(t)$ and indicate how $S_{n}(t)$ for higher $n$ can be obtained. This calculation shows that structure functions in both case A and B show 
the anomalous scaling associated with intermittency, but differ in their
functional form. As in sec. \ref{sec:sec3}, the analysis is simplified by taking the continuum limit in space. Details of these calculations are presented below.    
\subsection{Statics}
\label{sec:sec4.1}
As in section \ref{sec:sec3.1}, it is possible to write time evolution equations for the distributions $P_{i,j}(M,t)$. These yield the following
 recursion relations for the steady state distributions $P_{i,j}(M)$:
\begin{subequations}
\begin{equation}
P_{i,j+1}(M)+P_{i,j-1}(M)+P_{i-1,j}(M)+P_{i+1,j}(M)-4P_{i,j}(M)=0 \qquad i>0,j>i+1,j<L
\label{eq4.1.1a}
\end{equation}
\begin{equation}
\begin{split}
a[(1-\delta_{M,0})P_{0,j}(M-1)-P_{0,j}(M)]+D[P_{0,j+1}(M)+P_{0,j-1}(M)+P_{1,j}& (M)-3P_{0,j}(M)]=0 \\
& i=0,j>i+1,j<L\\
\end{split}
\label{eq4.1.1b}
\end{equation}
\begin{equation}
P_{i,L-1}(M)+P_{i-1,L}(M)+P_{i+1,L}(M)-3P_{i,L}(M)=0 \qquad  i>0,j>i+1,j=L
\label{eq4.1.1c}
\end{equation}
\begin{equation}
2\delta_{M,0}+P_{i,i+2}(M)+P_{i-1,i+1}(M)-4P_{i,i+1}(M)=0 \qquad  i>0,j=i+1,j<L
\label{eq4.1.1d}
\end{equation} 
\label{eq4.1.1}
\end{subequations}
Unlike eq. \eqref{eq3.1.4b} in section 3, the $i=0$ equation in this case does not decouple from the
other equations. Hence, we need to solve the full system of equations in eq. \eqref{eq4.1.1} to obtain the distribution of total mass.
To do this, we define the generating function $G_{i,j}(z)=\sum\limits_{M=1}^{\infty} P_{i,j}(M)z^{M}$ and take the continuum limit: $i/L\rightarrow x$, 
$j/L\rightarrow y$, $G_{i,j}(z)\rightarrow G(x,y,z)$. The function $G(x,y,z)$ then satisfies:
\begin{subequations}
 \begin{equation}
 \frac{\partial^{2}{G(x,y,z)}}{\partial{x^{2}}}+\frac{\partial^{2}{G(x,y,z)}}{\partial{y^{2}}}=0 \qquad x<y
\label{eq4.1.2a}
\end{equation}
\begin{equation}
 G(x,x,z)=0,  \qquad  \qquad \left. \frac{\partial{G}}{\partial{y}}\right|_{y=1}=0, \qquad  \qquad \left. \frac{\partial{G}}{\partial{x}}\right|_{x=0}=
\frac{aL}{D}\left[(1-z)G(0,y,z)-zP_{0,j=yL}(0)\right]
\label{eq4.1.2b}
\end{equation}
\label{eq4.1.2}
\end{subequations}
The boundary condition at $x=0$ in this case is a Robin boundary condition, for which we have not been able to solve the corresponding Laplace equation on a square. However, the functions
 $g_{n}(x,y)=\partial^{n}{G}/\partial{z^{n}}|_{z=1}$ satisfy equations that are easier to solve. The functions $g_{n}(x,y)$,
in turn give $\langle M^{n}_{xy} \rangle$ which are just the continuum limit versions of the moments $\langle M^{n}_{i,j} \rangle$. The equations for $g_{n}(x,y)$ 
[obtained by differentiating eq. \eqref{eq4.1.2}] are:
\begin{subequations}
 \begin{equation}
 \frac{\partial^{2}{g_{n}(x,y)}}{\partial{x^{2}}}+\frac{\partial^{2}{g_{n}(x,y)}}{\partial{y^{2}}}=0 \qquad x<y
\label{eq4.1.3a}
\end{equation}
\begin{equation}
\begin{split}
& \qquad \qquad \qquad g_{n}(x,x)=0, \qquad \left. \frac{\partial{g_{n}(x,y)}}{\partial{y}}\right|_{y=1}=0, \qquad 
\left. \frac{\partial{g_{n}(x,y)}}{\partial{x}}\right|_{x=0}=-\zeta ng_{n-1}(0,y) \\
& \text{where } \qquad  \zeta=\frac{aL}{D}\qquad \text{and} \quad g_{0}(0,y)=1 \\
\end{split}
\label{eq4.1.3b}
\end{equation}
\label{eq4.1.3}
\end{subequations}
Note that unlike the generating function $G(x,y,z)$, the functions $g_{n}(x,y)$ obey Neumann boundary conditions at $x=0$ and $y=1$. We solve the
 above system of equations recursively, with the solution of $g_{n}(x,y)$ providing the boundary conditions for $g_{n+1}(x,y)$ and so on. 
 The expressions for $\langle M^{n}_{xy} \rangle$,
thus obtained (details in appendix \ref{appendix2.1}), retaining only leading order terms in $\zeta=aL/D$ are:
\begin{subequations}
\begin{equation}
 \langle M_{xy} \rangle = \frac{\zeta}{2}(x-y)(x+y-2)
\label{eq4.1.4a}
\end{equation}
 \begin{equation}
\begin{split}
\langle M_{xy}^{2}\rangle=& \frac{\zeta^{2}}{3}(x-y)(x+y-2) +\\
& 4\zeta^{2}\sum\limits_{n=1}^{\infty} \frac{1}{(n\pi)^{3}}\left[\frac{\cos[n\pi x]\cosh[n\pi (1-y)]-\cos[n\pi y]\cosh[n\pi (1-x)]}{\sinh[n\pi]}\right] \\
\end{split}
\label{eq4.1.4b}
\end{equation}

 \begin{equation}
\begin{split}
\langle M_{xy}^{3}\rangle=&\frac{2\zeta^{3}}{5}(x-y)(x+y-2)+ \\ 
& 12\zeta^{3}\sum\limits_{n=1}^{\infty} \left[\frac{\cosech[n\pi]}{(n\pi)^{4}}\left(\coth[n\pi]-\frac{1}{n\pi}\right)\right]
\left[\frac{\cos[n\pi x]\cosh[n\pi (1-y)]-\cos[n\pi y]\cosh[n\pi (1-x)]}{\sinh[n\pi]}\right]
\end{split}
\label{eq4.1.4c}
\end{equation}

\begin{equation}
\begin{split}
\langle M_{xy}^{4}\rangle=&8\zeta^{4}\left[\frac{1}{15}-3\sum\limits_{n=1}^{\infty}\left(\frac{1}{(n\pi)^{6}}-\frac{2\coth[n\pi]}{(n\pi)^{5}}\right)\right](x-y)(x+y-2) \\ 
& +16\zeta^{4}\sum\limits_{n=1}^{\infty}\vast\{\left[\frac{\cosech[n\pi]}{(n\pi)^{3}}\left(12\sum\limits_{m=1}^{\infty}\frac{\coth[m\pi]}{m\pi(m^{2}\pi^{2}+n^{2}\pi^{2})}
 +\frac{6\coth^{2}[n\pi]-1}{(n\pi)^{2}}-\frac{3}{(n\pi)^{4}}\right)\right]\times\\
&\left[\frac{\cos[n\pi x]\cosh[n\pi (1-y)]-\cos[n\pi y]\cosh[n\pi (1-x)]}{\sinh[n\pi]}\right]\vast\}
\end{split}
\label{eq4.1.4d}
\end{equation}
\label{eq4.1.4}
\end{subequations}

Equation \eqref{eq4.1.4} also gives the moments of the total mass $M=M_{01}$ in the system. To leading order in $\zeta=aL/D$, these are:
\begin{subequations}
\begin{equation}
\langle M\rangle =\zeta/2, \qquad \langle M^{2}\rangle\sim 0.461\zeta^{2}, \qquad \langle M^{3}\rangle\sim 0.615\zeta^{3}, 
\qquad \langle M^{4}\rangle\sim 1.074\zeta^{4}
\label{eq4.1.5a}
\end{equation}
\begin{equation}
\langle [M-\langle M\rangle]^{2}\rangle \sim 0.211\zeta^{2}, \qquad \langle [M-\langle M\rangle]^{3}\rangle \sim 0.173\zeta^{3}, \qquad  \langle [M-\langle M\rangle]^{4}\rangle \sim 0.349\zeta^{4}
\label{eq4.1.5b}
\end{equation}
\label{eq4.1.5}
\end{subequations}

\begin{wrapfigure}{r}{0.4\textwidth}
\begin{center}
\vspace{-0.5cm}
\includegraphics[width=0.4\textwidth]{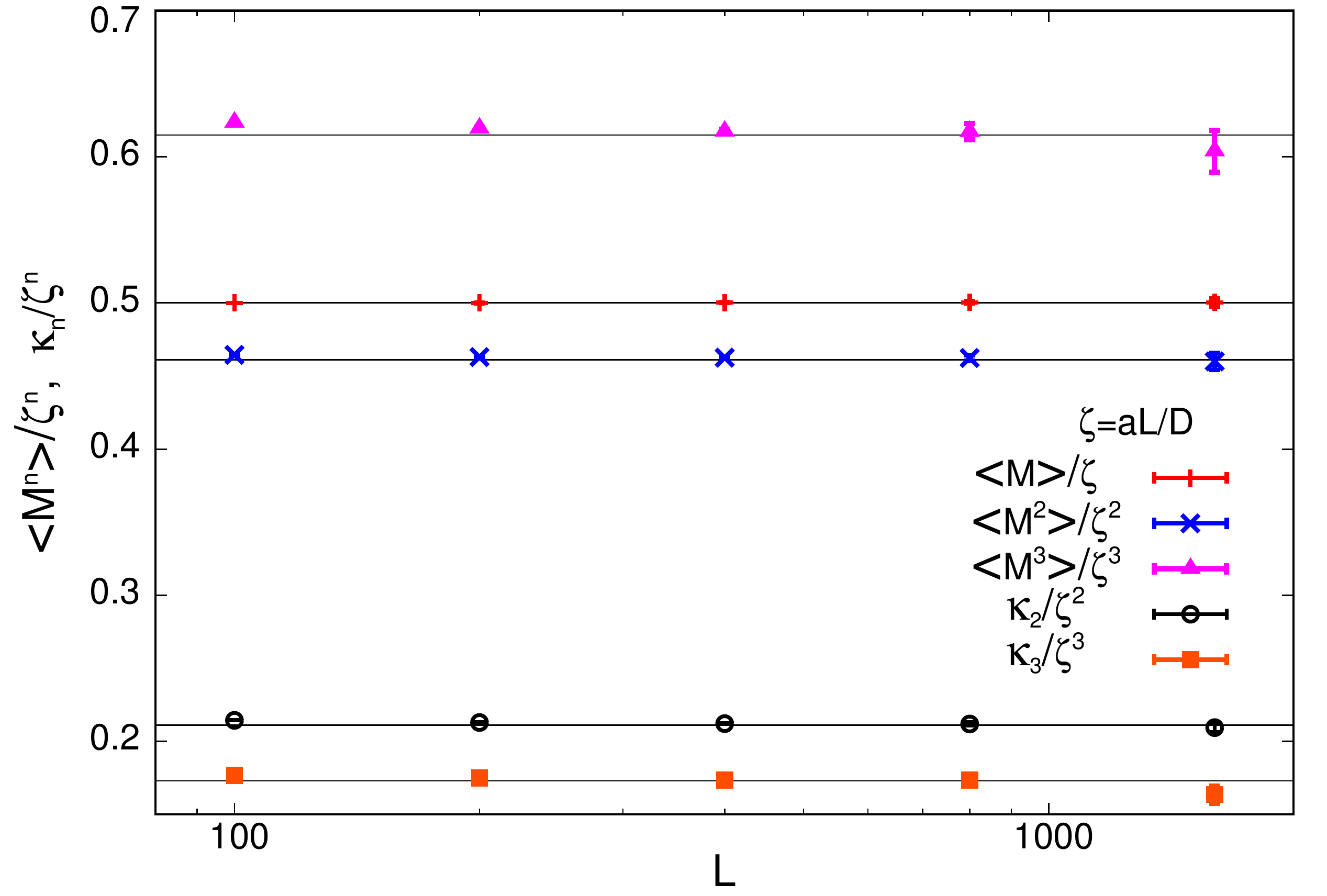}
\end{center}
\vspace{-0.3cm}
\caption{$\langle M^{n}\rangle/\zeta^{n}$ vs. $L$ for $n=1,2,3$. Cumulants $\kappa_{2}=\langle[M-\langle M\rangle]^{2}\rangle$ and
 $\kappa_{3}=\langle[M-\langle M\rangle]^{3}\rangle$ plotted as $\kappa_{n}/\zeta^{n}$ vs. $L$. Solid lines are the analytical predictions given by eq. \eqref{eq4.1.5}}
\label{fig6}
\end{wrapfigure}

\paragraph*{}
 Note that the various moments and cumulants of the total mass in this case scale as the same power of $L$ as in model A in sec. \ref{sec:sec3} (with $1/L$ injection 
rates). In particular, the rms fluctuations of the total mass are again anomalously large, scaling as $L$ rather
 than $\sqrt{L}$, so that $\Delta M/\langle M\rangle$ is finite, even in the limit $L\rightarrow \infty$.
The large fluctuations of total mass in this model also have their origin in the formation (and exit) of a macroscopic aggregate with a characteristic mass proportional
to $L$.  A comparison with numerics (fig. \ref{fig6}) shows that the numerical prefactors obtained in eq. \eqref{eq4.1.5} are close to exact. Note that these 
are different from the corresponding pre-factors in eq. \eqref{eq3.1.14} for model A in sec. \ref{sec:sec3.1}.

\paragraph*{}
The method that we have used to derive $\langle M^{n}_{xy}\rangle$ can also be generalised to calculate moments of the $r$-sector probability 
$P_{x_{1},x_{2},...x_{r+1}}\left(M_{1},M_{2},.. M_{r}\right)$ which is the joint probability of finding masses $M_{1}$, $M_{2}$,\ .. $M_{r}$ in $r$ contiguous stretches 
$x_{1}$ to $x_{2}$, $x_{2}$ to $x_{3}$, ... $x_{r}$ to $x_{r+1}$ of the lattice respectively. Moments of such $r$-sector probability distributions will be required in
sec. \ref{sec:sec4.2} for the computation of structure functions of order 3 or more. For example, we need $\langle M_{xy}^{2}M_{y1}\rangle$ in order to calculate $S_{3}(t)$
 and moments such as $\langle M_{xy}^{2}M_{yz}M_{z1}\rangle$ to calculate $S_{4}(t)$ and so on.
\paragraph*{}
We indicate below how such multi-sector moments can be calculated. For simplicity, consider the two-sector distribution $P_{i,j,k}(M_{1},M_{2})$. This evolves in time by exchange
 of mass at sites $i$, $j$ and $k$, and in steady state, satisfies the discrete Laplace equation in three variables. We can, as before, use the generating function 
approach, take the continuum limit in space, and obtain the differential equation satisfied by the moment $\langle M_{xy}^{2}M_{yz}\rangle$ where $0\leq x\leq y\leq z\leq 1$.
 This equation is just the three dimensional Laplace equation on a pyramidal region with mixed boundary conditions (Neumann and Dirichlet) on various faces of the pyramid.
 It can be solved in the same way as the 2D equation considered so far, i.e. by mapping it onto a Laplace equation on a cube with appropriately chosen boundary conditions 
(see appendix \ref{appendix2.1} for details). This method can also be generalised to higher $r$: in general, to obtain moments of 
$P_{x_{1},x_{2},...x_{r+1}}\left(M_{1},M_{2},.., M_{r}\right)$, it is necessary to solve the $r+1$ dimensional Laplace equation (on an $r+1$ dimensional hyper-pyramid). 
However, while the generalisation to $r\geq 2$ is conceptually straightforward, the actual calculations become very cumbersome and we do not pursue this further here.
 
\subsection{Dynamics}
\label{sec:sec4.2} 
We now turn to the computation of dynamical structure functions $S_{n}(t)$ for this model. We will explicitly calculate only $S_{2}(t)$, and indicate how higher order
structure functions can be obtained.
\paragraph*{}
As discussed in sec. \ref{sec:sec3.2}, the structure function of order $n$ can be obtained from the corresponding
 autocorrelation functions of the total mass. Thus, to calculate $S_{2}(t)$, we need $\langle M(0)M(t)\rangle-\langle M(0)\rangle\langle M(t)\rangle$,
 which is  a special case of the general spatio-temporal correlation function $\langle M_{0,L}(0)M_{i,j}(t)\rangle-\langle M_{0,L}(0)\rangle\langle M_{i,j}(t)\rangle$
 or its continuum limit version: $H_{2}(x,y,t)=\langle M_{01}(0)M_{xy}(t)\rangle-\langle M_{01}(0)\rangle\langle M_{xy}(t)\rangle$. 
\paragraph*{}
By following a procedure similar to that followed in section \ref{sec:sec3.2}, we obtain the following time-dependent equation for 
$H_{2}(x,y,t)$:
\begin{subequations}
\begin{equation}
\frac{\partial{H_{2}(x,y,t)}}{\partial{t}}=\frac{D}{L^{2}}\left[\frac{\partial^{2}{H_{2}(x,y,t)}}{\partial{x^{2}}}+\frac{\partial^{2}{H_{2}(x,y,t)}}{\partial{y^{2}}}\right]
\label{eq4.2.1a}
\end{equation}
\begin{equation}
\left.\frac{\partial{H_{2}}}{\partial{x}}\right|_{x=0}=0, \qquad \left.\frac{\partial{H_{2}}}{\partial{y}}\right|_{y=1}=0, \qquad H_{2}(x,x,t)=0.
\label{eq4.2.1b}
\end{equation}
\begin{equation}
\begin{split}
 H_{2}(x,y,0)&=2\langle M_{01}M_{xy}\rangle-2\langle M_{01}\rangle\langle M_{xy}\rangle\\
& = 2\langle M_{0x}M_{xy}\rangle+2\langle M_{xy}^{2}\rangle+2\langle M_{xy}M_{y1}\rangle-2\langle M_{01}\rangle\langle M_{xy}\rangle\\
& =\langle M_{0y}^{2}\rangle-\langle M_{0x}^{2}\rangle+\langle M_{x1}^{2}\rangle-\langle M_{y1}^{2}\rangle-2\langle M_{01}\rangle\langle M_{xy}\rangle
\end{split}
\label{eq4.2.1c}
\end{equation}
\label{eq4.2.1}
\end{subequations}

This is the 2D heat equation on a right triangle with mixed boundary conditions and a specified initial state $H_{2}(x,y,0)$ which is expressible in terms of single-sector mass
 moments $\langle M_{xy} \rangle$ and $\langle M^{2}_{xy} \rangle$ [eq. \eqref{eq4.1.4}]. 
Equation \eqref{eq4.2.1} can be solved (details in appendix \ref{appendix2.3}); the $x=0$, $y=1$ solution then gives $\langle M(0)M(t)\rangle-\langle M(0)\rangle\langle M(t)\rangle$, from which we obtain the 
following expression for $S_{2}(t)$:

\begin{equation}
 S_{2}(t)=\sum\limits_{n=1,3,5...}^{\infty} \left\{\frac{16\zeta^{2}}{(n \pi)^{4}}\left(n\pi\coth\left[\frac{n\pi}{2}\right]-1\right) 
+\frac{4\zeta}{(n\pi)^{2}}\right\}\left\{1- \exp\left[-\frac{D\pi^{2} n^{2}t}{L^{2}}\right]\right\}, \qquad \zeta=\frac{aL}{D}
\label{eq4.2.2}
\end{equation}

\paragraph*{}
To study intermittency, we need to to extract the small $t$ behaviour of $S_{2}(t)$ from eq. \eqref{eq4.2.2}. This cannot be done by simply expanding 
$1- \exp\left[-\frac{D\pi^{2} n^{2}t}{2L^{2}}\right]$ as $\sim \frac{D\pi^{2} n^{2}t}{L^{2}}$, as the resultant sum over $n$ diverges due to the presence of the $\coth$ term.
Instead, we keep the exponential term as it is, and approximate the discrete sum in eq. \eqref{eq4.2.2} by an integral over $n$. On taking the limit $Dt/L^{2}\rightarrow0$
 of this integral (details in appendix \ref{appendix1.2}), we get the following small $t$ form of $S_{2}(t)$:
\begin{equation}
 S_{2}(t) \sim  -\frac{4}{\pi}\left(\frac{aL}{D}\right)^{2}\frac{Dt}{L^{2}}\log\left[A_{1}\frac{Dt}{L^{2}}\right] 
\label{eq4.2.4}
\end{equation}
 where $A_{1}$ is a constant which can be derived approximately using our analysis (see appendix \ref{appendix1.2}) but is more accurately obtained by fitting to 
numerical data.
\paragraph*{}
 Higher order structure functions can be calculated using a similar procedure but the calculations become progressively
 more cumbersome. To calculate $S_{3}(t)$, we need to calculate the function $H_{3}(x,y,t)=3\langle M_{01}^{2}(0)M_{xy}(t)\rangle-3\langle M_{01}^{2}\rangle\langle M_{xy}\rangle
-3\langle M_{01}(0) M_{xy}^{2}(t)\rangle+3\langle M_{01}\rangle\langle M_{xy}^{2}\rangle$ which satisfies the following 
time evolution equation:
\begin{subequations}
\begin{equation}
\frac{\partial{H_{3}(x,y,t)}}{\partial{t}}=\frac{D}{L^{2}}\left[\frac{\partial^{2}{H_{3}(x,y,t)}}{\partial{x^{2}}}+\frac{\partial^{2}{H_{3}(x,y,t)}}{\partial{y^{2}}}\right]
\label{eq4.2.3a}
\end{equation}
\begin{equation}
\left.\frac{\partial{H_{3}}}{\partial{x}}\right|_{x=0}=3\zeta H_{2}(0,y,t), \qquad \left.\frac{\partial{H_{3}}}{\partial{y}}\right|_{y=1}=0, \qquad H_{3}(x,x,t)=0.
\label{eq4.2.3b}
\end{equation}
\begin{equation}
\begin{split}
 H_{3}(x,y,0)&=3\langle M_{01}^{2}M_{xy}\rangle-3\langle M_{01}\rangle\langle M_{xy}\rangle
-3\langle M_{01}M_{xy}^{2}\rangle+3\langle M_{01}\rangle\langle M_{xy}^{2}\rangle\\
& = \langle M_{0y}^{3}\rangle -\langle M_{0x}^{3}\rangle+\langle M_{x1}^{3}\rangle-\langle M_{y1}^{3}\rangle-2\langle M_{xy}^{3}\rangle-3\langle M_{01}^{2}\rangle \langle M_{xy}\rangle
+3\langle M_{01}\rangle \langle M_{xy}^{2}\rangle\\
& \quad + 3\left[\langle M_{0x}^{2}M_{xy}\rangle + \langle M_{0y}^{2}M_{y1}\rangle -\langle M_{0x}^{2}M_{x1}\rangle -\langle M_{xy}^{2}M_{y1}\rangle\right]
\end{split}
\label{eq4.2.3c}
\end{equation}
\label{eq4.2.3}
\end{subequations}
An important difference between equations \eqref{eq4.2.1} and \eqref{eq4.2.3} is in the nature of the initial conditions: while $H_{2}(x,y,0)$ is expressible in terms of the moments 
 $\langle M^{2}_{xy}\rangle$ and $\langle M_{xy} \rangle$ of mass in a single sector, $H_{3}(x,y,0)$ also involves terms like $\langle M_{xy}^{2}M_{y1}\rangle$
which requires knowledge of two-sector probabilities i.e. the joint probability of finding mass $M_{1}$ and $M_{2}$ in two contiguous stretches of the lattice. Similarly, the 
expression for $H_{4}(x,y,0)$ involves terms like $\langle M_{xy}^{2}M_{yz}M_{z1}\rangle$ for which we need to calculate joint probabilities of mass in three
contiguous stretches. As discussed at the end of section \ref{sec:sec4.1}, the calculation of the moments of these multi-region joint probabilities is rather cumbersome. 
However, in principle, it can be done, thus giving the $t=0$ value of correlation functions such as $H_{3}(x,y,t)$. Once this $t=0$ value is known, eq. \eqref{eq4.2.3} can be
solved (details in appendix \ref{appendix2.3}) to obtain $S_{3}(t)$.
\paragraph*{}
We have also studied higher order structure functions using numerical simulations. Numerics show that they obey the scaling: $S_{n}(t)= L^{n} \ensuremath{\mathcal{F}}_{n}(Dt/L^{2})$, which is
consistent with eq. \eqref{eq4.2.4}. Motivated by the small $t$ expression for $S_{2}(t)$ [eq. \eqref{eq4.2.4}], we try to fit $\ensuremath{\mathcal{F}}_{n}$ to the form
 $\ensuremath{\mathcal{F}}_{n}(y)\sim y \ensuremath{\mathcal{G}}_{n}[\log(y)]$ for small $y$. Good fits are obtained for $S_{3}(t)$ and $S_{4}(t)$ by choosing a polynomial form for the function
 $\ensuremath{\mathcal{G}}_{n}$. Thus, in case B also, structure functions for $n\geq 2$ show intermittency, scaling as $t$ but with multiplicative $\log(t)$ corrections which result in a weak $n$-dependence of 
the scaling form. 
\paragraph*{}
As before, structure functions $\tilde{S_{n}}(t)=\langle|M(t)-M(0)|^{n}\rangle$ for non-integer $n$ cannot be calculated analytically by this approach and must be studied using numerical
simulations. Numerics show that unlike the case with spatially separated influx and outflux [sec. \ref{sec:sec3}], in this model, structure functions show self-similar scaling
$\tilde{S_{n}}(t)\sim t^{n/2}$ for $n\leq 2$ and anomalous scaling for $n\geq 2$. This behaviour can be explained heuristically by considering a toy model [similar to  fig. \ref{fig5} in
sec. \ref{sec:sec3.2}] in which $M(t)$ grows as $\sqrt{t}$ instead of $\sim t$ between successive crashes. This is a plausible approximation to the real time series $M(t)$, as the real
$M(t)$ evolves through a large number of $\ensuremath{\mathcal{O}\!(1)}$ changes, which may be positive or negative, thus
resulting in an effective random walk behaviour.  
\paragraph*{}
In conclusion, a comparison of this model [model (B)] with the model studied in sec. \ref{sec:sec3} [model (A)] shows that while the static properties of the total mass in both
 cases are qualitatively similar, significant differences appear in the dynamical properties. This is apparent even in a typical time series $M(t)$ in the two cases
 [figs. \ref{fig2a} and \ref{fig2b}], where the time series corresponding to case (B) shows sharp drops of $M(t)$ at many scales. This difference manifests itself 
in the multiplicative $log(t)$ terms in the structure functions. Moreover, the anomalous, intermittency-associated scaling of $S_{n}(t)$ in this case sets in only for $n\geq 2$, 
rather than for $n\geq1$ as in model (A).

\section{Case C: Driven movement; outflux from right boundary}
\label{sec:sec5}   
\paragraph*{}
In this section, we analyse the case where single particles are injected onto site $1$ at rate $a$, aggregates move unidirectionally to the right with rate $D$, and
exit from site $L$, also with rate $D$. Note that this is the same as monitoring only the first $L$ sites of a semi-infinite lattice with injection at the origin, and unidirectional movement
of aggregates everywhere. 
\paragraph*{}
While aggregating systems with translational invariance such as the Takayasu model show behaviour which is independent of bias, 
in systems with localised injection, various properties including mass distributions can be qualitatively different, depending on whether movement of aggregates is biased or not \cite{cheng}.  For example, by writing a continuity equation like eq.
 \eqref{eq4.0.1}, it is easy to see that the average mass at a site $\langle m_{i}\rangle=a/D$ is independent of $i$, unlike the linearly falling spatial profile in the
 unbiased models with current [secs. \ref{sec:sec3} and \ref{sec:sec4}]. The distribution $p_{i}(m)$ of mass at site $i$, which has been derived in \cite{jain}, also behaves differently, decaying as $\sim \exp(-D^{2}m^{2}/4a^{2}i)$ at large $m$, 
in contrast to the the non-Gaussian form in eq. \eqref{eq3.1.13}. In this section, we extend the analysis in \cite{jain} to calculate the distribution of mass in an 
arbitrary region of the system. We then use these results to compute the dynamical structure functions $S_{n}(t)$ and show that this system also exhibits strong temporal 
intermittency of the total mass, though at much shorter time scales than the unbiased system. Unlike the previous two sections, in this case, we work with recursion 
relations directly, without taking a continuum limit in space.
\subsection{Statics}
\label{sec:sec5.1}
\paragraph*{}
 For the calculations in this section, it turns out to be more convenient to define $M_{i,j}$ and
$P_{i,j}(M)$ in the following way:
\begin{equation}
M_{i,j}=\sum\limits_{l=i+1}^{i+j} m_{l} \qquad 
P_{i,j}(M)=\text{Prob}(M_{i,j}=M)
\label{eq5.1.1}
\end{equation}
Thus, $P_{i,j}(M)$ is now the probability of finding mass $M$ in the region [$i+1$, $i+j]$.
 As in the unbiased case with outflux only from site $L$ (section \ref{sec:sec3}),
the equations for  $P_{0,j}(M)$ decouple from $P_{i,j}(M)$ with $i>0$. In steady state, $P_{0,j}(M)$ satisfy the following recursion relations:
\begin{subequations}
 \begin{equation}
a[1-\delta_{M,0}]P_{0,j}(M-1)-aP_{0,j}(M)+D[P_{0,j-1}(M)-P_{0,j}(M)]=0 \qquad j>1
\label{eq5.1.2a}
\end{equation}
\begin{equation}
a[1-\delta_{M,0}]P_{0,1}(M-1)-aP_{0,1}(M)+D[\delta_{M,0}-P_{0,1}(M)]=0 \qquad j=1
\label{eq5.1.2b}
\end{equation}
\label{eq5.1.2}
\end{subequations}

From eq. \eqref{eq5.1.2}, it follows that the generating function $Q_{j}(z)=\sum\limits_{M=0}^{\infty}P_{0,j}(M)z^{M}$ satisfies:
\begin{equation}
 Q_{j}(z)=\frac{Q_{j-1}(z)}{1+\eta(1-z)}, \qquad Q_{0}(z)=1, \qquad \eta=a/D
\label{eq5.1.3}
\end{equation}
From eq. \eqref{eq5.1.3}, we can obtain the generating function $Q_{L}(z)$ of the probability distribution of mass in a system of $L$ sites:
\begin{equation}
 Q_{L}(z)=\frac{1}{[1+\eta(1-z)]^{L}}
\label{eq5.1.4}
\end{equation}
Inverting $Q_{L}(z)$ yields the following expression for the distribution $P(M)$ of mass $M$ in a system of $L$ sites:
\begin{equation}
 P(M)=\frac{\eta^{M}}{(1+\eta)^{L+M}}{{L+M-1}\choose M}
\label{eq5.1.5}
\end{equation}
Various moments and cumulants of $M$ can also be worked out from eq. \eqref{eq5.1.4}. They are:
\begin{subequations}
\begin{equation}
\langle M\rangle=\eta L  \qquad
\langle M^{2}\rangle=\eta^{2}L^{2}+\eta(1+\eta)L  \qquad
\langle M^{3}\rangle=\eta^{3}L^{3}+ 3\eta^{2}(1+\eta)L^{2}+\eta(1+\eta)(1+2\eta)L  
\label{eq5.1.6a}
\end{equation}
\begin{equation}
 \Delta M^{2}=\langle[M-\langle M\rangle]^{2}\rangle=\eta(1+\eta)L, \qquad \langle[M-\langle M\rangle]^{3}\rangle=\eta(1+\eta)(1+2\eta)L
\label{eq5.1.6b}
\end{equation}
\label{eq5.1.6}
\end{subequations}
 In fact, it follows from eq. \eqref{eq5.1.4} that all cumulants of the total mass scale as $L$. 
This calculation, thus demonstrates an important difference between the driven and diffusive cases: the total mass in the driven case \emph{does not have a broad distribution}.
 Specifically, the rms fluctuation scales as $\sqrt{L}$, and not $L$ as was found in the the diffusive case. Further, 
in the limit $L\rightarrow \infty$, the distribution $P(M)$ in eq. \eqref{eq5.1.15} approaches a Gaussian distribution for the variable $(M-\langle M\rangle)/\Delta M$. 
\paragraph*{}
Another point of difference from the diffusive case is that the recursion relation in eq. \eqref{eq5.1.3} cannot be solved by taking a
 continuum limit in space and replacing $Q_{j}-Q_{j-1}$ by 
$\partial Q/\partial x$.
 This follows from a self-consistency argument similar to that employed in sec. \ref{sec:sec3.1}. 
\footnote{A continuum approximation would be self-consistent if $\eta\propto L^{-\alpha}$, but this would result in sub-extensive total mass}
Thus, in the analysis of the driven case, we need to necessarily work with recursion
 relations rather than differential equations. 
\paragraph*{}
We now turn to the calculation of $P_{i,j}(M)$ for $i>0$. Most of this analysis has already been carried out in \cite{jain} but we briefly summarise it here in the interest of
continuity. In steady state, the probabilities $P_{i,j}(M)$ satisfy the following recursion relations:
\begin{subequations}
\begin{equation}
P_{i-1,j+1}(M)+P_{i,j-1}(M)-2P_{i,j}(M)=0 \qquad i>0,j>1
\label{eq5.1.7a}
\end{equation}
\begin{equation}
P_{i-1,2}(M)+\delta_{M,0}-2P_{i,1}(M)=0 \qquad j=1
\label{eq5.1.7b}
\end{equation}
\label{eq5.1.7}
\end{subequations}
The equation for the case $i=0$ [eq. \eqref{eq5.1.2}] decouples from these and has already been solved. From eq. \eqref{eq5.1.7} and \eqref{eq5.1.3}, it follows that the 
generating functions $F_{i,j}(M)=\sum\limits_{M=0}^{\infty}P_{i,j}(M)z^{M}$ satisfy:
\begin{subequations}
\begin{equation}
 F_{i-1,j+1}(z)+F_{i,j-1}(z)-2F_{i,j}(z)=0, \qquad i>0,j>1
\label{eq5.1.8a}
\end{equation}
\begin{equation}
 F_{i,0}(z)=1,   \qquad F_{0,j}(z)=Q_{j}(z)=\frac{1}{[1+\eta(1-z)]^{j}}, \quad \eta=a/D
\label{eq5.1.8b}
\end{equation}
\label{eq5.1.8}
\end{subequations}
The above set of equations can be solved using the functions:
\begin{equation}
G_{u,j}(z)=\sum\limits_{i=1}^{\infty}F_{i,j}(z)u^{i} \qquad H_{u,v}(z)=\sum\limits_{j=1}^{\infty}G_{u,j}(z)v^{j} 
\label{eq5.1.9}
\end{equation}
By performing the sums over both $i$ and $j$ on eq. \eqref{eq5.1.8}, the following equation for $H_{u,v}(z)$ is obtained:
\begin{equation}
H_{u,v}(z)=\frac{uvG_{u,1}(z)-\frac{uv^{2}}{1-u}-\frac{uv^{2}}{[1+\eta(1-z)][1-v+\eta(1-z)]}}{v^{2}-2v+u}
\label{eq5.1.10}
\end{equation}
\paragraph*{}
Equation \eqref{eq5.1.10} can be solved to get a closed form expression for $H_{u,v}(z)$ in the same way as in \cite{jain}: the denominator of the R.H.S of eq. \eqref{eq5.1.10} can be 
expressed as $v^{2}-2v+u=(v_{-}-v)(v_{+}-v)$ where $v_{\pm}=1\pm \sqrt{1-u}$. Since the denominator has a zero at $v=v_{-}$ which is less than $1$, it follows that the numerator must also 
have a zero at $v=v_{-}$, to ensure that the function $H_{u,v}(z)$ is analytic for $v<1$. This condition allows us to solve for $G_{u,1}(z)$:
\begin{equation}
G_{u,1}(z)=\frac{v_{-}}{1-u}+\frac{v_{-}}{[1+\eta(1-z)][1-v_{-}+\eta(1-z)]} \qquad \text{where $v_{-}=1- \sqrt{1-u}$}
\label{eq5.1.11}
\end{equation}
Note that inverting $G_{u,1}(z)$ w.r.t $u$ and $z$ would give the single site mass distribution $p_{i}(m)$. This was done in \cite{jain}, and it was established that the probability of 
finding an aggregate at a distance $i$ from the source (point of injection) decays as $1/\sqrt{i}$ while the \emph{typical} size of the aggregate scales as $\sqrt{i}$. 
Thus, the exit events in a system with $L$ sites typically involve aggregates with $\ensuremath{\mathcal{O}\!(\sqrt{L})}$ mass, and occur at intervals of 
$\ensuremath{\mathcal{O}\!(\sqrt{L})}$ duration.
\paragraph*{}
We now extend the above analysis to find $F_{i,j}(z)$. Substituting from eq. \eqref{eq5.1.11} into eq. \eqref{eq5.1.10} allows for the determination of $H_{u,v}(z)$:
\begin{equation}
 H_{u,v}(z)=\frac{uv}{1-v+\sqrt{1-u}}\left[\frac{1}{1-u}+\frac{1}{[1-v+\eta(1-z)][\sqrt{1-u}+\eta(1-z)]}\right]
\label{eq5.1.12} 
\end{equation}
 $H_{u,v}(z)$ can be Taylor expanded in powers of $v$ to recover $G_{u,j}(z)$:
\begin{equation}
 G_{u,j}(z)=\left(\frac{u}{1-u}\right)\left(\frac{\eta^{2}(1-z)^{2}}{\eta^{2}(1-z)^{2}-(1-u)}\right)
\left[[1+\sqrt{1-u}]^{-j}-\left(\frac{1-u}{\eta^{2}(1-z)^{2}}\right)[1+\eta(1-z)]^{-j}\right]
\label{eq5.1.13}
\end{equation} 

Using $[1+\sqrt{1-u}]^{-j}=\sum\limits_{k=0}^{\infty}2^{-2k-j}\frac{j}{j+2k}{{j+2k}\choose{k}} u^{k}$, it is possible to invert $G_{u,j}(z)$ and get a formal expression for $F_{i,j}(z)$:
\begin{equation}
 F_{i,j}(z)=[1+\eta(1-z)]^{-j-i}[1-\eta(1-z)]^{-i}+j2^{-j}\sum\limits_{k=0}^{i-1}\left[\frac{2^{-2k}}{j+2k}{{j+2k}\choose{k}}\left\{1-[1-\eta^{2}(1-z)^{2}]^{k-i}\right\}\right]
\label{eq5.1.14}
 \end{equation}
Formal expressions for the moments $\langle M^{n}_{i,j}\rangle$ of the mass in the region [$i+1$, $i+j$] can be obtained by differentiating eq. \eqref{eq5.1.14}.
 In the limit $i\rightarrow \infty$, $j\rightarrow \infty$ with $j/\sqrt{i}$ finite, however, simpler asymptotic  expressions for 
$\langle M^{n}_{i,j}\rangle$ emerge.  An analysis of eq. \eqref{eq5.1.13} [details in appendix \ref{appendix1.3}] shows that in this limit, $\langle M^{n}_{i,j}\rangle$ satisfy the
simple scaling form $\langle M_{i,j}^{n}\rangle \sim i^{n/2}\ensuremath{\mathcal{R}}_{n}\left(\frac{j}{\sqrt{i}}\right)$:
 \begin{subequations}
\begin{equation}
\langle M_{i,j}\rangle=\eta j 
\label{eq5.1.15a}
\end{equation}
\begin{equation}
\langle M_{i,j}^{2}\rangle \sim i \ensuremath{\mathcal{R}}_{2}\left(\frac{j}{\sqrt{i}}\right) \sim \frac{4\eta^{2}i}{\sqrt{\pi}}\left[\left(\frac{j}{\sqrt{i}}\right)+\frac{1}{12}\left(\frac{j}{\sqrt{i}}\right)^{3}+....\right]
\label{eq5.1.15b}
\end{equation}
\begin{equation}
\langle M_{i,j}^{3}\rangle \sim i^{3/2} \ensuremath{\mathcal{R}}_{3}\left(\frac{j}{\sqrt{i}}\right) \sim 6\eta^{3}i^{3/2}\left[\left(\frac{j}{\sqrt{i}}\right)+\frac{1}{6}\left(\frac{j}{\sqrt{i}}\right)^{3}+....\right]
\label{eq5.1.15c}
\end{equation}
\begin{equation}
\langle M_{i,j}^{4}\rangle \sim i^{2} \ensuremath{\mathcal{R}}_{4}\left(\frac{j}{\sqrt{i}}\right) \sim \frac{32\eta^{4}{i^{2}}}{\sqrt{\pi}}\left[\left(\frac{j}{\sqrt{i}}\right)+\frac{1}{4}\left(\frac{j}{\sqrt{i}}\right)^{3}+....\right]
\label{eq5.1.15d}
\end{equation}
\label{eq5.1.15}
\end{subequations}
As discussed in the next section, this scaling limit also turns out to be crucial for the analysis of the intermittency properties of the system. 

\subsection{Dynamics}
\label{sec:sec5.2}
To calculate structure functions, we follow the general approach of sections \ref{sec:sec3.2} and \ref{sec:sec4.2}, i.e. we first find the correlation functions defined
 in eq. \eqref{eq3.2.3} and then relate them to structure
functions using eq. \eqref{eq3.2.4}. Let $f_{j}(t)$, $g_{j}(t)$ and $h_{j}(t)$ denote the following correlation functions: 
\begin{subequations}
\begin{equation}
f_{j}(t)=2C_{2}(j,t) 
\label{eq5.2.1a}
\end{equation}
\begin{equation}
g_{j}(t)=3C_{31}(j,t)-3C_{32}(j,t)  
\label{eq5.2.1b}
\end{equation}
\begin{equation}
h_{j}(t)= 4C_{41}(j,t)-6C_{42}(j,t)+4C_{43}(j,t)
\label{eq5.2.1c}
\end{equation}
\label{eq5.2.1}
\end{subequations}
where $C_{2}(j,t)$, $C_{31}(j,t)$ etc. are defined in eq. \eqref{eq3.2.3}. By solving for these correlation functions and using the $j=L$ solution, various structure functions can
 be obtained from: 
\begin{subequations}
\begin{equation}
 S_{2}(t)=2\langle M_{0,L}^{2}\rangle-2\langle M_{0,L}\rangle^{2}- f_{L}(t)
\label{eq5.2.2a}
\end{equation}
\begin{equation}
 S_{3}(t)= g_{L}(t)
 \label{eq5.2.2b}
\end{equation}
\begin{equation}
 S_{4}(t)= 2\langle M_{0,L}^{4}\rangle-8\langle M_{0,L}^{3}\rangle\langle M_{0,L}\rangle+6\langle M_{0,L}^{2}\rangle^{2}-h_{L}(t)
\label{eq5.2.2c}
\end{equation}
\label{eq5.2.2}
\end{subequations}
The time evolution equations satisfied by $f_{j}(t)$, $g_{j}(t)$ and $h_{j}(t)$ can be derived from the time evolution equation for $M_{0,j}(t)$ by following the same procedure as in
 sec. \ref{sec:sec3.2}. These equations are:
\begin{subequations}
\begin{equation}
f_{j}(t+1)=(1-D)f_{j}(t)+Df_{j-1}(t)  
\label{eq5.2.3a}
\end{equation}
\begin{equation}
g_{j}(t+1)=(1-D)g_{j}(t)+Dg_{j-1}(t)-3af_{j}(t)  
\label{eq5.2.3b}
\end{equation}
\begin{equation}
h_{j}(t+1)=(1-D)h_{j}(t)+Dh_{j-1}(t)-4ag_{j}(t)+6af_{j}(t)  
\label{eq5.2.3c}
\end{equation}
\label{eq5.2.3}
\end{subequations}
 
These equations can be solved for $t<j$ by induction to obtain:
\begin{subequations}
\begin{equation}
f_{j}(t)=\sum_{k=0}^{t}{t\choose k}(1-D)^{t-k}D^{k}f_{j-k}(0)  
\label{eq5.2.4a}
\end{equation}
\begin{equation}
g_{j}(t)=\sum_{k=0}^{t}\left[{t\choose k}(1-D)^{t-k}D^{k}g_{j-k}(0)\right]  -3atf_{j}(t-1)
\label{eq5.2.4b}
\end{equation}
\begin{equation}
h_{j}(t)=\sum_{k=0}^{t}\left[{t\choose k}(1-D)^{t-k}D^{k}h_{j-k}(0)\right]  -at[4g_{j}(t-1)-6f_{j}(t-1)]-12a^{2}\frac{t(t-1)}{2}f_{j}(t-2)
\label{eq5.2.4c}
\end{equation}
\label{eq5.2.4}
\end{subequations}
Equation \eqref{eq5.2.4} expresses $f_{j}(t)$, $g_{j}(t)$ and $h_{j}(t)$ in terms of their $t=0$ values. The $t=0$ values $f_{j}(0)$, $g_{j}(0)$ and $h_{j}(0)$ can be expressed in terms of
$\langle M_{0,j}^{n}\rangle$ and $\langle M_{j,L-j}^{n}\rangle$ as follows:
\begin{subequations}
 \begin{equation}
\begin{split}
f_{j}(0)=\langle M_{0,L}^{2}\rangle+\langle M_{0,j}^{2}\rangle-\langle M_{j,L-j}^{2}\rangle-2\langle M_{0,j}\rangle\langle M_{0,L}\rangle
\end{split}
\label{eq5.2.5a}
\end{equation}
\begin{equation}
\begin{split}
g_{j}(0)& =\langle M_{0,L}^{3}\rangle-\langle M_{0,j}^{3}\rangle-\langle M_{j,L-j}^{3}\rangle-3\langle M^{2}_{0,L}\rangle\langle M_{0,j}\rangle+
3\langle M_{0,L}\rangle\langle M_{0,j}^{2}\rangle
\end{split}
\label{eq5.2.5b}
\end{equation}
\begin{equation}
\begin{split}
h_{j}(0)& =\langle M_{0,L}^{4}\rangle+\langle M_{0,j}^{4}\rangle-\langle M_{j,L-j}^{4}\rangle-4\langle M_{0,L}^{3}\rangle\langle M_{0,j}\rangle+
6\langle M_{0,L}^{2}\rangle\langle M_{0,j}^{2}\rangle-4\langle M_{0,L}\rangle\langle M_{0,j}^{3}\rangle.
\end{split}
\label{eq5.2.5c}
\end{equation}
\label{eq5.2.5}
\end{subequations}
Note that this is just eq. \eqref{eq3.2.8}, if we take into account the difference in the way $M_{i,j}$ is defined in sections \ref{sec:sec3} and \ref{sec:sec5}.
It is now possible to obtain the structure functions by using eqs. \eqref{eq5.2.4} and \eqref{eq5.2.5}. For example, $S_{2}(t)$ can be expressed as:
\begin{equation}
\begin{split}
S_{2}(t)&=2[\langle M_{L}^{2}\rangle-\langle M_{L}\rangle^{2}]-2\sum_{k=0}^{t}{t\choose k}(1-D)^{t-k}D^{k}[\langle M_{L-k}M_{L}\rangle-\langle M_{L-k}\rangle\langle M_{L}\rangle]\\
&=2[\langle M_{L}^{2}\rangle-\langle M_{L}\rangle^{2}]-\sum_{k=0}^{t}{t\choose k}(1-D)^{t-k}D^{k}[\langle M_{L}^{2}\rangle+\langle M_{L-k}^{2}\rangle-\langle M_{L-k,k}^{2}\rangle-
2\langle M_{L-k}\rangle\langle M_{L}\rangle]\\
\end{split}
\label{eq5.2.6}
\end{equation}
Since the time interval between successive exit events is $\ensuremath{\mathcal{O}\!(\sqrt{L})}$, we expect the system to show temporal intermittency in the limit where
$t /\sqrt{L}$ is small. Thus, we need to investigate the behaviour of $S_{2}(t)$ in the scaling limit $L\rightarrow \infty$, $t \rightarrow \infty$, $t/\sqrt{L}$ finite but small.  
In this limit, the dominant contribution to the sum over $k$ in eq. \eqref{eq5.2.6} must come from $k \sim \ensuremath{\mathcal{O}\!(\sqrt{L})}$. Thus, term such as 
$\langle M_{L-k,k}^{2}\rangle$ can be approximated as $\langle M_{L,k}^{2}\rangle$ for which we can use the asymptotic ($k\rightarrow\infty$, $L\rightarrow\infty$, 
$k/\sqrt{L}$ finite) expressions in eq. \eqref{eq5.1.15}.  Using the exact expressions for $\langle M_{L}^{2}\rangle$, 
$\langle M_{L-k}^{2}\rangle$ etc.  and the asymptotic expression for $\langle M_{L-k,k}^{2}\rangle$ [eq. \eqref{eq5.1.15}] in eq. \eqref{eq5.2.6} gives:
\begin{equation}
S_{2}(t)\sim \sum_{k=0}^{t}{t\choose k}(1-D)^{t-k}D^{k}\left[\frac{4\eta^{2}k\sqrt{L}}{\sqrt{\pi}}-\eta^{2}k^{2}+\frac{\eta^{2}k^{3}}{3\sqrt{\pi}\sqrt{L}}+...\right]
\label{eq5.2.7}
\end{equation}
On performing the sum over $k$, an expression in powers of $t$ is obtained. Since we are interested in the thermodynamic limit $L\rightarrow\infty$, 
we need to retain only the leading order term in $L$ in the
coefficient of each power of $t$. This finally gives:
\begin{equation}
S_{2}(t)\sim\frac{4\eta^{2}L}{\sqrt{\pi}}\left[\left(\frac{Dt}{\sqrt{L}}\right)-\frac{\sqrt{\pi}}{4}\left(\frac{Dt}{\sqrt{L}}\right)^{2}+\frac{1}{12}\left(\frac{Dt}{\sqrt{L}}\right)^{3}+...\right]
\label{eq5.2.8}
\end{equation}

\paragraph*{}
It is possible to compute higher order structure functions in exactly the same way, the only additional feature being that now the calculation of $g_{j}(t)$ involves expressions for 
$f_{j}(t)$ and so on [see eq. \eqref{eq5.2.4}].  

In the scaling limit $t\rightarrow \infty$, $L\rightarrow \infty$, $t/\sqrt{L}\rightarrow 0$, we finally obtain: 
\begin{equation}
  S_{n}(t)\equiv\langle[M_{L}(t)-M_{L}(0)]^{n}\rangle \sim L^{n/2}\ensuremath{\mathcal{G}}_{n}\left(\frac{Dt}{\sqrt{L}}\right)
\label{eq5.2.9}
\end{equation}
 where:
\begin{subequations}
\begin{equation}
\ensuremath{\mathcal{G}}_{2}(q) \sim\frac{4\eta^{2}}{\sqrt{\pi}}q\left[1-\frac{\sqrt{\pi}}{4}q+\frac{1}{12}q^{2}+...\right]
\label{eq5.2.10a}
\end{equation}
\begin{equation}
\ensuremath{\mathcal{G}}_{3}(q) \sim -6\eta^{3} q\left[1-\frac{2}{\sqrt{\pi}}q+\frac{1}{2}q^{2}+...\right]
\label{eq5.2.10b}
\end{equation}
\begin{equation}
\ensuremath{\mathcal{G}}_{4}(q) \sim\frac{32\eta^{4}}{\sqrt{\pi}}q\left[1-\frac{3\sqrt{\pi}}{4}q+q^{2}+...\right]
\label{eq5.2.10c}
\end{equation}
\label{eq5.2.10}
\end{subequations}
\begin{SCfigure}
\centering
\includegraphics[width=0.5\textwidth]{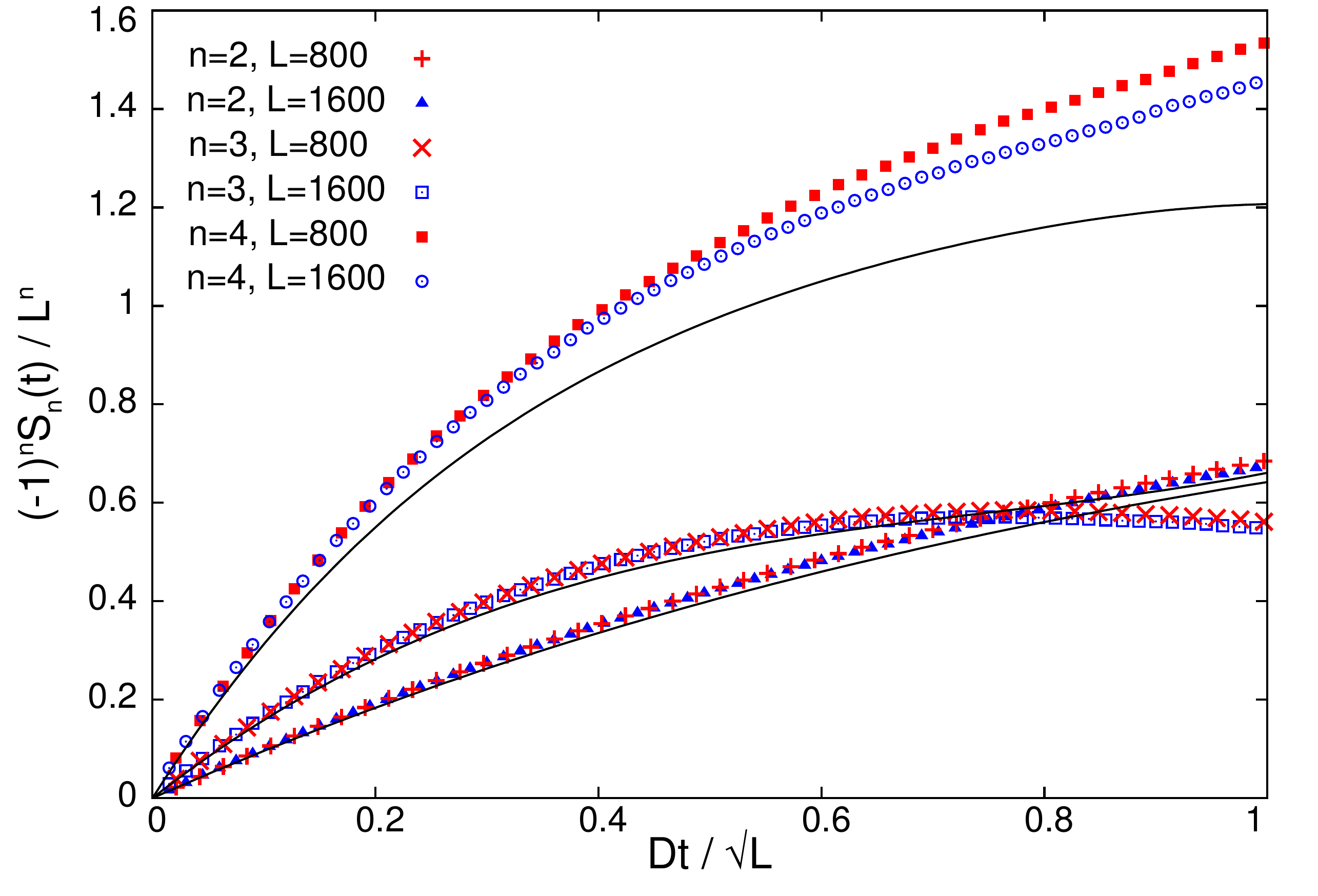}
\caption{Structure functions:  $(-1)^{n}S_{n}(t)/L^{n/2}$
 vs. $Dt/\sqrt{L}$ for $n=2,3,4$ and two different $L$. Solid lines represent the small $t$ analytical expressions in eq. \eqref{eq5.2.10}. Deviation of analytical predictions from
numerics due to corrections to scaling which decay very slowly as $1/\sqrt{L}$ with system size $L$}
\label{fig7}
\end{SCfigure}
Thus, this system also shows strong intermittency, but at time scales that are $\ensuremath{\mathcal{O}\!(\sqrt{L})}$, rather than $\ensuremath{\mathcal{O}\!(L^{2})}$. 
Figure \ref{fig7} shows $S_{2}(t)$, $S_{3}(t)$ and $S_{4}(t)$, as obtained from numerics along with the analytical predictions of eq. \eqref{eq5.2.10}. The deviation of the analytical curves
from the numerical data is due to corrections to scaling which decay very slowly as $1/\sqrt{L}$ with system size $L$. If we keep track of these sub-leading terms in our analysis,
then the resultant expressions for $S_{n}(t)$ show good agreement with numerics.
The flatness $\kappa(t)$ in the limit $t/\sqrt{L}\rightarrow 0$, as obtained from eq. \eqref{eq5.2.10} is:
\begin{equation}
\kappa(t)\sim 2\sqrt{\pi}\left(\frac{Dt}{\sqrt{L}}\right)^{-1}\left[1-\frac{\sqrt{\pi}}{4}\left(\frac{Dt}{\sqrt{L}}\right)+\left(\frac{40-9\pi}{48}\right)\left(\frac{Dt}{\sqrt{L}}\right)^{2}+...\right]
\label{eq5.2.11}
\end{equation}
This also shows an $L$-dependent divergence as $t\rightarrow 0$, but with a weaker dependence on $L$ than in the diffusive case.
\paragraph*{}
In conclusion, in this section we showed that the steady state distribution of total mass $M$ has a Gaussian form, with rms fluctuations that scale as $\sqrt{L}$ with system size $L$. 
These $\sqrt{L}$ fluctuations are, however, not a consequence of uncorrelated addition of random $\ensuremath{\mathcal{O}\!(1)}$
 variations in $M$ at every time step, but rather, of 
 sudden $\ensuremath{\mathcal{O}\!(\sqrt{L})}$ changes occurring sporadically at  $\ensuremath{\mathcal{O}\!(\sqrt{L})}$ time separations. 
The intermittent behaviour of the dynamical structure functions captures this distinction. 
\section{Conclusions}
\label{sec:sec6}
\paragraph*{}
In this paper, we have studied a simple one-dimensional model with diffusion and aggregation of particles in the bulk, in conjunction with influx of single particles at one boundary and 
outflux of aggregates at one or both boundaries. We have analytically calculated various static and dynamical properties, focusing, in particular, on the computation of dynamical 
structure functions, in order to probe turbulence-like behaviour in the system. 
These calculations demonstrate that the system shows giant number fluctuations and temporal intermittency of the total mass in the case where aggregates undergo unbiased diffusion. 
We rationalise these results in terms of the formation and exit of a macroscopic aggregate containing a finite fraction of the total mass. Further, we show that this system is sensitive to
 boundary conditions: subtle differences arise if both influx and outflux are allowed at the same boundary, as opposed to opposite ends of the 1D system. 
We also analyse the case with fully biased movement. In this case, the typical mass of aggregates exiting the system scales as $\sqrt{L}$, which gives rise to
normal, Gaussian fluctuations of the total mass. Nevertheless, the passage of such aggregates through the system is captured by 
dynamical structure functions which still exhibit intermittency, but at a different characteristic time scale.
\paragraph*{}
The analytic approach used in this paper hinges mainly on the closure of the equations for the probability distribution of mass $M_{i,j}$ in a single stretch of the lattice. 
Such closure properties form the basis of many exact calculations for one-dimensional reaction diffusion systems \cite{doering,takayasu,majumdar2}.
The present work extends this sort of analysis to calculate various spatio-temporal correlation functions, from which we then obtain structure functions $S_{n}(t)$ of the total mass 
for integer order $n$. There are, however, several questions that we are not able to address, at least in a simple way, using this approach. One such question
is related to the properties of the macroscopic aggregate in the system. While we infer its presence in the unbiased case from the form of the universal tail of
various probability distributions, we are not able to directly calculate its properties, for instance, by analysing the properties of the largest aggregate in the system. 
Apart from yielding more detailed information about this system, such a line of 
enquiry may also reveal interesting connections with earlier work which relates the properties of macroscopic
 aggregates to extreme value statistics \cite{evans}.
\paragraph*{}
The model studied in this paper is a special limit of a more general model \cite{sachdeva} which also allows for fragmentation of single particles from aggregates.  
In \cite{sachdeva}, it was shown using numerical simulations that for non-zero fragmentation rates smaller than a critical value, the total mass in the system shows giant fluctuations and temporal 
intermittency. This general model is, however, not easy to study analytically, as the equations for $M_{i,j}$ in a single stretch are no longer closed 
in the presence of fragmentation. Analysing the general model and obtaining the critical fragmentation rate, at which the large fluctuations and
intermittency disappear, thus remains an interesting open question.
\paragraph*{}
Another important open question is related to the behaviour of the system in higher dimensions. As discussed in sec. \ref{sec:sec2.3}, aggregation models typically have an upper critical
dimension equal to two, above which the role of fluctuations becomes unimportant, and a mean field description suffices. It would be interesting to investigate 
how the intermittency properties of the present model change with dimension, and whether this model also has the same upper critical dimension. As with other reaction-diffusion models in higher 
dimensions \cite{RG,howard,connaughton}, field theoretic approaches may provide some insight into these questions.
\paragraph*{}
An interesting direction for future study would be to explore whether other systems with cluster or aggregate formation also show temporal intermittency of mass or particle number.
A broader question is whether there is a more general connection between temporal intermittency of particle number and giant number fluctuations, which often arise due to a
 `clustering' tendency in a system. This question could be relevant in many systems with giant fluctuations such as self-propelled particles \cite{ramaswamy,chate} and
 sliding particles on fluctuating interfaces \cite{das}. Dynamical structure functions of particle number and the time dependence of flatness could be useful probes of 
turbulence-like behaviour in these systems as well.
\paragraph*{Acknowledgements}
 We thank R. Dandekar, D. Dhar, and J. Krug for useful discussions.

\appendix
\section{Appendix 1: Asymptotics:}
\label{appendix1}
\subsection{$P(M)$ for large $M$ and large $L$ [eq. \eqref{eq3.1.8}, sec. \ref{sec:sec3.1}]:}
\label{appendix1.1}
The generating function $Q(z)=\sum\limits_{M=0}^{\infty} P(M)z^{M}$ of the probability distribution $P(M)$ in sec. \ref{sec:sec3.1} can be obtained by setting $y=1$ in eq. 
\eqref{eq3.1.7}. $Q(z)$ can be inverted to obtain $P(M)$ as follows:
\begin{equation}
 \begin{split}
Q(z) &= \sech\left[\sqrt{\beta(1-z)}\right] \qquad \text{where} \qquad \beta=\frac{2\tilde{a}L}{D}\\
&=\sum\limits_{n=0}^{\infty} \frac{E_{2n}}{(2n)!}(1-z)^{n}\beta^{n} \qquad \text{where $E_{2n}$ are Euler numbers}\\
& =\sum\limits_{n=0}^{\infty} \frac{E_{2n}}{(2n)!}\beta^{n}\sum\limits_{M=0}^{n} {n\choose M}(-1)^{M}z^{M}\\
& = \sum\limits_{M=0}^{\infty} \sum\limits_{n=M}^{\infty} (-1)^{M}z^{M}\frac{E_{2n}}{(2n)!}\beta^{n}{n\choose M}
\end{split}
\label{app1.1.1}
\end{equation}

Thus, $P(M)$ is given by:
\begin{equation}
 P(M)= (-1)^{M}\sum\limits_{n=M}^{\infty}\frac{E_{2n}}{(2n)!}\beta^{n}{n\choose M}
\label{app1.1.2}
\end{equation}

This is exact so far, but can be further simplified for large $M$ by using the large $n$ form of $E_{2n}$:
\begin{equation}
 E_{2n} \sim 8(-1)^{n}\sqrt{\frac{n}{\pi}}\left(\frac{4n}{\pi e}\right)^{2n} 
\sim (-1)^{n}\frac{4}{\pi} \left(\frac{4}{\pi^{2}}\right)^{n} (2n)! \qquad \text{using Stirling's approximation}
\label{app1.1.3}
\end{equation}

Substituting from eq. \eqref{app1.1.3} into eq. \eqref{app1.1.2}, we get:
\begin{equation}
\begin{split}
P(M)&\sim\frac{4}{\pi}(-1)^{M}\sum\limits_{n=M}^{\infty}{n\choose M}\left(\frac{-4\beta}{\pi^{2}}\right)^{n}\\
&= \frac{4}{\pi}\left(\frac{4\beta}{\pi^{2}}\right)^{M}\sum\limits_{n=0}^{\infty}{n+M\choose n}\left(\frac{-4\beta}{\pi^{2}}\right)^{n}\\
&=\frac{\frac{4}{\pi}\left(\frac{4\beta}{\pi^{2}}\right)^{M}}{\left(1+\frac{4\beta}{\pi^{2}}\right)^{M+1}}\\
& \sim \frac{\pi}{\beta}\left(1-\frac{\pi^{2}}{4\beta}\right)^{M} \qquad \text{for $M\gg 1$ and $\frac{\pi^{2}}{4\beta}\ll 1$}\\
& \sim \frac{\pi}{\beta}\exp\left(-\frac{\pi^{2}M}{4\beta}\right) 
\end{split}
\label{app1.1.4}
\end{equation}
which is the same as eq. \eqref{eq3.1.8}.

\subsection{Small $t$ behaviour of $S_{2}(t)$ from eq. \eqref{eq4.2.2} [sec. \ref{sec:sec4.2}]:}
\label{appendix1.2}
Equation \eqref{eq4.2.2} expresses the structure function $S_{2}(t)$ as the following infinite sum (ignoring the sub-leading terms in $L$):
\begin{equation}
 S_{2}(t)=16\zeta^{2}\sum\limits_{n=1,3,5...}^{\infty} \left\{\frac{1}{(n \pi)^{4}}\left(n\pi\coth\left[\frac{n\pi}{2}\right]-1\right)\right\}
\left\{1- \exp\left[-\frac{D\pi^{2} n^{2}t}{L^{2}}\right]\right\} \qquad \zeta=aL/D
\label{app1.2.1}
\end{equation}
To extract the small $t$ ($Dt/L^{2}\ll1$) behaviour of this expression, we approximate the sum over $n$  by an integral, and then take the $Dt/L^{2}\rightarrow 0$ limit 
of this integral, so that:
\begin{equation}
\begin{split}
 S_{2}(t) & \sim 16\zeta^{2}\int\limits_{0}^{\infty}\frac{1-e^{-\pi^{2}(2l+1)^{2}\tau}}{\pi^{3}(2l+1)^{3}} \vast\{\coth \left[\frac{\pi}{2}(2l+1)\right]
-\frac{1}{\pi(2l+1)}\vast\}dl
  \qquad \text{where} \quad \tau=\frac{Dt}{L^{2}} \\
& = \frac{8\zeta^{2}\tau}{\pi}\int\limits_{\pi\sqrt{\tau}}^{\infty}\left[\frac{1-e^{-x^{2}}}{x^{3}}\right]\vast\{ \coth \left[\frac{x}{2\sqrt{\tau}}\right]-
\frac{\sqrt{\tau}}{x}\vast\}dx  \qquad x=(2l+1)\pi\sqrt{\tau}
\end{split}
\label{app1.2.2}
\end{equation}
In the limit $\tau\rightarrow 0$, we have $\coth \left[x/2\sqrt{\tau}\right]\sim 1$, so that $S_{2}(t)$ becomes:
\begin{equation}
\begin{split}
S_{2}(t) & \sim \frac{8\zeta^{2}\tau}{\pi}\int\limits_{\pi\sqrt{\tau}}^{\infty}\left[\frac{1-e^{-x^{2}}}{x^{3}}\right]\left[1-\frac{\sqrt{\tau}}{x}\right]dx\\
& = \frac{8\zeta^{2}\tau}{\pi}\vast\{\left[\frac{1}{2\pi^{2}\tau}-\frac{e^{-\pi^{2}\tau}}{2\pi^{2}\tau}-\frac{1}{2}Ei[-\pi^{2}\tau]\right]+
\left[ \frac{e^{-\pi^{2}\tau}(1-2\pi^{2}\tau)-(1-2\pi^{7/2}\tau^{3/2}\Erfc[\pi \sqrt{\tau}])}{3\pi^{3}\tau}\right]\vast\}\\
&\qquad \text{where} \quad Ei[-\pi^{2}\tau]=-\int\limits_{\pi^{2}\tau}^{\infty} \frac{e^{-u}}{u}du \quad \text{and}\quad 
\Erfc[\pi \sqrt{\tau}]=(2/\sqrt{\pi})\int\limits_{\pi\sqrt{\tau}}^{\infty} e^{-u^{2}} du\\
\end{split}
\label{app1.2.3}
\end{equation}
In the limit $\tau \rightarrow 0$, the complementary error function behaves as $\Erfc[\pi\sqrt{\tau}]\sim 1-2\sqrt{\pi \tau}$ and the exponential integral 
 has the asymptotic form $Ei[-\pi^{2}\tau]\sim \gamma +\log[\pi^{2}\tau]$  where $\gamma$ is the Euler-Mascheroni constant. Thus, as $\tau \rightarrow 0$,
the expression in eq. \eqref{app1.2.3} tends to:
\begin{equation}
 S_{2}(t) \sim \frac{4\zeta^{2}\tau}{\pi}\left[1-\gamma-\frac{2}{\pi}-\log[\pi^{2}\tau]\right]
\label{app1.2.4}
\end{equation}
This expression is not in very good agreement with numerics because of the significant corrections that appear while approximating the  discrete sum with an integral. 
By taking these correction terms into account using the Euler-Maclaurin formula, better agreement with numerics is obtained.
These correction terms basically modify the constants inside the square bracket in eq. \eqref{app1.2.4}, so that the small $t$ form of $S_{2}(t)$ is still given by:  
\begin{equation}
 S_{2}(t) \sim \frac{4\zeta^{2}\tau}{\pi}(A_{0}-\log[\tau])= -\frac{4}{\pi}\left(\frac{aL}{D}\right)^{2}\frac{Dt}{L^{2}}\log\left[A_{1}\frac{Dt}{L^{2}}\right] 
\label{app1.2.5}
\end{equation}
where the simplest way of obtaining the constant $A_{1}$ is  by fitting to numerical data.

\subsection{Asymptotic expressions for $\langle M_{i,j}^{n}\rangle$ [eq. \eqref{eq5.1.15}, sec. \ref{sec:sec5.1}]:}
\label{appendix1.3}
We start with the generating function $G_{u,j}(z)$ in eq. \eqref{eq5.1.13},
\begin{equation}
\begin{split}
 G_{u,j}(z)&=\sum\limits_{i=1}^{\infty}F_{i,j}(z)u^{i}=\sum\limits_{i=1}^{\infty}\left[\sum\limits_{M=0}^{\infty}P_{i,j}(M)z^{M}\right]u^{i}\\
&=\left(\frac{u}{1-u}\right)\left(\frac{\eta^{2}(1-z)^{2}}{\eta^{2}(1-z)^{2}-(1-u)}\right)
\left[[1+\sqrt{1-u}]^{-j}-\left(\frac{1-u}{\eta^{2}(1-z)^{2}}\right)[1+\eta(1-z)]^{-j}\right]
\end{split}
\label{apeq1}
\end{equation} 
The generating function $\sum\limits_{i=1}^{\infty}\langle M_{i,j}^{2}\rangle u^{i}$ 
can be obtained by differentiating the above expression w.r.t. $z$ and then setting $z=1$
\begin{equation}
\sum\limits_{i=1}^{\infty}\langle M_{i,j}^{2}\rangle u^{i}=\frac{\eta^{2}u}{(1-u)^{2}}\left[2-2\{1+\sqrt{1-u}\}^{-j}+j(1+j)(1-u)\right]+\frac{\eta u j}{1-u}
\label{apeq2}
\end{equation}
Since we are interested in $\langle M_{i,j}^{2}\rangle$ in the limit $i\rightarrow \infty$, we consider the  $u\rightarrow 1$ limit of the above equation.  
This can be obtained by Taylor expanding eq. \eqref{apeq2} in powers of the small parameter $1-u$ and retaining only the first few terms 
(terms that become asymptotically large) in $1-u$. This gives:
\begin{equation}
\sum\limits_{i=1}^{\infty}\langle M_{i,j}^{2}\rangle u^{i}\sim  u\left[\frac{2\eta^{2}j}{(1-u)^{3/2}} + \frac{\eta j}{1-u} + \frac{\eta^{2}j(j^{2}+3j+2)}{3(1-u)^{1/2}}+...\right]
\label{apeq3}
\end{equation}
Each of the above terms can be now Taylor expanded about $u=0$ to give:
\begin{equation}
\langle M_{i,j}^{2}\rangle \sim 2\eta^{2}j(2i)\left(\frac{(2i)!}{2^{2i} (i!)^{2}}\right)+\eta j+\frac{\eta^{2}j(j^{2}+3j+2)}{3}\left(\frac{2i}{2i-1}\right)\left(\frac{(2i)!}{2^{2i} (i!)^{2}}
\right)+..
\label{apeq4}
\end{equation}
By taking the limit $i\rightarrow \infty$ and using Stirling's approximation for $i!$, eq. \eqref{apeq4} becomes:
\begin{equation}
\langle M_{i,j}^{2}\rangle \sim \frac{4\eta^{2}j\sqrt{i}}{\sqrt{\pi}}+ \eta j +\frac{\eta^{2}j(j^{2}+3j+2)}{3\sqrt{\pi}\sqrt{i}}+....
\label{apeq5}
\end{equation}
The same procedure can be followed to compute $\langle M_{i,j}^{3}\rangle$ and $\langle M_{i,j}^{4}\rangle$ in the $i\rightarrow \infty$ limit: first,
 $\sum\limits_{i=1}^{\infty}\langle M_{i,j}^{3}\rangle u^{i}$ and $\sum\limits_{i=1}^{\infty}\langle M_{i,j}^{4}\rangle u^{i}$ are expanded in powers of $1-u$ and next, 
each of the $(1-u)^\alpha$ terms in this expansion is further expanded about $u=0$. Finally, by using Stirling approximation for $i!$ etc. in the $i\rightarrow \infty$ 
limit, we get:
\begin{subequations}
\begin{equation}
\langle M_{i,j}^{3}\rangle\sim 6\eta^{3}ji+\frac{12\eta^{2}}{\sqrt{\pi}}j\sqrt{i}+\eta j+\eta^{3}j(j^{2}+3j+2)+ \frac{\eta^{2}j(j^{2}+3j+2)}{\sqrt{\pi}\sqrt{i}}+...
\label{apeq7a}
\end{equation}
\begin{equation}
 \langle M_{i,j}^{4}\rangle\sim \frac{32\eta^{4}}{\sqrt{\pi}}(ji^{3/2})+36\eta^{3}(ji) + 
\frac{28\eta^{2} j+8\eta^{4}j(j^{2}+3j+2)}{\sqrt{\pi}}\sqrt{i} + 6\eta^{3}j(j^{2}+3j+2)+\eta j+...
\label{apeq7b}
\end{equation}
\label{apeq7}
\end{subequations}
The expressions for $\langle M_{i,j}^{n}\rangle$ simplify further in the limit $j\rightarrow \infty$, $i\rightarrow \infty$ with $j/\sqrt{i}$ finite. In this limit, 
terms that are $\ensuremath{\mathcal{O}\!(j/i)}$ etc. go to zero, so that $\langle M_{i,j}^{n}\rangle$ are given by:
\begin{subequations}
\begin{equation}
\langle M_{i,j}^{2}\rangle \sim i \ensuremath{\mathcal{R}}_{2}\left(\frac{j}{\sqrt{i}}\right) \sim \frac{4\eta^{2}i}{\sqrt{\pi}}\left[\left(\frac{j}{\sqrt{i}}\right)+\frac{1}{12}\left(\frac{j}{\sqrt{i}}\right)^{3}+....\right]
\label{apeq8a}
\end{equation}
\begin{equation}
\langle M_{i,j}^{3}\rangle \sim i^{3/2} \ensuremath{\mathcal{R}}_{3}\left(\frac{j}{\sqrt{i}}\right) \sim 6\eta^{3}i^{3/2}\left[\left(\frac{j}{\sqrt{i}}\right)+\frac{1}{6}\left(\frac{j}{\sqrt{i}}\right)^{3}+....\right]
\label{apeq8b}
\end{equation}
\begin{equation}
\langle M_{i,j}^{4}\rangle \sim i^{2} \ensuremath{\mathcal{R}}_{4}\left(\frac{j}{\sqrt{i}}\right) \sim \frac{32\eta^{4}{i^{2}}}{\sqrt{\pi}}\left[\left(\frac{j}{\sqrt{i}}\right)+\frac{1}{4}\left(\frac{j}{\sqrt{i}}\right)^{3}+....\right]
\label{apeq8c}
\end{equation}
\label{apeq8}
\end{subequations}

\section{Appendix 2: Solution of various partial differential equations:}
\label{appendix2}
\subsection{Laplace equation on a right isosceles triangle [eqs. \eqref{eq3.1.9} and \eqref{eq4.1.3}]:}
\label{appendix2.1}
\begin{figure}[h]
\centering
\includegraphics[width=0.6\textwidth]{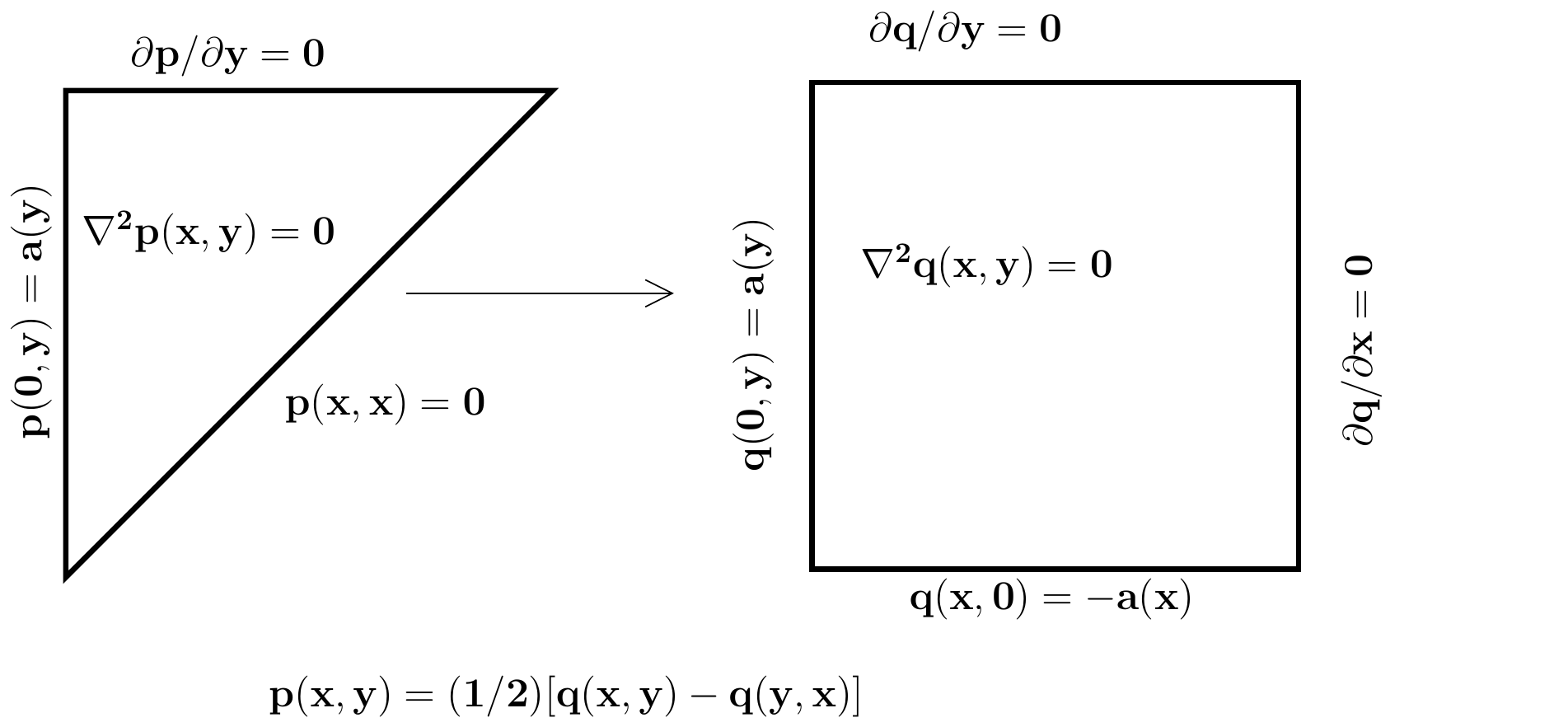}
\caption{Folding transformation used to solve Laplace equation on the right isosceles triangle}
\label{figapp}
\end{figure}

Consider a function $p(x,y)$ which:
\begin{enumerate}
\renewcommand{\theenumi}{\roman{enumi}}
\item satisfies the Laplace equation on a right isosceles triangle with vertices $(0,0)$, $(0,1)$ and $(1,0)$. 
\item is equal to zero uniformly on the hypotenuse.
\footnote{If the function is equal to some constant $c$ on the hypotenuse, then we can define a new function $p(x,y)-c$ which satisfies all the three conditions (i)-(iii)
and can be solved for using the method described subsequently.} 
\item satisfies some specified boundary conditions [mixed B.C. for eq. \eqref{eq3.1.9} and Neumann B.C. for eq. \eqref{eq4.1.3}] on the other two sides of the triangle.
\end{enumerate}
\paragraph*{}
The key step in solving this equation is the folding transformation introduced in \cite{prager} for the Laplace equation on the equilateral triangle, and adapted to 
the right isosceles triangle in \cite{damle}. This transformation, as applied to eqs. \eqref{eq3.1.9} and \eqref{eq4.1.3}, is described below and also shown schematically
 in fig. \ref{figapp}.
\paragraph*{}
We first consider the case, where the boundary conditions at $x=0$ and $y=1$ are of the mixed kind, as in eq. \eqref{eq3.1.9}:
\begin{subequations}
 \begin{equation}
 \frac{\partial^{2}{p(x,y)}}{\partial{x^{2}}}+\frac{\partial^{2}{p(x,y)}}{\partial{y^{2}}}=0 , \qquad 0\leq x\leq y\leq 1
\end{equation}
\begin{equation}
 p(x,y=x)=0,  \qquad  \qquad \left.\frac{\partial{p}}{\partial{y}}\right|_{y=1}=0, \qquad \qquad p(0,y)= a(y)
\label{app2.1.1b}
\end{equation}
\label{app2.1.1}
\end{subequations}
Consider another function $q(x,y)$ on the square [$0\leq x\leq 1$, $0\leq y\leq 1$] which satisfies:
\begin{subequations}
 \begin{equation}
 \frac{\partial^{2}{q(x,y)}}{\partial{x^{2}}}+\frac{\partial^{2}{q(x,y)}}{\partial{y^{2}}}=0
\end{equation}
\begin{equation}
q(0,y)= a(y) \qquad q(x,0)= -a(x) \qquad \left.\frac{\partial{q}}{\partial{y}}\right|_{y=1}= \left.\frac{\partial{q}}{\partial{x}}\right|_{x=1}=0
\end{equation}
\label{app2.1.2}
\end{subequations} 

Then, it can be seen that the function $w(x,y)=\frac{1}{2}[q(x,y)-q(y,x)]$ is the required solution of eq. \eqref{app2.1.1} in the triangular region as:
\begin{subequations}
 \begin{equation}
 \nabla^{2}w(x,y)=\frac{1}{2}[\nabla^{2}q(x,y)-\nabla^{2}q(y,x)]=0
\end{equation}
\begin{equation}
\begin{split}
&w(x,y=x)=\frac{1}{2}[q(x,x)-q(x,x)]=0\\ 
&\left.\frac{\partial{w}}{\partial{y}}\right|_{y=1}=\frac{1}{2}\left.\left(\frac{\partial{q(x,y)}}{\partial{y}}-\frac{\partial{q(y,x)}}{\partial{y}}\right)\right|_{y=1}
=\frac{1}{2}\left(\left.\frac{\partial{q(x,y)}}{\partial{y}}\right|_{y=1}-\left.\frac{\partial{q(x,y)}}{\partial{x}}\right|_{x=1}\right)=0 \\
&w(0,y)= \frac{1}{2}[q(0,y)-q(y,0)]=\frac{1}{2}[a(y)-(-a(y))]=a(y)
\end{split}
\end{equation}
\label{app2.1.3}
\end{subequations}

\paragraph*{}
Equation \eqref{app2.1.2} can be solved by a standard application of the superposition method \cite{haberman} i.e.  by decomposing $q(x,y)$ as 
$q(x,y)=u(x,y)+v(x,y)$, such that:
\begin{subequations}
 \begin{equation}
  \nabla^{2} u(x,y)=0 \qquad \left.\frac{\partial{u}}{\partial{y}}\right|_{y=1}= \left.\frac{\partial{u}}{\partial{x}}\right|_{x=1}=0 \qquad u(x,0)=0 \qquad u(0,y)=a(y)
 \end{equation}
\begin{equation}
  \nabla^{2} v(x,y)=0 \qquad \left.\frac{\partial{v}}{\partial{y}}\right|_{y=1}= \left.\frac{\partial{v}}{\partial{x}}\right|_{x=1}=0 \qquad v(x,0)=-a(x) \qquad v(0,y)=0
 \end{equation}
\end{subequations}
Each of the functions $u(x,y)$ and $v(x,y)$ can be solved for by separation of variables. Their sum gives the function $q(x,y)$, which in turn gives $p(x,y)$. 

\paragraph*{}
The case where both the $x=0$ and $y=1$ boundary conditions are of the Neumann kind [as in eq. \eqref{eq4.1.3}] can be dealt with similarly using the folding transformation.
Suppose, the boundary condition at $x=0$ in eq. \eqref{app2.1.1b} is given by $\left.\frac{\partial{p}}{\partial{x}}\right|_{x=0}=b(y)$, then 
 the corresponding Laplace equation on the square that needs to be solved is:  
\begin{equation}
\nabla^{2} q(x,y)=0, \qquad \left.\frac{\partial{q}}{\partial{y}}\right|_{y=1}= \left.\frac{\partial{q}}{\partial{x}}\right|_{x=1}=0,
\qquad \left.\frac{\partial{q}}{\partial{x}}\right|_{x=0}=b(y), \qquad 
\left.\frac{\partial{q}}{\partial{y}}\right|_{y=0}=-b(x)
\label{app2.1.5}
\end{equation}
\paragraph*{}
Note that \emph{all} the boundary conditions for this equation are of the Neumann kind. The Laplace equation with Neumann boundary conditions has
 a solution only if $\int\limits_0^1 dy\left.\frac{\partial{q}}{\partial{x}}\right|_{x=1} - \int\limits_0^1 dy\left.\frac{\partial{q}}{\partial{x}}\right|_{x=0} 
+\int\limits_0^1 dx\left.\frac{\partial{q}}{\partial{y}}\right|_{y=1} -\int\limits_0^1 dx\left.\frac{\partial{q}}{\partial{y}}\right|_{y=0} =0$ i.e. there is no net flux 
through the boundaries \cite{haberman,pinchover}. This is simply because the solution of the Laplace equation is the steady state solution of a heat equation and for a 
steady state solution to exist, the net heat flux through the 
boundaries must be zero. Thus, while decomposing $q(x,y)$ as $q(x,y)=u(x,y)+v(x,y)$, both $u(x,y)$ and $v(x,y)$ must 
individually satisfy the condition of no net flux through the boundaries. This can be done \cite{pinchover} by defining a new function:
\begin{equation}
s(x,y)=q(x,y)+(K/2)\left[(1-x)^{2}-(1-y)^{2}\right] 
\end{equation}
 The function $s(x,y)$ now satisfies:\\
\begin{equation}
 \nabla^{2}s(x,y)=0, \qquad
\left.\frac{\partial{s}}{\partial{y}}\right|_{y=1}= \left.\frac{\partial{s}}{\partial{x}}\right|_{x=1}=0, \qquad \left.\frac{\partial{s}}{\partial{x}}\right|_{x=0}=b(y)-K,
\qquad \left.\frac{\partial{s}}{\partial{y}}\right|_{y=0}=-b(x)+K
\end{equation}
Choosing $K=\int\limits_0^1 b(y)dy$ ensures that the net flux of $s(x,y)$ through each side of the square is zero.
Then $s(x,y)$ can be decomposed as $s(x,y)=u(x,y)+v(x,y)$ such that:
\begin{subequations}
\begin{equation}
\nabla^{2} u(x,y)=0, \qquad \left.\frac{\partial{u}}{\partial{y}}\right|_{y=1}= \left.\frac{\partial{u}}{\partial{x}}\right|_{x=1}=\left.\frac{\partial{u}}{\partial{y}}\right|_{y=0}=0,
\qquad \left.\frac{\partial{u}}{\partial{x}}\right|_{x=0}=b(y)-K
\label{app2.1.6a}
\end{equation}
\begin{equation}
\nabla^{2} v(x,y)=0, \qquad \left.\frac{\partial{v}}{\partial{y}}\right|_{y=1}= \left.\frac{\partial{v}}{\partial{x}}\right|_{x=1}=\left.\frac{\partial{v}}{\partial{x}}\right|_{x=0}=0,
\qquad \left.\frac{\partial{v}}{\partial{y}}\right|_{y=0}=-b(x)+K
\label{app2.1.6b}
\end{equation}
\label{app2.1.6}
\end{subequations}
The functions $u(x,y)$ and $v(x,y)$ both satisfy the condition of zero net flux at the boundaries. Thus, solutions to eqs. \eqref{app2.1.6a} and \eqref{app2.1.6b} exist 
and can be found by separation of variables.
\paragraph*{}
The folding transformation described above, can also be used to solve the 3D Laplace equation on the region $0\leq x\leq y \leq z\leq 1$, which comes up in the calculation 
of multi-sector moments such as $\langle M^{2}_{xy}M_{yz}\rangle$ [see sec. \ref{sec:sec4.1}]. For example, suppose $g(x,y,z)$ satisfies the Laplace 
equation on the region $0\leq x\leq y \leq z\leq 1$ with the boundary conditions $g(x,x,z)=g(x,y,y)=0$ and some specified Dirichlet or Neumann boundary conditions on the
 other two surfaces. Then, we can find $g(x,y,z)$ by solving for a function $h(x,y,z)$ which satisfies the Laplace equation inside a cube 
[$0\leq x \leq 1$, $0\leq y \leq 1$, $0\leq z \leq 1$] and appropriately chosen boundary conditions on the six faces. If $h(x,y,z)$ is known, 
then $g(x,y,z)$ can be obtained as the antisymmetric combination:\\
\begin{center}
\vspace{-0.2cm}
 $g(x,y,z)=(1/6)[h(x,y,z)-h(y,x,z)+h(y,z,x)-h(z,y,x)+h(z,x,y)-h(x,z,y)]$
\end{center}
This procedure can be generalised to solve Laplace equations on higher dimensional regions as well.

\subsection{Inhomogeneous heat equation in 1 spatial dimension [eqs. \eqref{eq3.2.7} and \eqref{eq3.2.8}]}
\label{appendix2.2}
Equations \eqref{eq3.2.7} and \eqref{eq3.2.8} constitute a set of equations of the type:
\begin{equation}
 \begin{split}
&\frac{\partial{p(x,t)}}{\partial{t}}=\gamma\frac{\partial^{2}{p(x,t)}}{\partial{x^{2}}}+a(x,t)\\
&p(0,t)=0, \qquad \left.\frac{\partial{p}}{\partial{x}}\right|_{x=1}=0, \qquad p(x,0)=b(x)
\end{split}
\label{app2.3.1}
\end{equation}
The inhomogeneous heat equation can be solved \cite{haberman} by expressing $p(x,t)$ as the sum of the complementary and particular solutions. 
The complementary solution, which is the solution of the homogeneous equation corresponding to eq. \eqref{app2.3.1}, has a variable separable form given by:
\begin{equation}
p_{c}(x,t)=\sum\limits_{n=0}^{\infty}u_{n}\exp(-\gamma \alpha_{n}^{2}t)\sin[\alpha_{n}x] \qquad \text{where $\alpha_{n}=\left(n+\frac{1}{2}\right)\pi$ }
\label{app2.3.2}
\end{equation}
By assuming the particular solution to be of the form $\sum\limits_{n=0}^{\infty}B_{n}(t)\sin[\alpha_{n}x]$, the general solution
 can be expressed as $p(x,t)=\sum\limits_{n=0}^{\infty}[B_{n}(t)+u_{n}\exp(-\gamma \alpha_{n}^{2}t)]\sin[\alpha_{n}x]
=\sum\limits_{n=0}^{\infty}C_{n}(t)\sin[\alpha_{n}x]$. The inhomogeneous source term $a(x,t)$ in eq. \eqref{app2.3.1} can also be written in the same eigenbasis as 
$a(x,t)=\sum\limits_{n=0}^{\infty}A_{n}(t)\sin[\alpha_{n}x]$.
Then, it follows that $C_{n}(t)$ satisfies:
\begin{equation}
\dot{C}_{n}(t)=-\gamma \alpha_{n}^{2}C_{n}(t)+A_{n}(t)
\label{app2.3.3}
\end{equation}
This can be solved to give $C_{n}(t)= C_{n}(0)\exp[-\gamma \alpha_{n}^{2}t]+\int\limits_0^t A_{n}(t')\exp[-\gamma \alpha_{n}^{2}(t-t')]dt'$ where $C_{n}(0)$ can be obtained from the
initial condition $p(x,0)=\sum\limits_{n=0}^\infty C_{n}(0)\sin[\alpha_{n}x]=b(x)$.

\subsection{Heat equation in 2 spatial dimensions on a right isosceles triangle with Neumann boundary conditions [eq. \eqref{eq4.2.1} and eq. \eqref{eq4.2.3}]:}
\label{appendix2.3}
Consider a function $p(x,y,t)$ which satisfies the heat equation on the triangular region $0\leq x\leq y\leq 1$:
\begin{subequations}
 \begin{equation}
 \frac{\partial{p(x,y,t)}}{\partial{t}}=\gamma\left[\frac{\partial^{2}{p(x,y,t)}}{\partial{x^{2}}}+\frac{\partial^{2}{p(x,y,t)}}{\partial{y^{2}}}\right]
\end{equation}
\begin{equation}
 p(x,y=x,t)=0, \qquad \left.\frac{\partial{p}}{\partial{y}}\right|_{y=1}=0, \qquad \left.\frac{\partial{p}}{\partial{x}}\right|_{x=0}=a(y,t) \qquad p(x,y,0)=b(x,y)
\end{equation}
\label{app2.4.1}
\end{subequations}

As in the case of the Laplace equation, this can be solved by first solving the corresponding heat equation on a square:
\begin{subequations}
 \begin{equation}
 \frac{\partial{q(x,y,t)}}{\partial{t}}=\gamma\left[\frac{\partial^{2}{q(x,y,t)}}{\partial{x^{2}}}+\frac{\partial^{2}{q(x,y,t)}}{\partial{y^{2}}}\right]
\end{equation}
\begin{equation}
\left.\frac{\partial{q}}{\partial{y}}\right|_{y=1}=\left.\frac{\partial{q}}{\partial{x}}\right|_{x=1}=0, \qquad 
\left.\frac{\partial{q}}{\partial{x}}\right|_{x=0}=a(y,t)\qquad \left.\frac{\partial{q}}{\partial{y}}\right|_{y=0}=-a(x,t)
\end{equation}
\label{app2.4.2}
\end{subequations}
where the initial condition at $t=0$ is unspecified. If we can solve eq. \eqref{app2.4.1} (upto undetermined constant coefficients corresponding to the unknown initial condition),
then the function $p(x,y,t)$ can be obtained from this solution using $p(x,y,t)=\frac{1}{2}[q(x,y,t)-q(y,x,t)]$. The constant coefficients can now be determined from the 
initial condition $p(x,y,0)=b(x,y)$ by using appropriate Fourier transforms.
\paragraph*{}
If $a(x,t)=0$ [as in eq. \eqref{eq4.2.1}], then the boundary conditions are homogeneous and eq. \eqref{app2.4.2} can be solved simply by 
separation of variables. If however, $a(x,t)$ is non-zero [as in eq. \eqref{eq4.2.3}], then an additional transformation is required. 
This involves defining a new function:
\begin{equation}
 s(x,y,t)=q(x,y,t)+\frac{(1-x)^{2}}{2}a(y,t)-\frac{(1-y)^{2}}{2}a(x,t)
\label{app2.4.3}
\end{equation}
It can be checked that the function $s(x,y,t)$ now satisfies an inhomogeneous heat equation (with a source term) and homogeneous boundary conditions:
\begin{subequations}
\begin{equation}
\begin{split}
 \frac{\partial{s(x,y,t)}}{\partial{t}}=&\gamma\left[\frac{\partial^{2}{s(x,y,t)}}{\partial{x^{2}}}+\frac{\partial^{2}{s(x,y,t)}}{\partial{y^{2}}}\right]\\
& +\frac{(1-x)^{2}}{2}\left[\frac{\partial{a(y,t)}}{\partial{t}}-\gamma\frac{\partial^{2}{a(y,t)}}{\partial{y^{2}}}\right]   
-\frac{(1-y)^{2}}{2}\left[\frac{\partial{a(x,t)}}{\partial{t}}-\gamma\frac{\partial^{2}{a(x,t)}}{\partial{x^{2}}}\right]+\gamma[a(x,t)-a(y,t)]    
\end{split} 
\label{app2.4.4a}
\end{equation}
\begin{equation}
\left.\frac{\partial{s}}{\partial{y}}\right|_{y=1}=\left.\frac{\partial{s}}{\partial{x}}\right|_{x=1}=
\left.\frac{\partial{s}}{\partial{y}}\right|_{y=0}=\left.\frac{\partial{s}}{\partial{x}}\right|_{x=0}=0
\label{app2.4.4b}
\end{equation}
\label{app2.4.4}
\end{subequations}
The transformation in eq. \eqref{app2.4.3} is not a general prescription for solving eq. \eqref{app2.4.2} for an arbitrary function $a(x,t)$. It works only when 
 $\left.\frac{\partial{a(x,t)}}{\partial{x}}\right|_{x=1}=\left.\frac{\partial{a(x,t)}}{\partial{x}}\right|_{x=0}=0$, which is the case for eq. \eqref{eq4.2.3}.
\paragraph*{}
Equation \eqref{app2.4.4} is the heat equation with an inhomogeneous term. It can be solved, as in the 1D case, by first obtaining the complementary solution of the corresponding
homogeneous equation and then obtaining the particular solution by appropriate Fourier transforms of the inhomogeneous source terms.


\begin{thebibliography}{}
\bibitem{dey}  Dey, S., Das, D., Rajesh, R.: Phys. Rev. Lett \textbf{108}, 238001 (2012).
\bibitem{chate} Chat\'{e}, H., Ginelli, F., Gr\'{e}goire, G., Raynaud, F.: Phys. Rev. E \textbf{77}, 046113 (2008).
\bibitem{ramaswamy} Ramaswamy, S., Simha, R.A., Toner, J.: Europhys. Lett. \textbf{62}, 196 (2003).
\bibitem{narayan} Narayan, V., Ramaswamy, S., Menon, N., Science 317, 105 (2007).
\bibitem{goldhirsch} Goldhirsch, I., Zanetti, G.: Phys. Rev. Lett. \textbf{70}, 1619 (1993). 
\bibitem{das} Das, D., Barma, M.: Phys. Rev. Lett. \textbf{85}, 1602 (2000); Nagar, A., Barma, M., Majumdar, S.N.: Phys. Rev. Lett. \textbf{94}, 240601 (2005).
\bibitem{bec} Bec, J., Khanin, K.:  Phys. Rep. \textbf{447}, 1 (2007).
\bibitem{frisch} Frisch, U.: Turbulence: The Legacy of A. N. Kolmogorov, (Cambridge Univ. Press, Cambridge, 1995).
\bibitem{majumdar2} Majumdar, S.N., Sire, C.: Phys. Rev. Lett. \textbf{71}, 3729 ͑(1993͒).
\bibitem{rajesh} Rajesh, R., Majumdar, S.N.: Phys. Rev. E \textbf{62}, 3186 (2000).
\bibitem{kang} Kang, K., Redner, S.: Phys. Rev. A \textbf{30}, 2833 (1984). 
\bibitem{takayasu} Takayasu, H.: Phys. Rev. Lett. \textbf{63}, 2563 (1989); Takayasu, H., Nishikawa, I., Tasaki, H.: Phys. Rev. A \textbf{37}, 3110 (1988).
\bibitem{connaughton} Connaughton, C., Rajesh, R., Zaboronski, O.: Phys. Rev. Lett. \textbf{94}, 194503 (2005); Physica D \textbf{222}, 97 (2006).
\bibitem{sachdeva} Sachdeva, H., Barma, M., Rao, M.: Phys. Rev. Lett. \textbf{110}, 150601 (2013).
\bibitem{leyvraz} Leyvraz, F.: Phys. Rep. \textbf{383}, 95 (2003).
\bibitem{doering} Doering, C.R., ben-Avraham, D.: Phys. Rev. Lett. \textbf{62}, 2563 (1989) 
\bibitem{cheng} Cheng, Z., Redner, S., Leyvraz, F.: Phys. Rev. Lett. \textbf{62}, 2321 (1989).
\bibitem{derrida} Derrida, B.,  Hakim, V., Pasquier, V.: Phys. Rev. Lett. \textbf{75}, 751 (1995); Derrida, B.: J. Phys. A \textbf{28}:1481 (1995). 
\bibitem{hinrichsen} Hinrichsen, H., Rittenberg, V., Simon, H.: J. Stat. Phys. \textbf{86}, 1203 (1997).
\bibitem{howard} Howard, M., Godr\`{e}che, C.: J. Phys. A \textbf{31}, L209 (1998). 
\bibitem{spouge}  Spouge, J.L.: Phys. Rev. Lett. \textbf{60}, 871 (1988); J. Phys. A \textbf{21}, 4183 (1988). 
\bibitem{doering2} Doering, C.R., ben-Avraham, D.: Phys. Rev. A \textbf{38}, 3035 (1988).
\bibitem{majumdar1}  Majumdar, S.N., Krishnamurthy, S., Barma, M.: Phys. Rev. Lett. \textbf{81}, 3691 (1998).
\bibitem{majumdar3} Majumdar, S.N., Krishnamurthy, S., Barma, M.: Phys. Rev. E \textbf{61}, 6337 (2000).
\bibitem{jain} Jain, K., Barma, M: Phys. Rev. E \textbf{64}, 016107 (2001).
\bibitem{reuveni} Reuveni, S., Eliazar, I., Yechiali, U.: Phys. Rev. E \textbf{84}, 041101 (2011); Phys. Rev. Lett. \textbf{109}, 020603 (2012); 
Reuveni, S., Hirschberg, O., Eliazar, I.,  Yechiali, U.: arxiv:1309.2894v1.
\bibitem{racz} R\'{a}cz, Z.: Phys. Rev. Lett. \textbf{55}, 1707 (1985).
\bibitem{vicsek} Vicsek, T., Meakin, P., Family, F.: Phys. Rev. A \textbf{32}, 1122 (1985)
\bibitem{ball} Ball, R.C., Connaughton, C., Jones, P.P.,  Rajesh, R., Zaboronski, O.: Phys. Rev. Lett. \textbf{109}, 168304 (2012)
\bibitem{majumdar_rev} Majumdar, S.N.: Les Houches (2008) lecture notes, arXiv:0904:4097
\bibitem{evans_rev} Evans, M.R., Hanney, T.: J. Phys. A \textbf{38}, R195 (2005).
\bibitem{RG} Droz, M., Sasv\'{a}ri, L., Phys. Rev. E \textbf{48} R2343 (1993); Peliti, L., J. Phys. A \textbf{19}, L365 (1986); Zaboronski, O.: Phys. Lett. A \textbf{281}, 119 (2001) 
\bibitem{damle} Damle, A., Peterson, G.C.: SIAM Undergraduate Research Online, Volume 3, Issue 1, http://www.siam.org/students/siuro/vol3/S01061.pdf, 2010, pp. 187-208.
\bibitem{prager} Pr\'{a}ger, M.: Appl. Math. \textbf{43}(4), 311 (1998).
\bibitem{haberman} Haberman, R.: Applied partial differential equations: with Fourier series and boundary value problems, 
(Pearson Prentice Hall, 2004).
\bibitem{evans} Evans, M.R., Majumdar, S.N.: J. Stat. Mech: Theory Exp. \textbf{P05004} (2008).
\bibitem{pinchover} Pinchover, Y., Rubinstein, J.: An Introduction to Partial Differential Equations, (Cambridge Univ. Press, Cambridge, 2005).
\end{thebibliography}
\end{document}